\documentclass{ws-rv9x6}
\usepackage{subfigure}   
\usepackage{ws-rv-thm}   
\usepackage{ws-rv-van}   
\makeindex
\DeclareMathAlphabet{\mathitbf}{T1}{cmr}{bx}{it} 

\begin{document}

\chapter[Nature of the Spin Glass Phase in Finite
Dimensional (Ising) Spin Glasses]{Nature of the Spin Glass Phase in Finite
Dimensional (Ising) Spin Glasses }\label{ra_ch1}

\author[J.J. Ruiz-Lorenzo]{Juan J. Ruiz-Lorenzo\footnote{ruiz@unex.es}}

\address{Departamento de F\'isica,\\ Universidad de Extremadura, 06006
  Badajoz, Spain}
\address{Instituto de Computaci\'on Cient\'ifica Avanzada de Extremadura (ICCAEx),\\ Universidad de Extremadura, 06006 Badajoz, Spain}
\address{Instituto de Biocomputaci\'on y F\'{\i}sica de los Sistemas Complejos (BIFI),\\ 50018 Zaragoza, Spain}

\begin{abstract}
Spin glasses are the paradigm of complex systems. These materials
present really slow dynamics. However, the nature of the spin glass
phase in finite dimensional systems is still controversial.  Different
theories describing the low temperature phase have been proposed:
droplet, replica symmetry breaking and chaotic pairs.
We present analytical studies of critical properties of spin glasses, in particular, 
critical exponents at and below the phase
transition, existence of a phase transition in a magnetic field, 
computation of the lower critical dimension (in presence/absence of a magnetic field).
We also introduce some rigorous results based on the concept of metastate.
Finally, we report some numerical results regarding the
construction of the Aizenman-Wehr metastate, scaling of the
correlation functions in the spin glass phase and existence of a
phase transition in a field, confronting these results with
the predictions of  different theories.
  
\end{abstract}
\body

\tableofcontents

\section{~Introduction}
\label{sec:intro}

Spin glasses are often considered as the paradigm of complex
systems.\index{complex system} They show frustration and randomness\index{frustration} which are now ubiquitous
characteristics in nature. In addition, the role played by spin
glasses in magnetism is huge: spin glass behavior jointly with
ferrromagnetism and antiferromagnetism are the three most frequent
forms of ``magnetic order''. On the theoretical side, radical
approaches to describe them have been developed and in some cases are
still needed.  Finally, there is great interplay among spin glasses
and other systems, from the molecular evolution to
astrophysics.\cite{mydosh:93}

This chapter is devoted to the study of properties of spin glasses in
finite dimensions (mainly in three dimensions) using analytical and
numerical approaches. The main goal is to determine existence of a
spin glass phase in finite dimensions and, if it exists, to
characterize its physical properties.

We start by describing some basic properties of spin glasses in
Sec.~\ref{sec:introSG}. Next, this book chapter continues in two main parts.

In the first one, we report the main theoretical results
(Sec.~\ref{sec:theory}), starting with the mean-field solution in
Sec.~\ref{sec:MF}, which already provides a complex picture even in
infinite dimensions. Sec.~\ref{sec:FD} is devoted
to study to what extent this complex picture survives in finite
dimensions.  In Sec.~\ref{sec:droplet} we report the droplet theory, a
phenomenological theory that can also be formulated in terms of the
Migdal-Kadanoff approximation of the renormalization group. Hereafter,
we resort to the ``standard'' approach based on a field theory
approach built on the complex mean-field solution, describing its main
findings both in absence (Sec.~\ref{sec:fieldtheory}) and in presence
of a magnetic field (Sec.~\ref{sec:infield}). We finish this part by
introducing in Sec.~\ref{sec:Metastate} an important concept and
tool of the metastate.

In the second part of the chapter, we describe numerical
simulations at equilibrium (Sec~\ref{sec:numres}). We start to
report some important numerical facts in absence of a magnetic
field. First, we show in Sec.~\ref{sec:exponentsh0} the existence of a phase
transition in three dimensions and how its universality class has been
characterized. Once we know there is a spin glass phase in three
dimensions, in Sec.~\ref{sec:cor} we present numerical simulations
that try to characterize the properties of this phase, in particular,
we focus on the behavior of the conditional correlation functions. 
Next, we continue by showing a numerical construction of the metastate and
properties of the spin glass phase one can draw from this powerful
tool (Sec.~\ref{sec:MetastateNum}). We close this part with the study
of spin glasses in a field. We focus on the simulations performed in
four dimensions and the rationale of the new numerical approaches that
have been useful to find the phase transition. However, the phase
transition in three dimensions in a field has been elusive even using
these new numerical tools (Sec.~\ref{sec:eqH}).

The book chapter finishes with the conclusions and two appendices. In the first one we report
the finite size scaling tools needed to analyze the critical behavior
of these systems, namely the quotient method and the analysis at
fixed coupling (Sec.~\ref{sec:quotient}).  Part of the numerical
simulations presented in this chapter have been obtained with the help
of  Janus I and II supercomputers. In the last appendix, we have
described the basic characteristics of these two dedicated computers
(Sec.~\ref{sec:janus}).

This book chapter is based on the lectures given by the author in Lviv
during the Ising Lectures 2019 and we have tried to report the
contents lectured there during two days. In these two
lectures, the focus was on equilibrium numerical simulations on finite
dimensional Ising spin glasses. Hence, we have not discussed in this
chapter important topics in spin glass physics as experiments and 
out-of-equilibrium simulations.

Finally, let us mention that it is also possible to study the
properties of the low temperature phase with $h=0$ and $h\neq 0$ by
simulating $D=1$ dimensional Ising spin glass with the coupling
decaying following a power law and it has been used for the study of
the spin glass phase inside and outside the mean-field region.
\cite{kotliar:83,leuzzi:99,leuzzi:09,leuzzi:11,larson:13,leuzzi:15,dilucca:20,katzgraber:03,
  katzgraber:05b,katzgraber:12,katzgraber:09,katzgraber:09b,aspelmeier:16}

\section{~A brief tour of spin glasses}
\label{sec:introSG}

In this section we describe the main physical properties of these
materials.

The main ingredients to obtain materials with a spin glass behavior
are magnetic interaction, randomness, frustration and
anisotropy.\cite{mydosh:93} However, materials with spin glass behavior can be
obtained in different ways, for example magnetic interaction is not
needed.\cite{binder:86,mydosh:93,young:98,dedominicis:06,fischer:93}

Metals with very diluted magnetic impurities are considered canonical
spin glasses. We can mention, for example, CuMn and Ag:Mn at $2.5\%$,
CdCr$_{1.7}$IN$_{0.3}$S$_4$ and  Fe$_{0.5}$Mn$_{0.5}$TiO$_3$. In these
materials one can identify magnetic interaction and randomness, via the
dilution of the magnetic moments.

Since the characteristic times associated with the magnetic impurities
are much bigger than the times associated with the electrons of the
metal, we can assume that the impurities are {\em
  quenched}. \index{quenched disorder} This approximation is similar
to that performed in molecular physics called Born-Oppenheimer approximation
(in molecules the nuclei play the role of the magnetic impurities in spin
glasses). It is possible to define another kind of disorder, the
so-called {\em annealed} one, in which the components in the material
(``normal'' and ``impurities'') have similar characteristic times and,
thus, the statistical mechanics considers all of them in the same way.

\begin{figure}[ht]
  \begin{center}
\includegraphics[width=0.9\columnwidth]{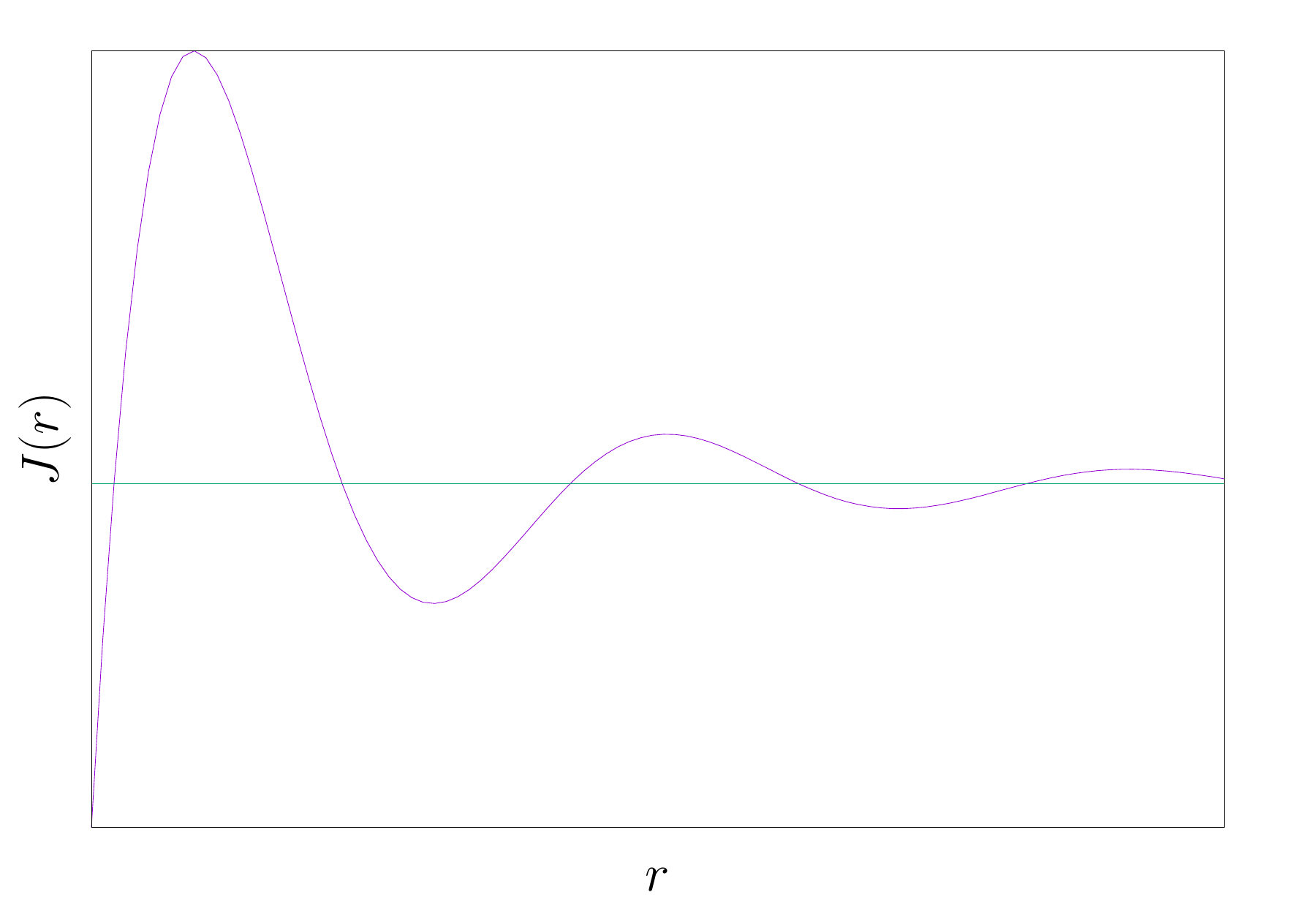}
    \end{center}
  \caption{Dependence  of coupling $J(r)$ on distance in the
    RKKY interaction. Notice the decay with distance, and most
    important fact, the oscillatory behavior, which induces frustration.}
  \label{fig:rkky}
\end{figure}

In spin glasses, the magnetic impurities do not interact following the
standard exchange interaction, rather, they interact among them via
the electrons moving in the conduction band of the metal, the
so-called RKKY interaction\index{interaction!RKKY} following the work
of Ruderman, Kittel, Kasuya and
Yosida.\cite{ruderman:54,kasuya:56,yosida:57} It has been shown
that the strength of the interaction, $J(r)$, between two magnetic
moments (impurities) sited at distance $r$ is given
by\cite{binder:86,mydosh:93}
\begin{equation}
J(r) \sim \frac{\cos (2 k_F
  r)}{r^3}\,,
\end{equation}
$k_F$ being the Fermi momentum. In addition to a power law decay on the
distance of the interaction, an oscillatory factor (the cosine)
appears:  depending on distance, sometimes it will induce a positive interaction and other times a
negative one (see Fig.~\ref{fig:rkky}).

\begin{figure}[ht]
  \begin{center}
 \includegraphics[width=0.9\columnwidth]{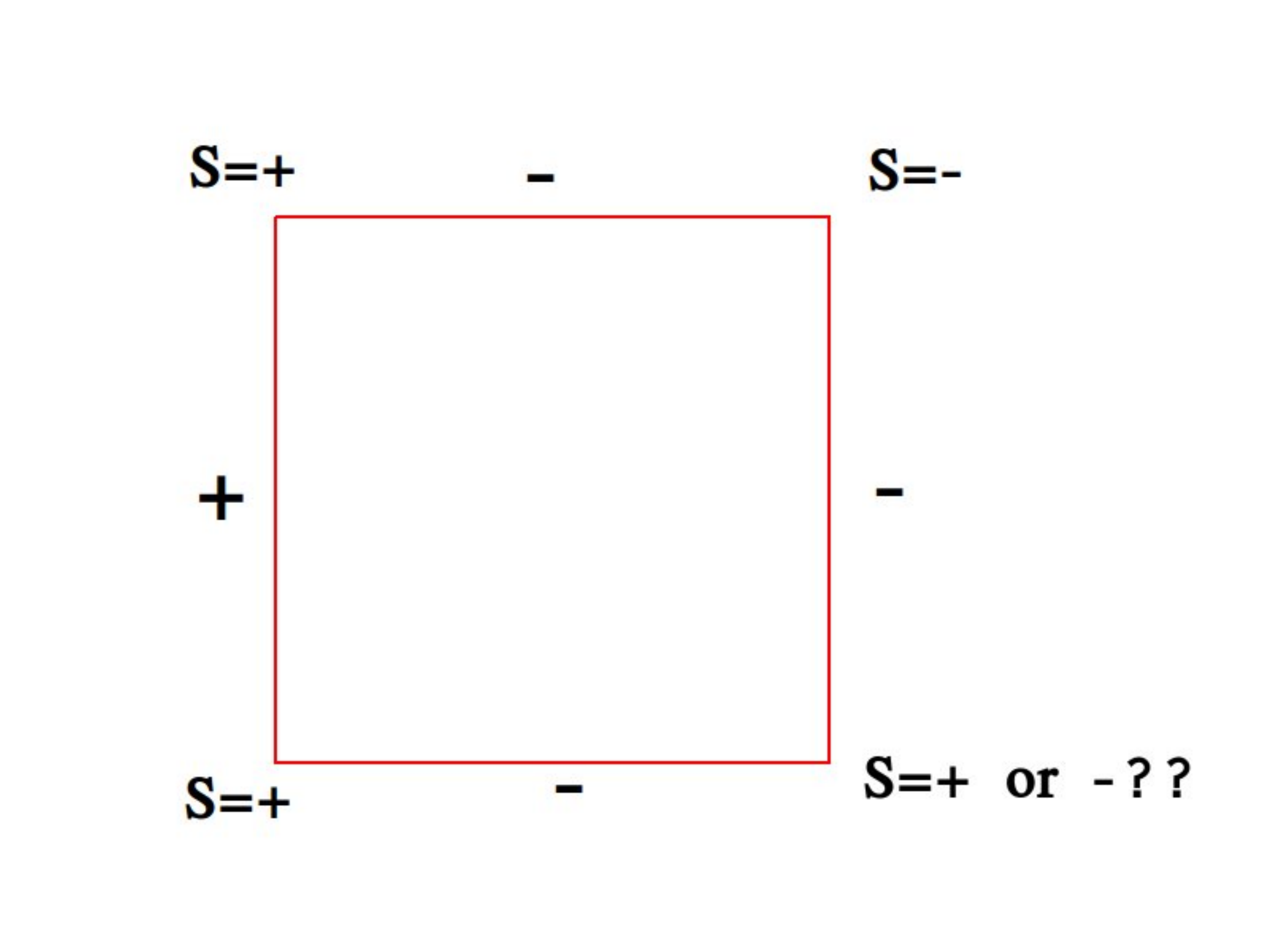}
    \end{center}
  \caption{Frustrated square. For a given choice of couplings
    (living on the links and their product being negative) and starting
    with $S=1$ for the top left spin, the value of the spin lying on
    the bottom right corner can be $+1$ or $-1$. Both values minimize
    the energy: frustration, the system has two options with the same
    ``cost''.}
  \label{fig:frustration}
\end{figure}

This change of the sign of the interaction produces the frustration in
the system. In Fig.~\ref{fig:frustration} we show a frustrated
square. In this square the product of the four couplings ($J_{ij}$
living in the links) is negative, and thus, there are different spin
configurations which provide the same energy: frustration.\index{frustration}

\begin{figure}[ht]
  \begin{center}
    \includegraphics[width=0.9\columnwidth]{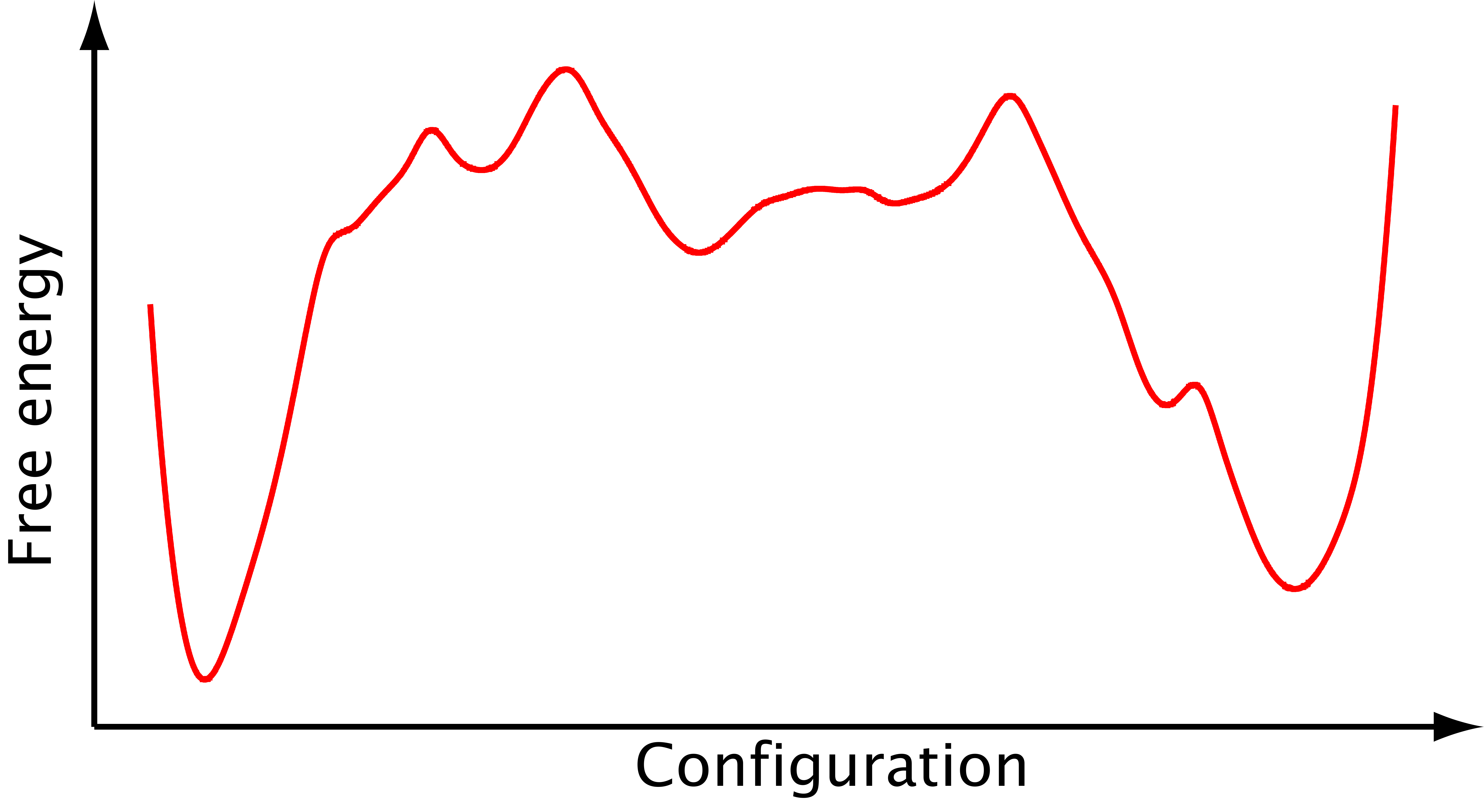}
  \end{center}
  \caption{Free energy landscape for a spin glass: notice the
    large number of minima, absolute and relative, and the diversity
    of free energy barriers separating them.}
  \label{fig:landscape}
\end{figure}

The joint effect of disorder and frustration usually produces a very
complicated landscape of free energy,
and in particular, a very slow dynamics. The landscape depicted in
Fig.~\ref{fig:landscape} is typical of glassy systems, showing a great
number of relative maxima and minima separated by high free energy
barriers.

The last ingredient to build the Ising spin glass model is the
anisotropy.\cite{mydosh:93} For instance Ag:Mn and CdCr$_{1.7}$IN$_{0.3}$S$_4$ are
well described by Heisenberg spins, although the description of
Fe$_{0.5}$Mn$_{0.5}$TiO$_3$ is based on Ising spins. However,
  some results obtained in experiments performed on films by the Texas
  group using Heisenberg spin glasses (as 
  CuMn)\cite{guchhait:14,guchhait:15a,guchhait:15b,guchhait:17,joh:99,zhai:17}
  have been confronted with numerical results simulating Ising spins
  showing a very good quantitative
  agreement.\cite{janus:17b,janus:18,fernandez:19b}

The previous discussion allows us  to write the following Edwards-Anderson
Hamiltonian\index{model!Edwards-Anderson} which describes the Ising
spin glass in a magnetic field $h$\cite{edwards:75,binder:86,mydosh:93}
 \begin{equation}
{\cal H}_J= - \sum_{i,j} J_{ij} s_i s_j + h \sum_i s_i\,,
\label{eq:EA}
 \end{equation}
 where the {\em quenched} stochastic variables ($J_{ij}$) can be drawn
 from a bimodal distribution or from a Gaussian one, both with zero
 mean and unit variance and $s_i=\pm 1$ are Ising spins. In finite
 dimensional Ising spin glasses the sum is restricted over all pairs
 of nearest neighbors, and will be denoted as $ \sum_{<ij>}$.  In this way,
 the Edwards-Anderson Hamiltonian takes into account the oscillatory
 behavior of the interaction in the RKKY
 theory.\cite{binder:86,mydosh:93}

 Considering that the disorder is quenched, one needs to compute
 the free energy of the system, from which we can derive the full
 thermodynamic of the model, in a two step procedure.

 First, we compute the free energy for a given instance, $F_J$
 (realization  or sample) of the disorder,
 \begin{equation}
F_J=-\frac{1}{\beta} \log Z_J\,,
 \end{equation}
 with $\beta\equiv 1/(k_B T)$, where the partition function $Z_J$ for
 a given disorder realization is given by
  \begin{equation}
Z_J=\sum_{[s]} \exp(-\beta {\cal H}_J) \,,
 \end{equation}
  where $\sum_{[s]}$ denotes the trace on all the spins.
  
  Second, we take the average of the free energy (of a disorder
  instance) over the whole set of instances, distributed with the 
  probability density function $p[J]$:
 \begin{equation}
F=\int \mathrm{d}[J] p[J] F_J\,,
 \end{equation}
 with
 \begin{equation}
   \mathrm{d}[J] \equiv \prod_{i<j} \mathrm{d} J_{ij}\,,
 \end{equation}
 \begin{equation}
 p[J]=\exp\biggr(-\frac12 N \sum_{i<j} J_{ij}^2\biggr) \,,
 \end{equation}
$N$ being the number of spins.
 
 Notice that we need to take the average of a logarithm. This fact
introduces strong technical difficulties in the analytical solution of
the model.

At this point, we can describe what is the spin glass order. In a spin
glass phase, all possible  staggered magnetizations
\begin{equation}
  m_\mathitbf{p}=\frac{1}{N} \sum_j e^{i \mathitbf{p} \mathitbf{r}_j} \langle s_j \rangle
\end{equation}
vanish (for all the momenta $\mathitbf{p}$), including the standard one
(${\mathitbf p}=0$). Moreover $s_j\equiv s_{\mathitbf{r}_j}$.
Despite this fact, a spin glass phase presents a non zero
  {\em local} magnetization $\langle s_i \rangle \neq 0$.

The order parameter $q$, the overlap, is then
\begin{equation}
  \label{eq:overlap}
q^J= \frac{1}{N} \sum_i \langle s_i
  \rangle^2\,,
\end{equation}
\begin{equation}
    \label{eq:overlapD}
q= \overline{q^J}\,,
\end{equation}
where $\langle(\cdots) \rangle$ is the thermal average (fixed
disorder) and $\overline{(\cdots)}$ is the disorder
average.  The overlap $q$ is zero in the paramagnetic phase
and takes a non zero value below the phase transition, in the spin
glass phase.

In a numerical simulation is easy to compute the overlap: we simulate
in parallel two non-interacting replicas of the system, $\{s_i^{(1)}\}$ and
$\{s_i^{(2)}\}$, in presence of the same disorder. Notice that
\begin{equation}
  \langle s_i^{(1)} s_i^{(2)} \rangle= \langle  s_i^{(1)} \rangle
  \langle  s_i^{(2)} \rangle
  = \langle  s_i^{(1)} \rangle^2\,,
\end{equation}
for a given disorder realization. The total overlap per spin is defined as 
\begin{equation}
q^{12}=\frac{1}{N} \sum_i q_i^{12}\,,
\end{equation}
where
\begin{equation}
q_i^{12}=  s_i^{(1)}  s_i^{(2)},
\end{equation}
and we can compute the probability density function associated with
this observable
\begin{equation}
P_J(q) = \langle   \delta(q-q^{12}) \rangle ,
\end{equation}
and the  probability distribution averaged over the disorder
\begin{equation}
\label{eq:pnorep}
  P(q) =\int \mathrm{d}[J] p[J] P_J(q)= \overline{P_J(q)}\,.
\end{equation}
It is easy to show that
\begin{equation}
q=\int \mathrm{d} q'~ q' P(q')\,.
\end{equation}

However, the overlap cannot be measured in experiments where the
critical behavior of the material is extracted via the non-linear susceptibility
($\partial^4 f/\partial h^4$) which is proportional to the spin glass susceptibility
defined as
\begin{equation}
  \chi_{\mathrm SG}= N \overline{\langle (q^{12})^2\rangle}\,.
\end{equation}

To study spin glass phases with several pure states or phases due to
broken ergodicity\cite{fischer:93}, see Sec. \ref{sec:Metastate} for more details, we
need to generalize the above defined overlap by means:
\begin{equation}
q^J_{\alpha \beta}= \frac{1}{N} \sum_i \langle s_i
  \rangle_\alpha \langle s_i
  \rangle_\beta  \,,
\end{equation}
\begin{equation}
q_{\alpha \beta}= \overline{q^J_{\alpha \beta}}  \,,
\end{equation}
where $\alpha$ and $\beta$ are two different pure states and
$\langle(\cdots) \rangle_\gamma$ denotes the thermal average
restricted to the state $\gamma$. In general, it is possible to write
$\langle(\cdots) \rangle=\sum_\alpha \langle (\cdots)\rangle_\alpha$
with $\sum_{\alpha} w_\alpha=1$, where the sum runs over all the pure
states, and then one can write
\begin{equation}
P_J(q)=\sum_{\alpha \beta} w_\alpha w_\beta \delta(q-q^J_{\alpha \beta}) \,.
\end{equation}

The Edwards-Anderson overlap, the
maximum possible overlap in the system, is just the maximum overlap:
$q_{\alpha \alpha}$.

Despite the fact the magnetic interaction plays an important role in
spin glasses, there have been found materials, with a spin glass
behavior, where the magnetism is not present. For example, it is
possible to study dipolar and quadrupolar spin glasses, where the role
of magnetization is played by the polarization
vector.\cite{binder:86,mydosh:93}

Finally, it is interesting to report the existence of mathematical
problems, not related with physics, which present a behavior similar
to the one of spin glasses. We can mention optimization problems as the
traveling salesman problem, neural networks and biological
evolution.\cite{mydosh:93,mezard:87}

\section{~Some theoretical results}
\label{sec:theory}
In this section we discuss different analytical approaches to  tackle spin glass behavior. We start with
the mean-field approximation, which is exact in infinite dimensions,
and then we study finite dimensional spin glasses by
presenting the droplet model and the approach based on field theory.

\subsection{~Mean-field solution}
\label{sec:MF}

Let us summarize the solution of the Edwards-Anderson
model (for $h=0$) in the mean-field approximation, the so-called
Sherrington-Kirpatrick model\cite{kirkpatrick:78}.\index{model!Sherrington-Kirkpatrick}
The starting point is to use the replica trick. \index{replica trick} In this trick we
replace the logarithm entering the quenched average by the following
limit\cite{mezard:87,dotsenko:01}
\begin{equation}
\log x=\lim_{n\to 0} \frac{x^n-1}{n}\,.
\end{equation}
Applying this trick to the computation of the quenched free energy we obtain
 \begin{equation}
\log Z_J=\lim_{n\to 0} \frac{Z_J^n-1}{n} \,.
  \end{equation}
The average on the disorder of the partition function of $n$ non-interacting replicas, $\{s^a_i\}$ ($a=1,\ldots,n$),
can be written as
\begin{equation}
  \label{eq:Zrep}
Z_n = \overline{Z_J^n} = \sum_{\{s^a\}} \int\mathrm{d}[J] 
\exp\biggl(\beta \sum_{a=1}^n \sum_{i<j} J_{ij} s_i^a s_j^a
- \frac12 N \sum_{i<j} J_{ij}^2\biggr) \,.
\end{equation}
Now, we can compute the integral on the disorder, getting
\begin{equation}
Z_n = \sum_{\{s^a\}} \exp\left[ \frac14 \beta^2 Nn 
+ \frac12 \beta^2 N \sum_{a<b}^n \biggl(\frac1N \sum_i s_i^a s_i^b\biggr)^2\right].
\end{equation}
At this point, we can define a first effective Hamiltonian via
\begin{equation}
  Z_n\propto \sum _{\{s^a\}} \exp\left(-\beta {\cal H}_\mathrm{eff}\{s_i^a\}  \right)\,,
  \label{eq:Heff}
\end{equation}
with
\begin{equation}
 {\cal H}_\mathrm{eff}\{s_i^a\}\equiv -\frac12 \beta N \sum_{a<b}^n \biggl(\frac1N \sum_i s_i^a s_i^b\biggr)^2 \,.
\end{equation}
This effective Hamiltonian depends on all the replicas
$\{s_i^a\}$ with $i=1, \ldots, N$ and $a=1,\ldots, n$. At this point, the disorder has been integrated out.

The quadratic term in the exponential can be made linear by using the
Hubbard-Stratonovich trick at the price to introduce the replica
matrix $Q_{ab}$
\begin{equation}
  \label{eq:SG-Zn}
  Z_n = \int \mathrm{d}[Q] 
\sum_{\{s^a\}} \exp\left[ \frac14 \beta^2 Nn - \frac12 \beta^2 N 
\sum_{a<b}^n Q_{ab}^2 + \beta^2 \sum_{a<b}^n \sum_i Q_{ab}s_i^as_i^b\right].
\end{equation} 
 with
 \begin{equation}
   \mathrm{d}[Q] \equiv \prod_{a<b} \mathrm{d} Q_{ab}\,,
 \end{equation}

In this way we can write the second  effective Hamiltonian
\begin{equation}
Z_n = \int \mathrm{d}[Q_{ab}] e^{- \mathcal {\cal H}_n\{Q_{ab}\}},
\end{equation}
 where
\begin{equation}
\label{Hn}  
\mathcal {\cal H}_n\{Q_{ab}\} = 
-\frac{Nn}4 \beta^2 + \frac{N}{2} \beta^2 \sum_{a<b} Q_{ab}^2-N
\log\left[ \sum_{\{S^a\}} \exp\left(\beta^2 
\sum_{a<b} Q_{ab} S^a S^b\right)\right]\,,
\end{equation}
where $S^a=\pm 1$ are Ising spins. This Hamiltonian depends only on
the overlap matrix.

Taking into account that the argument of the exponential is
proportional to the number of spins, $N$, we can compute $Z_n$ using
the saddle point method. The stationary condition is
\begin{equation}
\frac{\delta \mathcal H_n}{\delta Q_{ab}}= 0\,,
\end{equation}
that can be written as
\begin{equation}
  \label{eq:SG-Qab}
  Q_{ab} = \langle S^a S^b
  \rangle_Q\,\,,\, a\neq b\, ,
\end{equation}
where 
\begin{equation}
\langle S^a S^b \rangle_Q \equiv \lim_{n\to 0} \frac{\sum_{\{S^a\}}  S^a S^b \exp\left(\beta^2 
\sum_{a<b} Q_{ab} S^a S^b\right)}{\sum_{\{S^a\}} \exp\left(\beta^2 
\sum_{a<b} Q_{ab} S^a S^b\right)}\,.
\end{equation}
Furthermore, $\langle S^a S^b\rangle_Q =\lim_{n\to 0}  \langle s_i^a s_i^b\rangle_{{\cal H}_\mathrm{eff}}$.
  
At this stage of the analytical computation, we need to do some
hypotheses on the structure of the matrix $Q$. The simplest Ansatz,
called 0-step, is\cite{mezard:87}
\begin{equation}
Q_{ab} = (1-\delta_{ab}) q\,,
\end{equation}
where $q$ can be computed in a self-consistent way using the saddle point
equation (Eq.~\ref{eq:SG-Qab}).

However, this solution leads to two main problems. The first one is
that the 0-step solution does not provide with the correct value of
the energy and the second one, is that the entropy of this solution is
negative. We can try to understand these problems in the framework of
field theory.

The effective Hamiltonian can be developed in powers of matrix $Q$
following the framework of the Landau theory of phase
transitions. This effective Hamiltonian describes the physics near the
critical point.\cite{cardy:96} For this analysis it is enough to keep terms 
up to the fourth order in the matrix $Q$. At this order the effective
Hamiltonian is
\begin{equation}
  {\cal H}_n=\int {\mathrm d}^D x \left[ \left(\partial_\mu Q_{ab}\right)^2+ \tau
    \mathrm{Tr} Q^2+  g_3 \mathrm{Tr}Q^3+ g_4 \mathrm{Tr} Q^4+ \lambda
    \sum Q_{ab}^4 \right]\,.
\label{eq:FT}
\end{equation}
If $\lambda = 0$ the symmetry group of this Hamiltonian is
$O(n)$. Although, the symmetry group is reduced to the symmetry group
$S_n$ (permutations of $n$ elements) as $\lambda\neq 0$.

Now, let us consider again the $\lambda=0$ case. The 0-step choice for $Q_{ab}$
{\em spontaneously} breaks the $O(n)$ symmetry and Goldstone bosons\cite{amit:05}
(particles or excitations of zero mass) will appear. By turning on
the $\lambda$ coupling, the group $O(n)$ is broken {\em explicitly} and the
Goldstone bosons are no longer  massless (a detailed computation shows that the mass
is negative), which clearly indicates that the 0-step solution is
unstable.\cite{parisi:88,amit:05,mezard:87,itzykson:89}

Hence, one needs to find different solutions to that of the 0-step to
parameterize the overlap matrix
$Q_{ab}$. G. Parisi\cite{parisi:79,parisi:79b,parisi:80,parisi:80b,parisi:80c,parisi:83}
proposed a general Ansatz for the matrix $Q_{ab}$ which breaks
the original $n\times n$ matrix in boxes and the boxes in
sub-boxes an so on, in an infinite number of steps.

\begin{figure}[th]
  \begin{center}
    \includegraphics[angle=0,width=0.95\columnwidth]{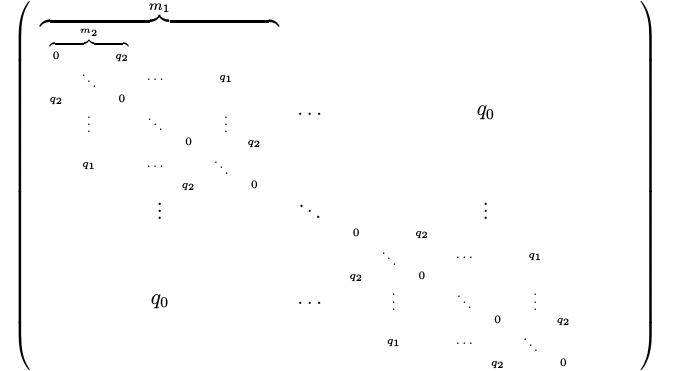}
    \end{center}
	\caption{2-step of the overlap matrix $Q_{ab}$. Notice that
          the matrix has been broken into two main  submatrices of sizes
          $m_1$ and $m_2$, taking values $q_0$, $q_1$ and $q_2$. From Ref.~[\refcite{monforte:13}].}
	\label{fig:2step}
\end{figure}

For example, in Fig.~\ref{fig:2step} is shown the 2-step level of the
Parisi solution (which has infinite levels). At this 2-step level
three real values of the overlap $q$ appear: $q_0$, $q_1$ and $q_2$ and
two integer numbers which determine the size of the sub-matrices (box
and sub-box), $m_1$ and $m_2$, such that $0<m_2<m_1<n$. Notice that
$m_2$ must divide $m_1$ and $m_1$ must divide $n$.\cite{mezard:87}

This scheme can be generalized, assuming that $n$, the number of
replicas, is large enough to allow a $k$-step level of breaking the
symmetry of the replicas, where $k$ could be arbitrarily
large. Finally, we need to do an analytic continuation to $n=0$ (to
comply with the replica trick).

Given matrix $Q_{ab}$ with the Parisi breaking scheme, one can
compute what is the probability to find a given value of $q$, denoted as
$p(q)$, assuming that all the matrix elements have the same
probability:\cite{mezard:87}
\begin{align} \nonumber
p(q) &= \frac{1}{n(n-1)} \sum_{a\neq b} \delta(Q_{ab} - q)\\ \nonumber
   &= \frac{n}{n(n-1)} \bigl[ (n-m_1) \delta(q-q_0)
+ (m_1-m_2) \delta (q-q_1)  \\
& \qquad  \qquad\qquad+(m_2-m_3) \delta(q-q_2) + \ldots
\bigr]\,,
\end{align}
with  $n>m_1>m_2>\ldots>1$.

Taking the limit $n \to 0$,\cite{mezard:87} one obtains,
\begin{equation}
p(q)  = m_1 \delta(q-q_0) + (m_2-m_1) \delta(q-q_1) 
+ (m_3-m_2) \delta(q-q_2) + \ldots
\label{eq:pdq}
\end{equation}
Notice that $p(q)$ is a probability density function composed by
sum of Dirac's deltas with different weights, hence, all these weights
must be positive and so  $0<m_1<m_2<\ldots<1$. Notice that the
limiting process ($n\to 0$) has inverted the order of the different
$m$'s.

At this point we can connect $p(q)$ with the pdf of the overlap
defined in Eq.~(\ref{eq:pnorep}), denoted as $P(q)$.

The overlap defined in Eqs. (\ref{eq:overlap}) and (\ref{eq:overlapD}) can be written in the framework of the
replica theory as ($a\neq b$)
\begin{equation}
q=  \overline{\langle s_i \rangle^2}=
  \overline{\langle s_i^a s_i^b \rangle}=
\overline{\left[
  \frac{\sum_{\{s^a, s^b\}} s_i^a s_i^b
    \exp\left(-\beta \sum_{k<l} J_{kl} (s_k^a s_l^a +
    s_k^b s_l^b) \right)}
  {\sum_{\{s^a, s^b\}}
    \exp\left(-\beta \sum_{k<l}J_{kl} (s_k^a s_l^a
    +
    s_k^b s_l^b) \right)}\right]}\,.
\end{equation}
Notice that the denominator, inside the average over the disorder,  is
just $Z_J^2$. We can introduce $n-2$ extra factors $Z_J$ in both
numerator and denominator,
obtaining in the limit $n\to 0$ (the final $Z_J^n$ in the denominator goes to one):\cite{fischer:93}
\begin{eqnarray}
 q & = & \overline{\langle s_i \rangle^2}=
 \lim_{n\to 0} \sum_{\{s^a\}} s_i^a s_i^b \exp\left(-\beta {\cal H}_\mathrm{eff}\{s_i^a\}\right)\\
 \label{eq:inter}
&=&\lim_{n\to 0} \frac{\sum_{\{s^a\}} s_i^a s_i^b \exp(-\beta {\cal H}_\mathrm{eff}\{s_i^a\})}
{\sum_{\{s^a\}} \exp(-\beta {\cal H}_\mathrm{eff}\{s_i^a\})}
=
\lim_{n\to 0} \langle s_i^a s_i^b \rangle_{\cal{H}_\mathrm{eff}}\\
&=&\langle S^a S^b \rangle_Q=\lim_{n\to 0} \frac{1}{n (n-1)} \sum_{a\neq b} Q_{ab}\,,
\end{eqnarray}
where, in the last equation, we must average over all the saddle-point
solutions (we also refer to the analytical computations we have done to
transform Eq.~(\ref{eq:SG-Zn}) into Eq.~(\ref{Hn})). In particular, in Eq.~(\ref{eq:inter}), we have used that in Eq.~(\ref{eq:Heff})  $Z_n\to 1$ as $n\to 0$.

Hence $p(q)$ and $P(q)$ have the same first momentum. It is possible
to generalize this computation and to show that
$p(q)=P(q)$.\cite{mezard:87} In the rest of this book chapter we will
denote the pdf of the overlap as $P(q)$.

We continue by studying the properties of $P(q)$. It is possible to
show that in the limit of an infinite number of breakings (full
replica symmetry breaking, RSB) the real parameters $q_k$ become a
continuous function $q_k\to q(x)$, with $x\in [0,1]$ and that the
function $x(q)$ (inverse of $q(x)$) is related with $P(q)$
via\cite{mezard:87}
\begin{equation}
\frac{\mathrm{d} x}{\mathrm{d} q} = P(q) \,.
\end{equation}

Before finishing this section we summarize some of the most important physical
and mathematical properties of the Parisi
solution:\cite{parisi:79,parisi:79b,parisi:80,parisi:80b,parisi:80c,parisi:83,mezard:84,mezard:84b,mezard:85,mezard:85b,mezard:86,mezard:87,marinari:00,dotsenko:01}

\begin{figure}[th]
  \begin{center}
    \includegraphics[angle=0,width=0.95\columnwidth]{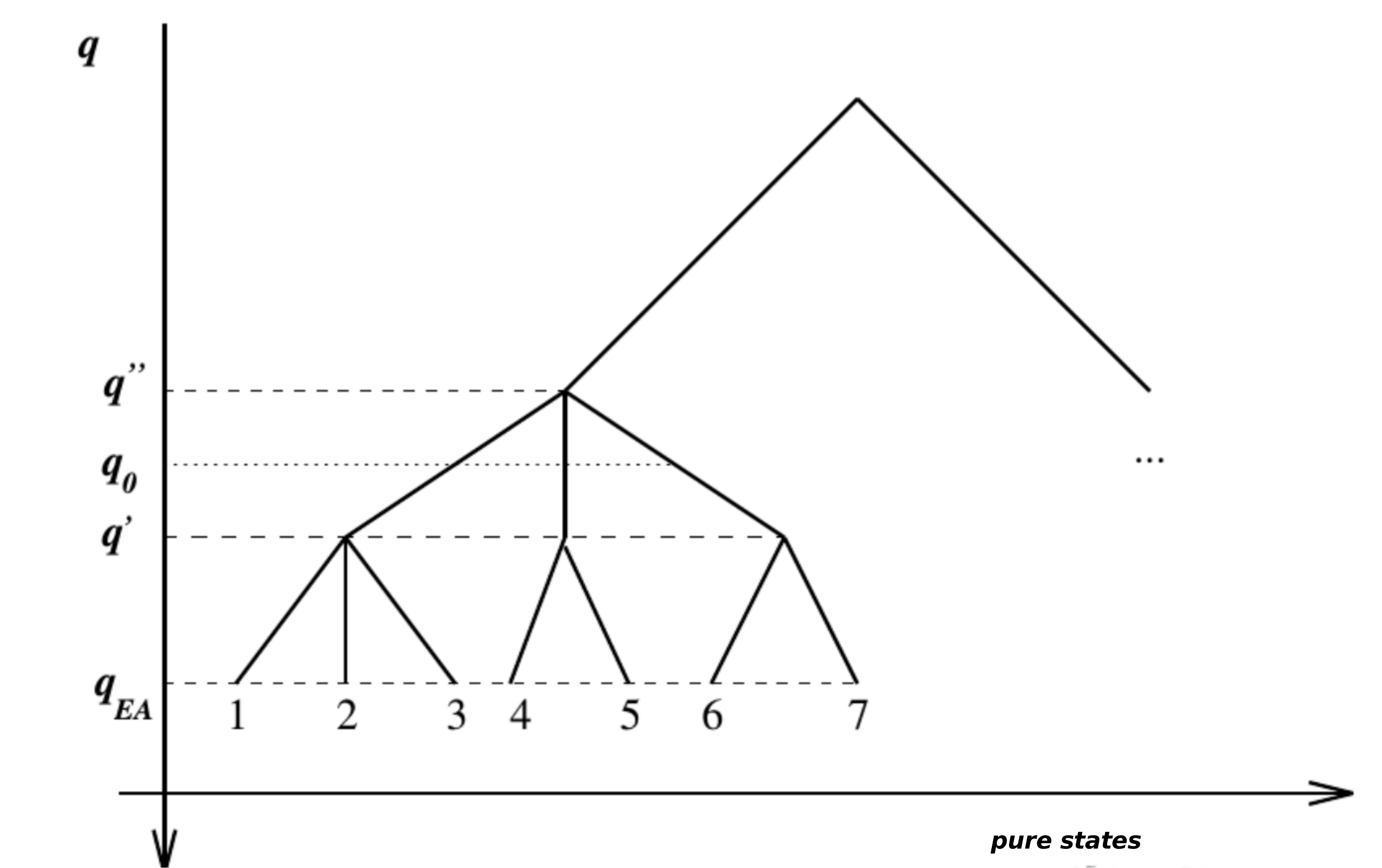}
    \end{center}
	\caption{ Ultrametric organization of the pure states in the Parisi solution.}
	\label{fig:ultra}
\end{figure}

\begin{itemize}

\item It is exact in infinite dimensions. This has rigorously been  shown
  in Refs.~[\refcite{guerra:03,talagrand:11,panchenko:13b}].

\item It shows an infinite number of pure states not related by any symmetry.

 \item These infinite pure states are organized in a ultrametric
   way\cite{rammal:86}\index{ultrametricity}. We can recall at this
   point, the definition of an ultrametric space.  A space is
   ultrametric if all the triplets of elements belonging to this space
   ($A$, $B$, $C$) satisfy the ultrametric inequality:
   $$d(A;B) \le \mathrm{max}(d(A,C),d(B,C))\,.$$ In spin glasses we
   can introduce a distance by using the overlaps (now the elements
   of this space are the pure states, see Sec. \ref{sec:Metastate})
   \begin{equation}
   d(\alpha,\beta)=\frac{1}{2} \left(q_\mathrm{EA}-q_{\alpha \beta}
   \right)\,.
   \end{equation}
 In Fig.~\ref{fig:ultra} we have drawn the ultrametic organization of
 the pure states: the end of the leaves are the pure states, having
 $q_\mathrm{EA}$ as their overlap.

 \item The spin glass phase is stable under small magnetic fields. A
   transition line in the temperature-magnetic field plane separates a
   paramagnetic phase from a spin glass one (the de Almeida-Thouless
   line\cite{dealmeida:78}).\index{de Almeida-Thouless line}

\item The excitations of the ground state are space filling,
     i.e. the dimension of the excitations is just that of the space,
     $D$.

   \item Overlap equivalence. All the possible definitions of the
     overlap, e.g. spin overlap or link overlap, provide the same
     information on the physical properties of the system. For example the link overlap is defined as
     \begin{equation}
q_l=\frac{1}{N D} \sum_{<ij>} s^{(1)}_i s^{(1)}_j  s^{(2)}_i s^{(2)}_j \,, 
     \end{equation}  
where $\{s^1_i\}$ and $\{s_i^2\}$ are two real replicas and $ \sum_{<ij>}$
denotes sum over all pairs of nearest neighbors.  In the SK model, one can show
that $q_l=q^2$.

     \item Stochastic stability. The Hamiltonian of spin glasses is
       {\em generic} against random perturbations.

\end{itemize}

\subsection{~Spin glasses in finite dimensions}
\label{sec:FD}
Once we have characterized the behavior of spin glasses in infinite
dimensions, the fully connected model, we want to understand what are
the properties of spin glasses in finite dimensions.
Different theories have been developed to describe the behavior of
spin glass in finite dimensions. Hereafter, in the next three
sections, we report the two most important approaches: the droplet
model and the RSB theory.

\subsection{~The droplet model}
\label{sec:droplet}

The droplet theory predicts that the spin glasses above the lower
critical dimension ($D_l$) show only two pure states in the spin glass
phase and the behavior of this spin glass phase is determined by
compact excitations on the ground state. The droplet model can be
formulated in terms of the Migdal-Kadanoff approximation of the
renormalization group\cite{wilson:74,wilson:75} (which is exact in
$D=1$)\cite{bray:87,bray:87b} or by means a phenomenological
theory,\cite{fisher:85, fisher:86, fisher:88, fisher:88b,fischer:93} both approaches being equivalent.
\index{theory!droplet}

The most important properties of this phenomenological theory are:
\begin{itemize}

\item The droplets are compact excitations with fractal dimension 
  $D_F$. The energy of a droplet of linear dimension $L$ grows as 
  $L^\theta$ with  $\theta<(D-1)/2<D-1<D_F<D$.
  
\item In the dynamics, the free energy barriers behave as 
  $L^\psi$, with    
  $\psi\ge \theta$.

\item The spin glass phase is unstable against the presence of a magnetic field.

\item There are two pure states (related by spin flip), and thus, the
  probability distribution of the overlap is trivial: sum of two
  Dirac's deltas.

\end{itemize}

Finally, there is a variation of the droplet model, known as the chaotic
pairs model. In this picture the system has two pure states (as in the
droplet model), but these two states vary chaotically with the size
of the system, see Sec.~\ref{sec:Metastate}.

\subsection{~Field theory ($h=0$)}
\label{sec:fieldtheory}

In this section, we address the problem of how to build a field theory using the RSB
solution as starting point in absence of a
magnetic field.  For a theoretical description of the
theory in presence of a magnetic field, see Sec.~\ref{sec:infield}.

We start considering the theory in infinite dimensions (using the mean-field approximation, which is exact in $D=\infty$). Next, the upper critical
dimension $D_u$ is computed. Above it, the predictions of the mean-field approximation hold and below, infrared divergences appear and we
need to resort to the renormalization group to tackle
them.\cite{amit:05,cardy:96,parisi:88,itzykson:89}

This is the standard approach and it has been applied (with a huge
success) in the study of a large number of models, for example, the
Ising model or models with $O(N)$ symmetry.

The upper critical dimension is determined by the dimension of the
cubic coupling, $g_3$, in the effective Hamiltonian, see
Eq.~(\ref{eq:FT}), and it turns to be $D_u=6$.  Below the upper
critical dimension $\tau$ and $g_3$ are the relevant parameters (using the
terminology of the renormalization group\cite{cardy:96}) and $g_4$ and
$\lambda$ are irrelevant, hence, we need to study the field theory of
a $\phi^3$ theory with tensor couplings. By using this theory, it is possible
to compute analytically the critical exponents using the
$\epsilon$-expansion (where $\epsilon=6-D$), see
Refs.~[\refcite{harris:76,alcantara:81,green:85,fischer:93}]. Moreover, by using 
this theoretical framework, it  has been possible to compute the
logarithmic corrections at the upper critical dimension in these models, see
Refs.~[\refcite{ruizlorenzo:98,kenna:06,kenna:06b,kenna:13,ruizlorenzo:17}].

To complete this discussion, let us remark that the lower critical
dimensions in absence of a magnetic field seems to be
$D_l=2.5$\cite{franz:92,boettcher:05,maiorano:18}. This issue has been studied
experimentally by studying spin glasses in film geometries, finding a strong evidence that
$2<D_l<3$.\cite{guchhait:14, guchhait:15a, guchhait:15b,guchhait:17}

The low temperature spin glass phase is critical (in this model the
$T=0$ critical point has infinite correlation length, as in the $O(N)$
model, $N>1$, $D>2$) and to perform a field theoretical analysis we
also need to consider the $g_4$ and $\lambda$ couplings. The mean
field solution, on which is based this approach, is very complicated
mathematically, as we have shown in Sec.~\ref{sec:MF}. These
computations have been partially performed\cite{dedominicis:93,
  dedominicis:98, dedominicis:06} and the behavior of the different
correlation functions (propagators) which depend on the overlap $q$
was obtained. All the connected correlation functions present an
algebraic decay (since the low temperature is critical) as in the
droplet model.\index{theory!replica symmetry breaking}

Thereupon, we summarize the main results.\cite{dedominicis:93, dedominicis:98,dedominicis:06}
Firstly, we define the (connected) correlation function 
\begin{equation}
  C_4(\mathitbf{r}|q)\equiv \frac{1}{V} \sum_{\mathitbf{x}}
  \overline{\langle s_\mathitbf{x}^{(1)} s_\mathitbf{x+r}^{(1)}
    s_\mathitbf{x}^{(2)} s_\mathitbf{x+r}^{(2)} \rangle}\,. 
\label{eq:Crq}
\end{equation}
$\langle (\cdots)\rangle$ denotes the thermal average,
$\overline{(\cdots)}$ is the average over the disorder, $V$ is the
volume of the system, $\{s_i^{(1)}\}$ and $\{s_i^{(2)}\}$ are, as
usual, two non-interacting real copies of the system (real replicas)
and the system is constrained to have a fixed overlap $q$
\begin{equation}
q=\frac{1}{V} \sum_{\mathitbf{x}}\overline{\langle s_\mathitbf{x}^{(1)} s_\mathitbf{x}^{(2)} \rangle}\,.
\end{equation}
This conditional correlation function decays algebraically  as
\begin{equation}
  C_4(\mathitbf{r}|q) \simeq q^2 + \frac{A(q}{ r^{\theta(q)}} \,.
\label{eq:Crqdecay}
\end{equation}

The following values for the exponent $\theta(q)$ were obtained:\cite{dedominicis:93, dedominicis:98, dedominicis:06,marinari:00}
\begin{itemize}
  
\item $\theta(q_\mathrm{EA})=D-2$. This results could be exact, a sort of Goldstone theorem.

\item $\theta(q)=D-3$ para $q_\mathrm{EA} > q > 0$. This exponent could be modified below $D_u=6$.
  
\item $\theta(q_m)=D-4$ for the smallest overlap   $q_m=0$. This mode is called replicon.\index{replicon}
\end{itemize}
The full (non-constrained) correlation function can be recovered with the help of $P(q)$, the probability to find a given overlap $q$, see Eq. (\ref{eq:pdq}), as
\begin{equation}
C_4(r)=\int {\mathrm d}q~ P(q)   C_4(r|q)\;.
\end{equation}

We can refer that the prediction of the droplet model is 
\begin{equation}
  C_4(r) \simeq q_\mathrm{EA}^2 + \frac{A}{r^{\theta}}\,,
\label{eq:C4-droplets}
\end{equation}
where  $\theta$ is the exponent which controls the thermal excitations of the system
(see Sec.~\ref{sec:droplet}).

In the droplet model there is only one correlation
function since there is only a pure state  having an equilibrium overlap
$q_\mathrm{EA}$, see Sec. \ref{sec:droplet}, whereas in the RSB theory it is possible to find two states
with overlap in the interval $[0,q_\mathrm{EA}]$, and therefore it is
possible to obtain a correlation function in which one replica belongs
to the state $\alpha$ and the other to the $\beta$ one having mutual
overlap $q_{\alpha \beta}=q$, measuring $C_4(r|q)$.

Finally, let us remark that it is possible to show in a rigorous way
that if the spin glasses in finite dimensions present ultrametricity,
the mathematical properties of this ultrametricity are the same as of the
ultrametricity found in the Parisi theory (RSB) which holds in infinite
dimensions.\cite{iniguez:96}\index{ultrametricity}

Furthermore, stochastic stability and replica equivalence or overlap
equivalence imply ultrametricity\cite{parisi:00} and numerical
simulations have provided strong numerical support for both stochastic
stability and overlap equivalence in three dimensional spin
glasses.\cite{janus:09b,janus:10,guerra:97,ghirlanda:98,contucci:07b}

\subsection{~Field theory ($h\neq 0$)}
\label{sec:infield}
\begin{figure}[ht]
  \begin{center}
  \includegraphics[width=0.95\columnwidth]{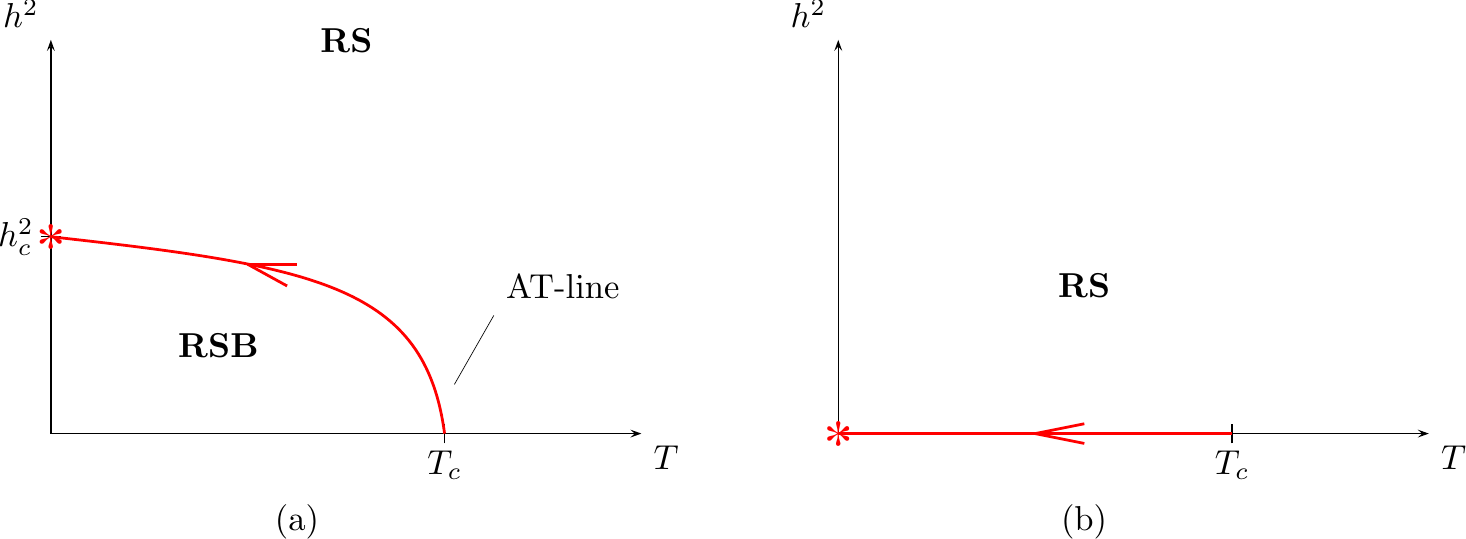}
    \end{center}
  \caption{Renormalization group flow for spin glasses in presence of
    a magnetic field for dimension $D_l<D<D_u$.\cite{parisi:12} On
    the left panel, we show the flow assuming replica symmetry
    breaking (phase transition in field). On the right the
    renormalization group flow assuming the spin glass phase is
    unstable under magnetic field (droplet model). The fixed points of
    the renormalization group equations for temperature and magnetic
    field are marked with a star. The continuous lines, with an arrow,
    mark the critical surfaces, in which the correlation length is
    infinite, and the arrows point to the directions in which the flow
    is moving as the scale (in space) of the renormalization group
    transformation grows.}
  \label{fig:RGH}

\end{figure}

\index{spin glass} The analytical study of spin
glasses in presence of a magnetic field below its upper critical
dimensions has been and still is a challenge. In Fig.~\ref{fig:RGH} we
show the renormalization group flow for spin glasses below the upper
critical dimension and in presence of a magnetic field assuming RSB or
droplet.  Despite all the difficulties, the following facts
are known:
  \begin{itemize}
  \item  The upper critical dimensions is  6.\cite{bray:11} 

  \item Due to the appearance of a dangerous irrelevant variable  the critical
    exponents for some observables change already in $D=8$.\cite{fisher:85}
    
  \item One can project the full theory on its most divergent
    components. Working with this projected theory\cite{bray:80} no
    fixed points were found at the first order in perturbation theory. The
    no existence of fixed point of the renormalization group in a
    theory is usually interpreted as
    \begin{enumerate}

    \item it is needed to work at higher order in perturbation theory
      in order to find it (or them);\cite{charbonneau:17}

    \item the fixed point is of non-perturbative nature and one needs
      to resort to non-perturbative methods;\cite{delamotte:07}

    \item the phase transition is first-order (runaway trajectories).

    \end{enumerate}
    
  \item By using the most general Hamiltonian in the replica symmetric
    phase and by relaxing the replica trick condition $n=0$, a fixed
    point below six dimensions was found.\cite{dedominicis:02}

  \item In Refs.~[\refcite{temesvari:08,parisi:12}] the de Almeida-Thouless transition 
  was found below the upper critical dimension ($D_U=6$)  using analytical prolongation.

  \end{itemize}
  
Finally, in Ref.~[\refcite{holler:19}] analytical computations were
performed suggesting the possible existence of a quasi-first-order
phase transition below the upper critical dimension.

\subsection{~Metastate}
\label{sec:Metastate}
\index{state!metastate}
Pure phases (or pure states) are macroscopic homogeneous states of matter,
for example, liquid, gaseous and solid phases. In principle, it is not difficult to 
determine if a lump of matter is in a given phase or not.
However, to define states in the thermodynamic limit (infinite volume) is not easy.

Rigorously, a state is a probability
distribution and with it we can compute averages: hence, we can see
this probability distribution as a linear functional.\cite{ruelle:69}

For example, in the two dimensional Ising model (with no disorder) one can
define three (states) phases: paramagnetic, ferromagnetic (with
positive magnetization) and  ferromagnetic (with negative
magnetization) phase. The two ferromagnetic phases can be defined using the
following limits
  \begin{equation}
  \langle(\cdots)\rangle_{+}=\lim_{h\to 0+} \lim_{L \to \infty} \langle(\cdots)\rangle_{(L,h)}\,\,,
  \end{equation}
  \begin{equation}
    \langle(\cdots)\rangle_{-}=\lim_{h\to 0-} \lim_{L \to \infty} \langle(\cdots)\rangle_{(L,h)}\,.
  \end{equation}

  In general, both in experiments and in numerical simulations, we
  obtain mixed states: i.e. numerical configurations or lumps with
  interfaces which split different pure states (or pure phases). These
  mixed states form a convex set in which its extremal points define
  the pure states. \index{state!pure}The mixed states can be written in terms of pure
  states in the following way (we particularize for the Ising model
  above one dimension and below the critical temperature):
    \begin{equation}
    \langle(\cdots)\rangle= w \langle(\cdots)\rangle_{+}+(1-w)\langle(\cdots)\rangle_{-} \,,
    \end{equation}
    where $0<w<1$ is the proportion of the positive magnetization phase.
    
  In a more general way, one can write an  expression of a given state
  $\Gamma$ in terms of pure states $\Gamma_\alpha$
\begin{equation}
 \Gamma = \sum_\alpha w_\alpha \Gamma_\alpha\,,
\end{equation}
with $\sum_\alpha w_\alpha=1$, $w_\alpha>0$ and the sum runs over all
the pure states. Another important property of pure states is that
intensive magnitudes do not fluctuate.

The generalization of the previous concepts and mathematical
tools to systems with quenched disorder is very complicated, since the
sequence of states $\Gamma_{L,J}$, $J$ denotes the quenched disorder, usually does not converge (chaotic
dependence on  system size).\cite{parisi:94,young:98}

Some models in which the sequence of states does not converge are:\cite{marinari:00}
\begin{enumerate}

\item Ising model ($D>1$) without disorder with spins on the fixed
  boundary conditions which change as it changes the size of the system.
  
\item Ising model ($D>2$) in presence of a random
  magnetic field with zero mean and unit variance. The magnetization
  of the ferromagnetic phase does not converge since it is given by
  the sign of $\sum_i h_i$, which is a stochastic variable.
  
\item Ising spin glasses in the chaotic pairs scenario. For each size
  the system presents two pure states (related by a global spin flip
  symmetry), however, these two states vary in a chaotic way as the
  lattice size grows.\cite{newman:92,newman:96b,newman:98}
\end{enumerate}

The concept of metastate was introduced to tackle the intrinsic lack of
convergence that show these
models\cite{newman:92,newman:96b,newman:98}. Despite the lack of
convergence of the sequence $\Gamma_{L,J}$, it is possible to compute
the frequency a given state appears in that sequence as $L\to
\infty$. The set of these frequencies defines the Newman-Stein  metastate.

\begin{figure}[ht]
  \begin{center}
    \includegraphics[angle=0,width=1.2\columnwidth]{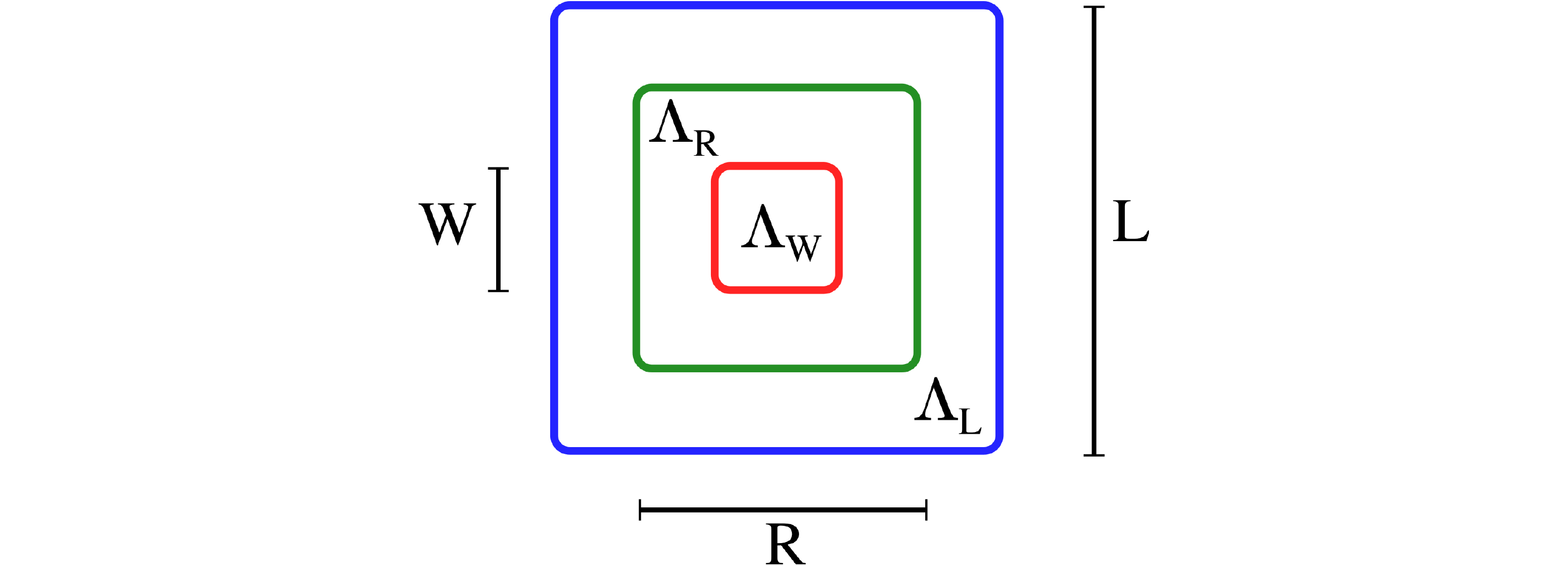}
    \end{center}
  \caption{Construction of the Aizenman-Wehr metastate. Notice the different boxes (squares in the figure) defined in order to compute the metastate.\cite{billoire:17}}
	\label{fig:defAWM}
\end{figure}

Before the introduction of the Newman-Stein metastate, another
definition of the metastate was given by Aizenman and
Wehr.\cite{aizenman:90} Although, a proof of  the equivalence
of both metastates is still lacking.

The definition of the Aizenman-Wehr allows for a numerical implementation. The construction of the 
Aizenman-Wehr is as follows:

\begin{enumerate}
\item For simplicity, let us consider a two-dimensional system, see
  Fig.~\ref{fig:defAWM}. In this figure we draw three different
  regions: $\Lambda_W$, $\Lambda_R$ and $\Lambda_L$.
\item We call the disorder inside the inner region $\Lambda_R$, 
  inner disorder, denoted as $\mathcal{I}$.
\item The disorder in the outer region  $\Lambda_L \backslash \Lambda_R$ is called an outer disorder and denoted as 
  $\mathcal{O}$.
  \item The measurements will be done in the region $\Lambda_W$. To
    avoid influences of the transition region between the inner and
    outer regions, we will try to implement the limit $\Lambda_W <<
    \Lambda_R << \Lambda_L$.
\end{enumerate}

In this framework we are interested to compute the following limit
\begin{equation}
\kappa_{\mathcal{I},R}(\Gamma) = \lim_{L\to\infty} \mathbb{E}_\mathcal{O}\Big[\delta^{(F)}\left(\Gamma - \Gamma_{L,J}\right)\Big]\,,
\end{equation}
where $\mathbb{E}_\mathcal{O}[\cdots]$ denotes the average over the outer disorder.

Assuming that the following limit exits
\begin{equation}
  \kappa(\Gamma) = \lim_{R\to\infty} \kappa_{\mathcal{I},R}(\Gamma),
\end{equation}
then it will not depend on the inner disorder $\mathcal{I}$ and will provide us the 
Aizenman-Wehr metastate.

We can compute averages over the metastate by using
\begin{equation}
\langle \cdots \rangle_\rho \equiv [\langle \cdots \rangle_\Gamma]_\kappa\,.
\end{equation}
Therefore, the metataste is also a state.

To characterize the metastate, it is very useful to compute the following correlation function
\begin{equation}
C_\rho(x) =\overline{\left[\langle s_\mathbf{0} s_\mathitbf{x} \rangle_\Gamma \right]_\kappa^2}
\sim |x|^{-\left(D-\zeta\right)}\;,
\label{eq:Cmeta}
\end{equation}
where $\zeta$ is a new exponent introduced by  Read.\cite{read:14}

The $\zeta$-exponent provides  important information about the structure of the states of the spin glass phase:
 \begin{itemize}
  \item $\log \mathcal{N}_\mathrm{states}(W)\sim W^{D-\zeta}$ (with
    $\zeta\ge 1$). Where $\mathcal{N}_\mathrm{states}(W)$ is the number of states in a system with size $W$.
    
  \item If $\zeta<D$ then metastate is disperse (therefore, the droplet theory does not hold).
    
  \item Read conjectured $\zeta=D-\theta(0)$, where, we recall,
    $\theta(0)$ is the exponent of the replicon mode, see Eq.~(\ref{eq:Crqdecay}) of Sec. \ref{sec:fieldtheory}.\cite{read:14}
    
\end{itemize}

\section{~Some numerical results}
\label{sec:numres}

After the discussion of some aspects of the theories aimed to explain
the properties of spin glasses in infinite and finite dimensions, we
report in the following sections numerical simulations at
equilibrium in presence/absence of the magnetic field.

\subsection{~Phase transition at $h=0$}
\label{sec:exponentsh0}
\index{phase transition!$h=0$}

Let us present numerical results showing the existence of a
phase transition in the three dimensional Ising spin glass at
$h=0$.

Usually, the numerical simulations are performed using the Monte Carlo
exchange method (also known as parallel tempering).\cite{hukushima:96,
  marinari:98b, marinari:98i,janke:13} Another popular approach in the last
years has been the population annealing
method.\cite{machta:10,wang:15b, wang:15c}

In this section, we closely follow Ref.~[\refcite{janus:13}] where the
supercomputer Janus I was used, see Sec.~\ref{sec:janus} for a
description of the most important characteristic of this computer.

As it has been stated in this chapter, the spin glass order parameter is based
on
the overlap. In numerical simulations, one can run in parallel different
non-interacting real replicas, denoted, as usual, as $s_\mathitbf{x}^a$. Usually, the
number of simulated real replicas is between two and six. We can recall
again the general definition of the overlap ($a\neq b$)
\begin{equation}
q_{\mathitbf{x}}^{ab}=s_{\mathitbf{x}}^a s_{\mathitbf{x}}^b\,.
\end{equation}
In the
simulations reported in this section\cite{janus:13} four copies of the system were
simulated with the same disorder (real replicas), therefore, one  can compute six different overlaps $q^{ab}$.

One can define the overlap-overlap correlation
function given by
\begin{equation}
  G(\mathitbf{r})=\frac{1}{V}\sum_{\mathitbf{x}} \overline{\langle q_{\mathitbf{x}
      +\mathitbf{r}}^{ab} q_{\mathitbf{x}}^{ab}\rangle}\, ,
\end{equation}
and finally the total overlap per volume of the system ($V$)
\begin{equation}
q^{ab}=\frac{1}{V}\sum_{\mathitbf{x}}  q_{\mathitbf{x}}^{ab}\,.
\end{equation}

Next, we compute the Fourier transform of the overlap-overlap
correlation function
\begin{equation}
  \widetilde G(\mathitbf{k})=\frac{1}{V}\sum_{\mathitbf{r}}G(\mathitbf{r})\,\mathrm{e}^{\mathrm{i}
    \mathitbf{k}\cdot\mathitbf{r}}\,,
\end{equation}
which allows us to compute the second-moment correlation length\cite{cooper:82,amit:05} 
\begin{equation}\label{eq:xi-second-moment}
\xi = \frac{1}{2 \sin (k_\mathrm{min}/2)} \sqrt{\frac{\widetilde G(0)}{\widetilde G(\mathitbf{k}_\mathrm{min})} -1},
\end{equation}
where ${\mathitbf{k}_\mathrm{min}}=(2\pi/L,0,0)$ or the other two
possible permutations. The total susceptibility is given by the
correlation function computed in the Fourier space at zero momentum
\begin{equation}
\chi_\mathrm{SG}= \widetilde G(0) = V \overline{\langle q^2\rangle}\,.
\end{equation}
Finally, the correlation length in units of system  size  $\xi/L$ is
\begin{equation}\label{eq:Rxi}
R_\xi = \xi/L\,.
\end{equation}
We recall that $\xi/L$ is universal at the critical point.

In addition to $\xi/L$ we can define the following observables which also take universal values at the critical point:
\begin{equation}
U_4=\frac{\overline{\langle q^4\rangle}}{\overline{\langle
    q^2\rangle}^2}\,,\label{eq:U4-def}
\end{equation}
\begin{equation}
U_{22}=\frac{\overline{\langle q^2\rangle^2}}{\overline{\langle
    q^4\rangle}}\,,\label{eq:U22-def}
\end{equation}
\begin{equation}
U_{111}=\frac{\overline{\langle
    q^{12}q^{23}q^{31}\rangle}^{\,4/3}}{\overline{\langle q^4\rangle}}\,,\label{eq:U111-def}
\end{equation}
\begin{equation}
U_{1111}=\frac{\overline{\langle q^{12}q^{23}q^{34} q^{41}\rangle}}{\overline{\langle q^4\rangle}}\,,\label{eq:U1111-def}
\end{equation}
\begin{equation}
  R_{12}=\frac{\widetilde G(2\pi/L,0,0)}{\widetilde G(2\pi/L,2\pi/L,0)},\label{eq:R12-def}
\end{equation}
  \begin{equation}
B_\chi = 3 V^2 \frac{\overline{\langle |\widetilde q
    (2\pi/L,0,0)|^4\rangle}}{[\widetilde G(2\pi/L,0,0)]^2}\,,\label{eq:Bchi-def}
\end{equation}
where
\begin{equation}
\widetilde q^{ab}(\mathitbf{k}) = \frac{1}{V} \sum_{\mathitbf{x}}\ q^{ab}_{\mathitbf{x}}\,
\mathrm{e}^{\mathrm{i} \mathitbf{k}\cdot\mathitbf{x}}\,.
\end{equation}

\begin{figure}[ht]
  \centering \includegraphics[width=0.65\columnwidth,
    angle=270]{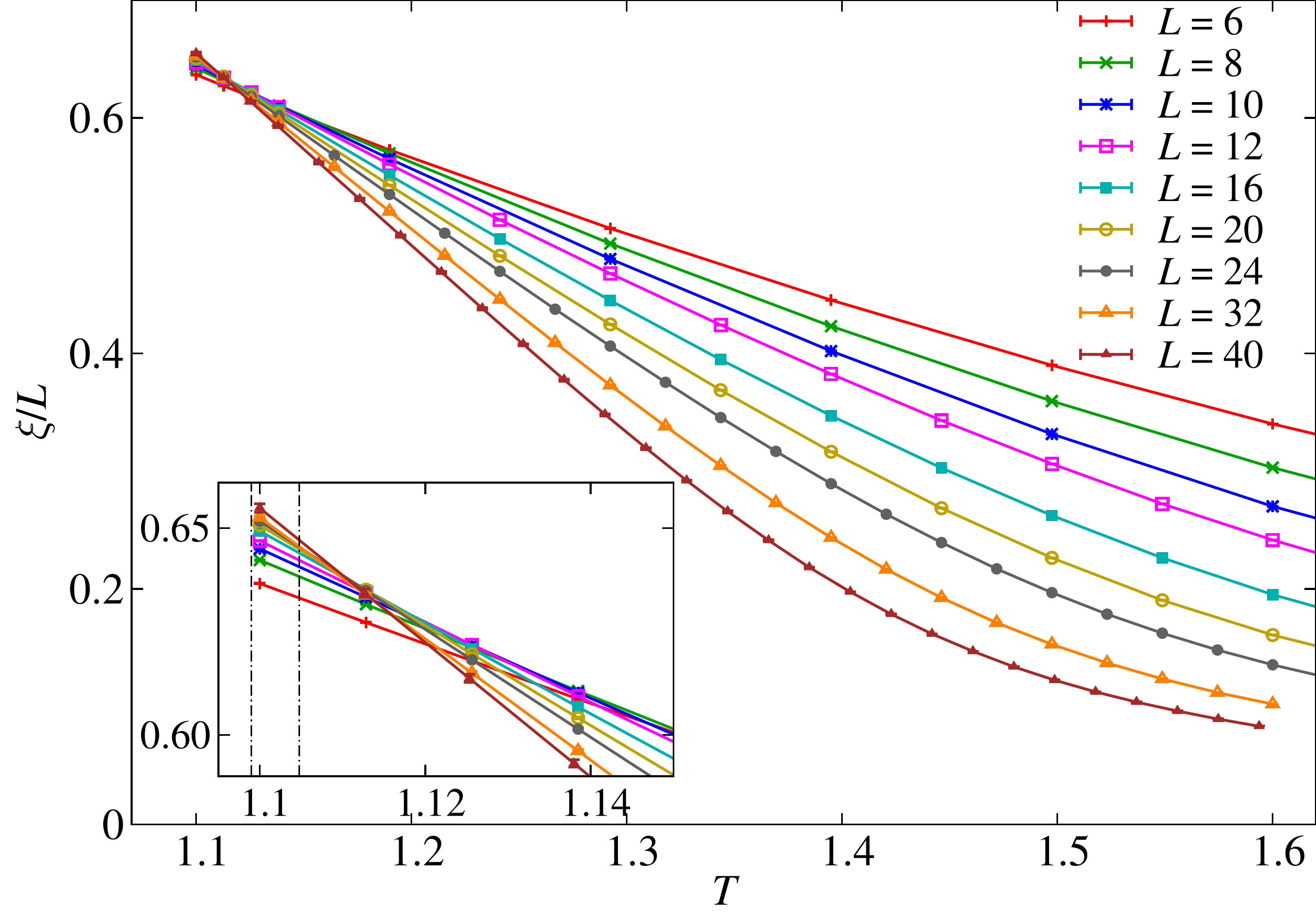}
  \caption{ Behavior of the second-moment
    correlation length as a function of the temperature for different
    values of the size of the system. In the inset, we show a zoom of
    the region in which $\xi/L$ curves cross. We mark with two
    vertical lines the final value of the critical temperature, and
    taking  into account the statistical error, one can
    quote: $T_\text{c} = 1.1019(29)$.\cite{janus:13}\label{fig:cortes}}
\end{figure}

We show in Fig.~\ref{fig:cortes} the different crossing points of the
$R_\xi$ as a function of the temperature and the size of the
system. The study of these crossing points by means of the quotient
method (see Sec.~\ref{sec:quotient} for a detailed description of this
method) allows to compute the critical exponents and other
universal\index{scaling!exponent} quantities.\cite{janus:13} The
results of this analysis can be read in
Table~\ref{tab:parameters}.\cite{janus:13}
\begin{table}[ht]
  \centering    
  \tbl{Summary of the cri\-ti\-cal exponents and the values (universal) of
    different cu\-mu\-lants at the critical point.\cite{janus:13}}
{
  \begin{tabular}{c}
  \hline
  \hline
Exponent/Observable \\
\hline
$\omega =1.12(10)  $  \\
$\eta = -0.3900(36) $ \\
$\nu = 2.562(42)    $  \\
$R_\xi^* = 0.6516(32)$  \\
$U_4^* = 1.4899(28)  $  \\
\hline
$\alpha =  -5.69(13)$  \\
$\beta = 0.782(10)$   \\
$\gamma = 6.13(11)$    \\
\hline
$T_{c} = 1.1019(29)$  \\
$\makebox[0.05\linewidth]{}U_{1111}^* =  0.4714(14)$\makebox[0.05\linewidth]{}\\
$U_{22}^* = 0.7681(16)$  \\
$U_{111}^* = 0.4489(15)$  \\
$B_\chi^* =   2.4142(51)$  \\
$R_{12}^* = 2.211\,\pm\, 0.006$  \\
  \hline
  \hline
  \end{tabular}
}
\label{tab:parameters}
\end{table}

In this part of the book chapter, we have only focused on the Ising
spin glass with binary couplings. The field theory of Ising spin glass
only depends on the first two cumulants of the probability
distribution of the couplings:  the other (infinite) cumulants of the
probability distribution of the couplings $J_{ij}$ only induce
irrelevant couplings in the field theory, and following the general
theory of the renormalization
group,\cite{wilson:74,wilson:75,amit:05,cardy:96,parisi:88,itzykson:89}
they only induce scaling corrections without changing the values of
the critical exponents (universality). This issue has been checked in
Refs.~[\refcite{palassini:99,jorg:08,jorg:08b,jorg:08c}]. However, some  
references, see
for example Ref.~[\refcite{bernardi:95}], have reported  violation of universality in Ising
spin glass models.

\subsection{~Behavior of the correlation functions in the spin glass phase}
\label{sec:cor}
\index{phase!spin glass}

Once we have characterized the critical point, we try to characterize
the properties of the low temperature spin glass phase.

\begin{figure}[t]
\centering
\includegraphics[angle=0,width=0.75\columnwidth]{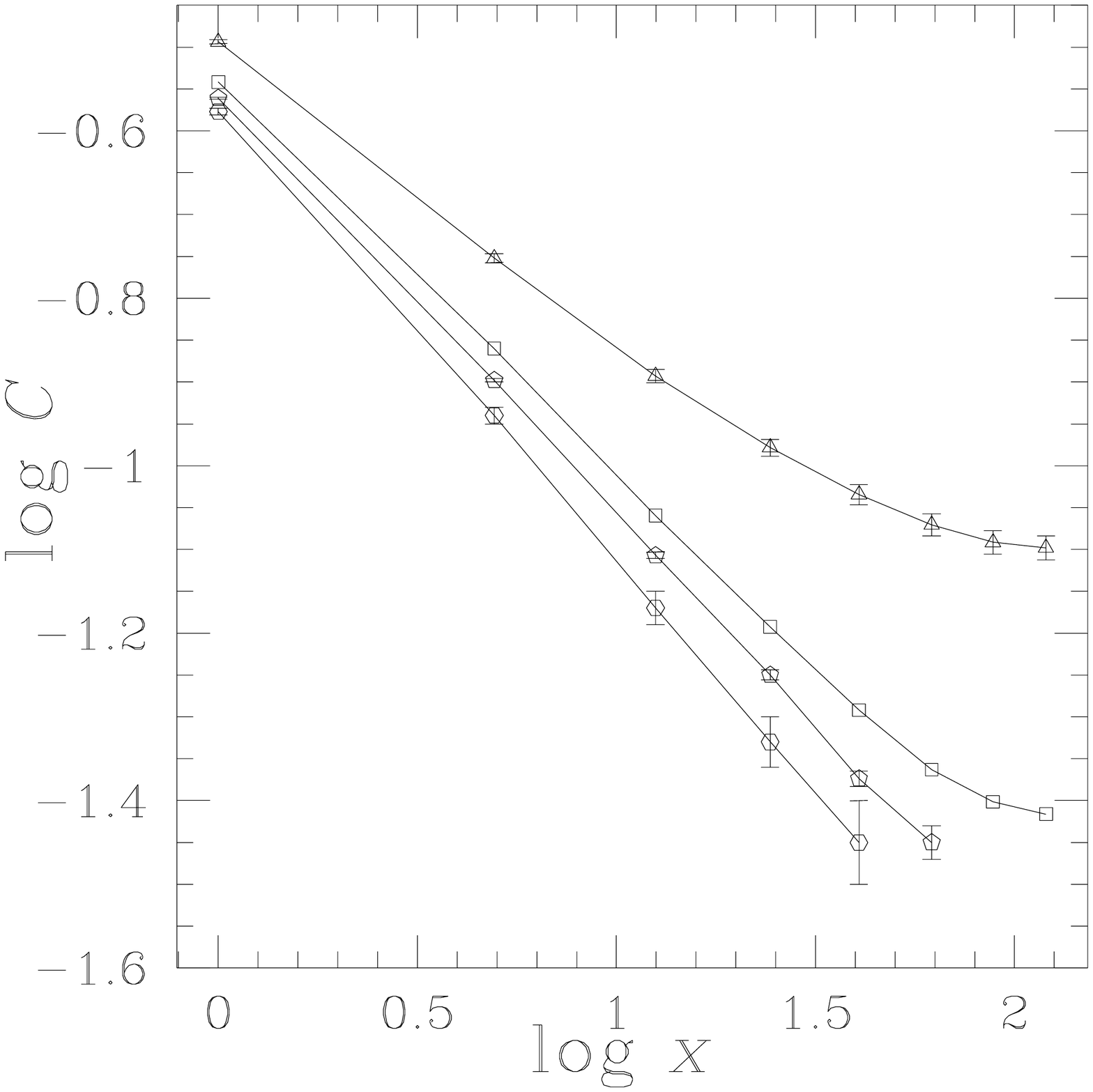}
\caption{Conditional overlap-overlap correlation function $C\equiv C_4(x|q)$ for
  the three-dimensional Ising spin glass with Gaussian
  couplings. $T=0.7\simeq 0.7 T_c$.
  The lower curve is the infinite time extrapolation of the non-equilibrium
  correlation function $C_4(x|q = 0)$ obtained by a sudden quench ($L=64$). The second curve
from the bottom is $C_4(x|q = 0)$ obtained by a slow annealing ($L=64$). The third curve
is the equilibrium correlation function computed with the constraint $|q|<0.01$ ($L=16$).
The upper curve is the full equilibrium correlation function, including all configurations
($L=16$). From
  Refs.~[\refcite{marinari:96,marinari:98d}].}
\label{fig:corR}
\end{figure}

\begin{figure}[t]
\centering
\includegraphics[height=\linewidth,angle=270]{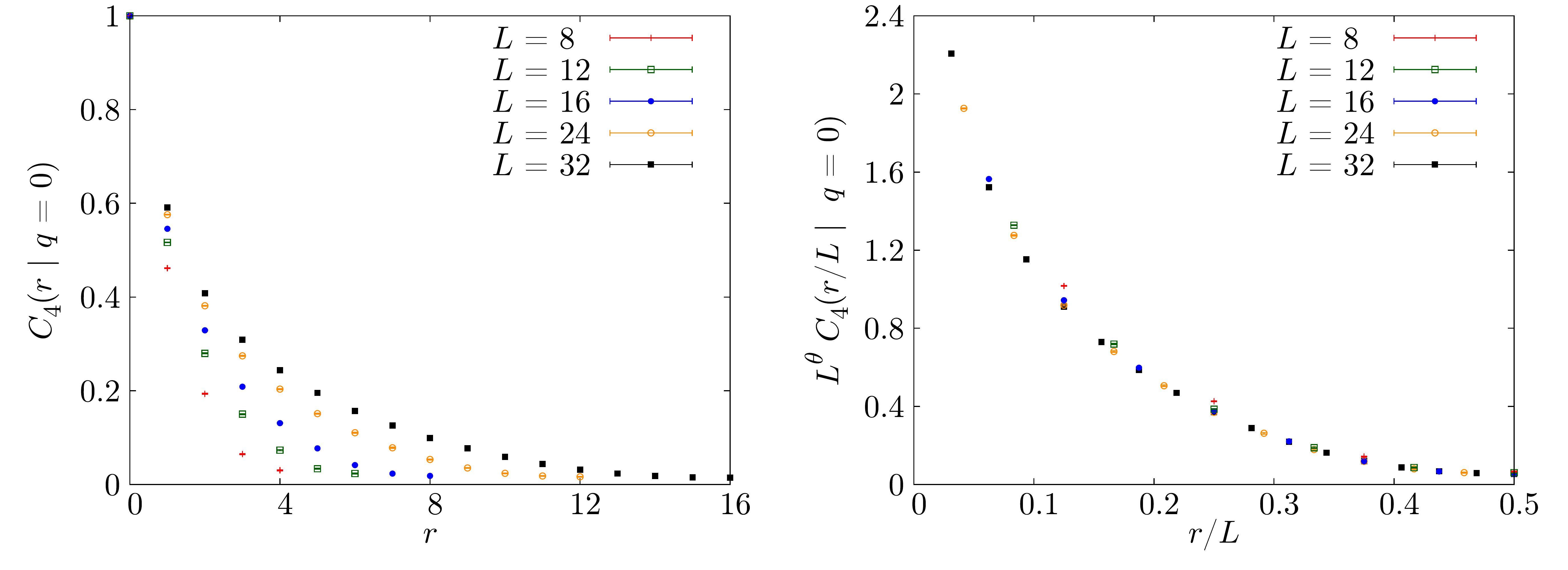}
\caption{$C_4(r|q=0)$ versus $r$ for  $T=0.703\simeq 0.7 T_c$ (left panel). 
We show in the right panel the scaling collapse of $L^{\theta(0)}C_4(r/L|q=0)$ using the replicon
exponent $\theta=0.38$ as a function of $r/L$.\cite{janus:09b}}
\label{fig:C4}
\end{figure}

Let us consider the conditional spatial correlation function
$C_4(r|q)$, see Eq.~(\ref{eq:Crq}). We focus on the
real space behavior of this conditional correlation function, a
detailed study of its Fourier transform can be found in Ref.~[\refcite{janus:10b}].

The first numerical computation of $C_4(r|q=0)$ was performed in
Ref.~[\refcite{marinari:96}] working in the out-of-equilibrium regime
(with an extrapolation to infinite time). Subsequently, this
out-equilibrium correlation function was confronted with an equilibrium  computation of the
same observable.\cite{marinari:98d} In
Fig.~\ref{fig:corR} we show these two $C_4(r|q=0)$ (lower curves of
this figure) together with the full correlation one (the upper curve of
the figure): notice the very good agreement between the
out-of-equilibrium and equilibrium computation of $C_4(r|q=0)$.

The conditional correlation function $C_4(r|q=0)$ corresponds with the
replicon correlation function in the framework of RSB theory and it
should decay algebraically with the exponent $\theta(0)$, see
Eq.~(\ref{eq:Crqdecay}) of Sec. \ref{sec:fieldtheory}. This power law
decay was corroborated in Refs.~[\refcite{marinari:96,marinari:98d}]
(see Fig.~\ref{fig:corR}). The replicon exponent was estimated to be
$\theta(0)=0.50(2)$ for the three dimensional Ising spin glass with
Gaussian couplings.\cite{marinari:96}

Most recent out-of-equilibrium numerical simulations performed on the
Janus I supercomputer have provided with an accurate value of the
replicon exponent: $\theta(0)=0.38(2)$.\cite{janus:09b,janus:10b}

In the next paragraphs, we report some results obtained at equilibrium
in the three dimensional Ising spin glass model with the help of Janus
I. In these numerical simulations, we were able to proceed up to $L\le
32$ inside the thermalized low temperature phase.\cite{janus:10}

\begin{figure}[t]
\centering
  \includegraphics[height=0.9\linewidth,angle=270]{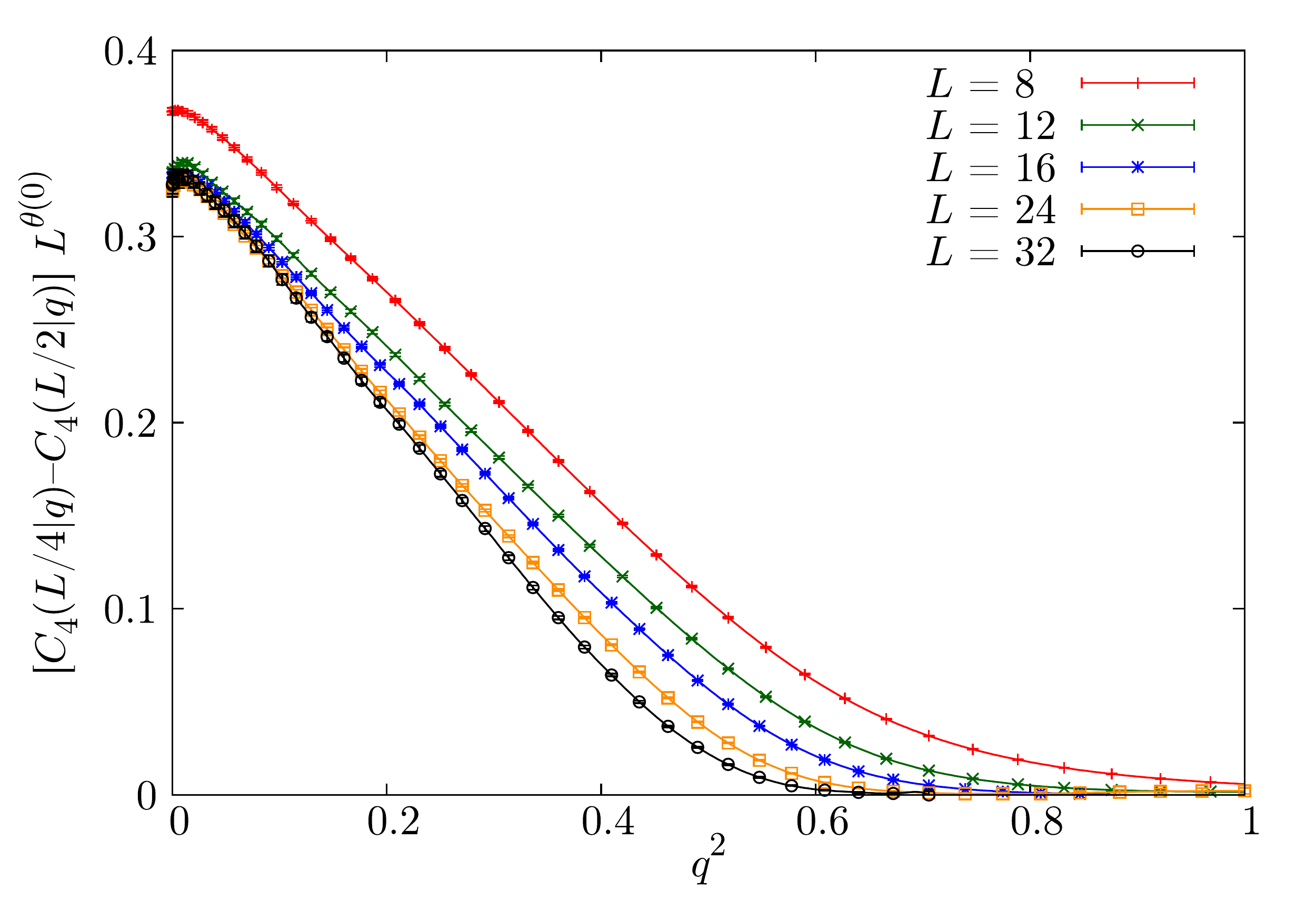}
\caption{Rescaled difference of the correlation function,
  Eq.~(\ref{eq:C4-substraida}), as
  function of $q^2$. We have used
  $\theta(0)=0.38(2)$ from a out-of-equilibrium numerical simulation.\cite{janus:09b}}\label{fig:C4-substraida}
\end{figure}

First, we study the $q=0$ sector in which the prediction of the
droplet model and the RSB theory are fully different. In left panel of
Fig.~\ref{fig:C4} we show $C_4(r|q=0)$ for $T=0.703\sim 0.7 T_c$ which
goes to zero for large $r$. In the right panel of this figure we show
the scaling plot of this observable by using the replicon exponent
$\theta=0.38(2)$ and not that predicted by the droplet theory
$y=\theta(q_\mathrm{EA})=0.2$.\cite{bray:87} Note that by using
$\theta=0.38(2)$ the curves collapse (see Fig.~\ref{fig:C4}-left panel).

Second, we consider the rescaled difference of the conditional
correlation function evaluated at $r=L/4$ and $r=L/2$, which following RSB should behave as
(see Eq.~(\ref{eq:Crqdecay}))
\begin{equation}\label{eq:C4-substraida}
L^{\theta(q)} \Big(C_4(r=L/4|q)-C_4(r=L/2|q) \Big) \sim 1\,,
\end{equation}
in order to remove the background, for large distances, in the
$C_4(r|q)$ correlation function (finite volume effect). 
This rescaled difference as a function of $q^2$ is shown in Fig. \ref{fig:C4-substraida}.
Notice that
a good scaling was found in the region $q^2<0.2$, this means that the correlation functions
$C_4(r|q)-q^2$ decay following a power law in the overlap interval (see also Ref.~[\refcite{contucci:09}]).

Moreover, for $q^2>0.2$, the scaling breaks down pointing to a value of
the $\theta$-exponent used to rescale in Eq.~(\ref{eq:C4-substraida})
that is larger than $\theta(0)$. A detailed analysis\cite{janus:10b}
suggests that $\theta(q_\mathrm{EA})\sim 0.6$ and that the crossover
between the small and large $q^2$ can be analyzed using finite size
scaling.\cite{janus:10b}

To close this section, we report the final values of these exponents
obtained in these numerical simulations at equilibrium:
$\theta(0)=0.377(14)$\cite{janus:10} and $\theta(q_\text{EA})\simeq
0.511(16)(60)$.\cite{janus:10} Finally we quote that an
out-of-equilibrium analysis provided $\theta(q_\text{EA})\simeq
0.78(10)$.\cite{janus:09b,janus:10b}

\subsection{~Metastate: ~Numerical results}
\label{sec:MetastateNum}
\index{state!metastate}

In this section, we build numerically the metastate following the
Aizenman and Wehr\cite{aizenman:90} and we present the main results
obtained in Ref.~[\refcite{billoire:17}].

We start with some definitions.  The average over the Gibbs state
$\langle\cdots\rangle_\Gamma$ is estimated via Monte Carlo thermal
averages $\langle\cdots\rangle$ at fixed disorder $J$.

The average over the metastate is given by
\begin{equation}
[\cdots]_\kappa = \frac{1}{\mathcal{N_O}} \sum_\mathtt{o} (\cdots) \,,
\end{equation}
and the one over the internal disorder by
\begin{equation}
  \overline{(\cdots)} = \frac{1}{\mathcal{N_I}} \sum_\mathtt{i} (\cdots)  \,.
\end{equation}
The indices $\mathtt{i}$ and $\mathtt{o}$ run over the number of inner and outer disorder realizations, denoted as 
$\mathcal{N_I}$ and $\mathcal{N_O}$, respectively.

For example, the metastate spin correlation function (see
Eq.~(\ref{eq:Cmeta})) can be explicitly computed as
\begin{eqnarray}
  \nonumber
  C_\rho(|\mathitbf{x}|) &=&\overline{\left[\langle s_\mathbf{0} s_\mathitbf{x} \rangle_\Gamma \right]_\kappa^2}\\ \nonumber
  &=&
  \frac{1}{\mathcal{N_I}} \sum_\mathtt{i} \left( \frac{1}{\mathcal{N_O}} \sum_\mathtt{o} \langle s_\mathbf{0}^\mathtt{i;o}
  s_\mathitbf{x}^\mathtt{i;o} \rangle \right)^2\\
&=& \frac{1}{\mathcal{N_I}} \sum_\mathtt{i} \frac{1}{\mathcal{N}_\mathcal{O}^{\,2}} \sum_\mathtt{o,o'} \langle s_\mathbf{0}^\mathtt{i;o} s_\mathitbf{x}^\mathtt{i;o} s_\mathbf{0}^\mathtt{i;o'} s_\mathitbf{x}^\mathtt{i;o'} \rangle \;.
\label{E_CX}
\end{eqnarray}

Inside $\Lambda_W$, see Fig.~\ref{fig:defAWM},  we can compute a generalized overlap using two non-interacting real replicas (Ising spins), denoted
as $\{\sigma \}$  and $\{\tau \}$. These two real replicas share the
same disorder, denoted by $\mathtt{i}$, with different or the same
outer disorder, denoted as $\mathtt{o}$, for $\sigma$,  and
$\mathtt{o'}$, for $\tau$:
\begin{equation}
  q_\mathtt{i;o,o'} \equiv \frac{1}{W^3}\sum_{\mathitbf{x}\in\Lambda_W} \sigma_\mathitbf{x}^\mathtt{i;o}
  \tau_\mathitbf{x}^\mathtt{i;o'}\;.
  \label{eq:qOOIrr}
\end{equation}
Notice the possible cases: $\mathtt{o}=\mathtt{o'}$ and
$\mathtt{o}\neq\mathtt{o'}$. Therefore, we can compute the
following probability distributions of the generalized overlap $q_\mathtt{i;o,o'}$: 
\begin{equation}
P_\mathtt{i}(q) = \frac{1}{\mathcal{N_O}} \sum_\mathtt{o} \langle \delta(q-q_\mathtt{i;o,o}) \rangle \,,
\end{equation}
\begin{equation}
P(q) = \frac{1} {\mathcal{N_I}}    \sum_\mathtt{i} P_\mathtt{i}(q) \,,
\end{equation}
\begin{equation}
P_{\rho,\mathtt{i}}(q) =\frac{1}{\mathcal{N}_\mathcal{O}^{\,2}} \sum_\mathtt{o,o'} \langle \delta(q-q_\mathtt{i;o,o'}) \rangle\,,
\end{equation}
\begin{equation}
P_\rho(q) = \frac{1}{\mathcal{N_I}} \sum_\mathtt{i} P_{\rho,\mathtt{i}}(q)\,.
\end{equation}

 \begin{figure}[ht]
\centering
\includegraphics[width=0.95\columnwidth]{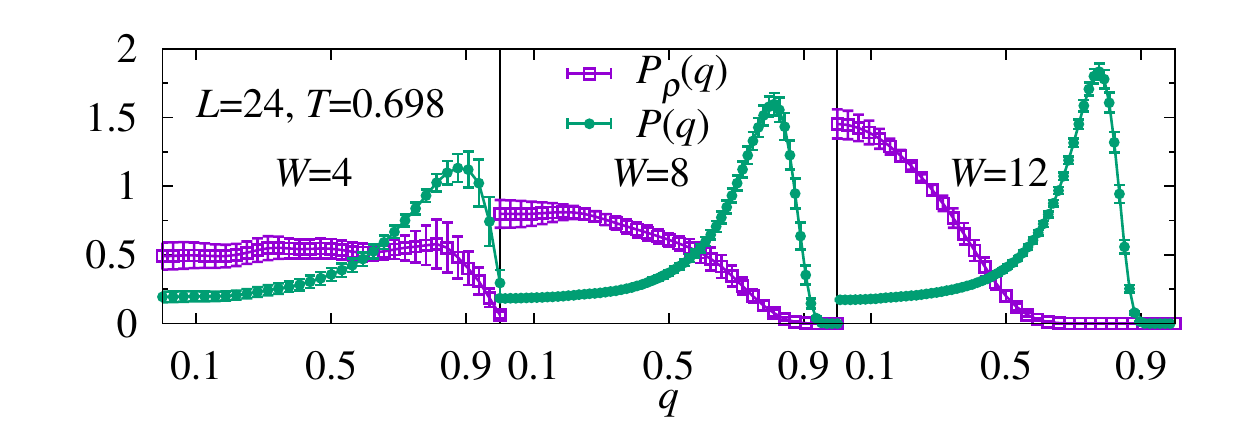}
\caption{The overlap probability distributions $P_\rho(q)$ and $P(q)$
  against the overlap for $L=24$, $R=L/2$, $T=0.698$ and for
  different values of the window size, $W=4,8$ and
  12.\cite{billoire:17}
\label{fig:AW4}}
\end{figure} 

$P(q)$ is just the standard overlap probability
distribution, but $P_\rho(q)$ is the probability distribution of the
overlap averaged over the metastate.

Regarding $P_\rho(q)$, despite having a trivial limit $P_\rho(q) \to
\delta(q)$ for $W\to\infty$,\cite{read:14} its variance for finite
values of $W$ provides with very useful information
\begin{equation}
\chi_\rho = \sum_{\mathitbf{x}\in\Lambda_W} C_\rho(x) =  W^d \int q^2 P_\rho(q)\,dq \sim W^\zeta\,,
\label{eq:chiA}
\end{equation}
which allows us to write the following scaling behavior
\begin{equation}
  \label{eq:scalingb}
\chi_\rho(W,R) = R^\zeta f(W/R) = W^\zeta g(W/R)\;.
 \end{equation}

 Finally, it is possible to show that  the re-scaled metastate-averaged probability distribution  is Gaussian:\cite{read:14}
$ P_\rho(q/(W^{-(\zeta-D)/2})\,.$

 In Fig.~\ref{fig:AW4}, we present the behavior of the standard
probability distribution of the overlap, $P(q)$, and the metastate
one. The first point is that both probabilities are completely
different. Moreover, for large values of $W$ the Gaussian shape of
$P_\rho(q)$ starts to be emerging.

\begin{figure}[ht]
\centering
\includegraphics[width=0.95\columnwidth]{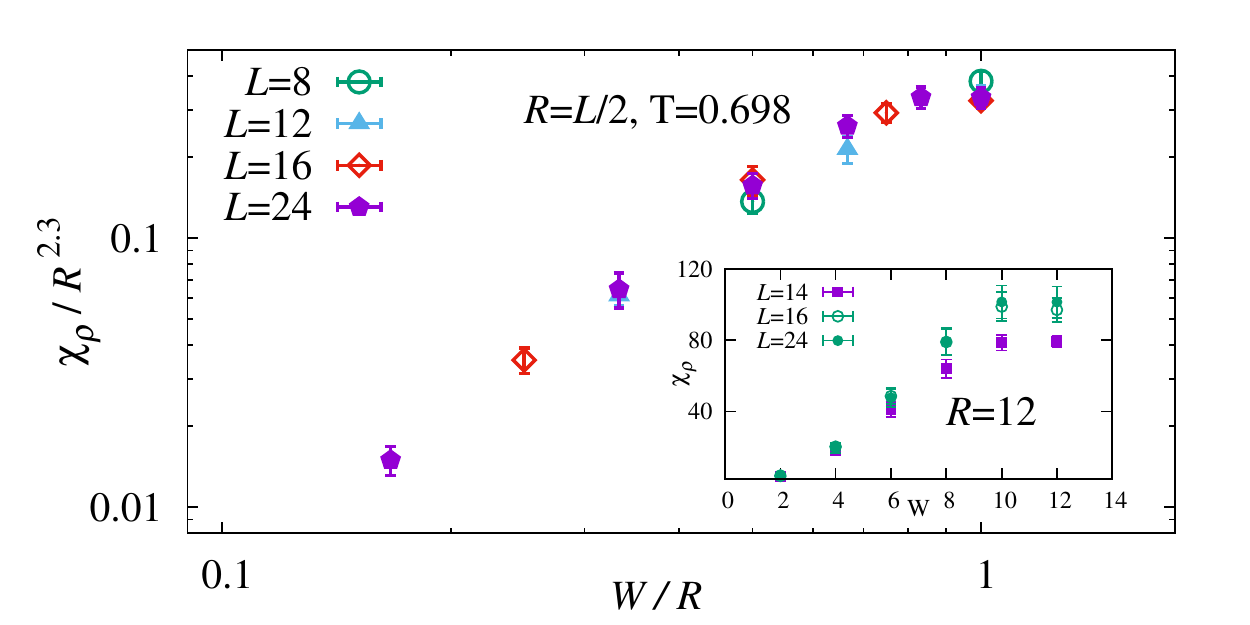}
\caption{ Scaling behavior of the metastate susceptibility $\chi_\rho$
  as a function of $W/R$ for $R=L/2$ at $T\simeq 0.64 T_c$. In the
  small panel we show the breakdown of the scaling behavior
  (asymptotic) which happens as $R/L>3/4$.\cite{billoire:17}
\label{fig:AW3}}
\end{figure} 

The scaling of the metastate susceptibility allows to compute the
$\zeta$-exponent obtained, $\zeta= 2.3(3)$. This value compares very
well with the replicon exponent $\zeta=2.62(2)$, providing support for
the Read conjecture regarding both exponents are the same. As a test
of the computed value, we show in Fig.~\ref{fig:AW3} a scaling plot of
the re-scaled susceptibility, see Eq.~(\ref{eq:scalingb}).

 \begin{figure}[ht]
\centering
\includegraphics[width=0.95\columnwidth]{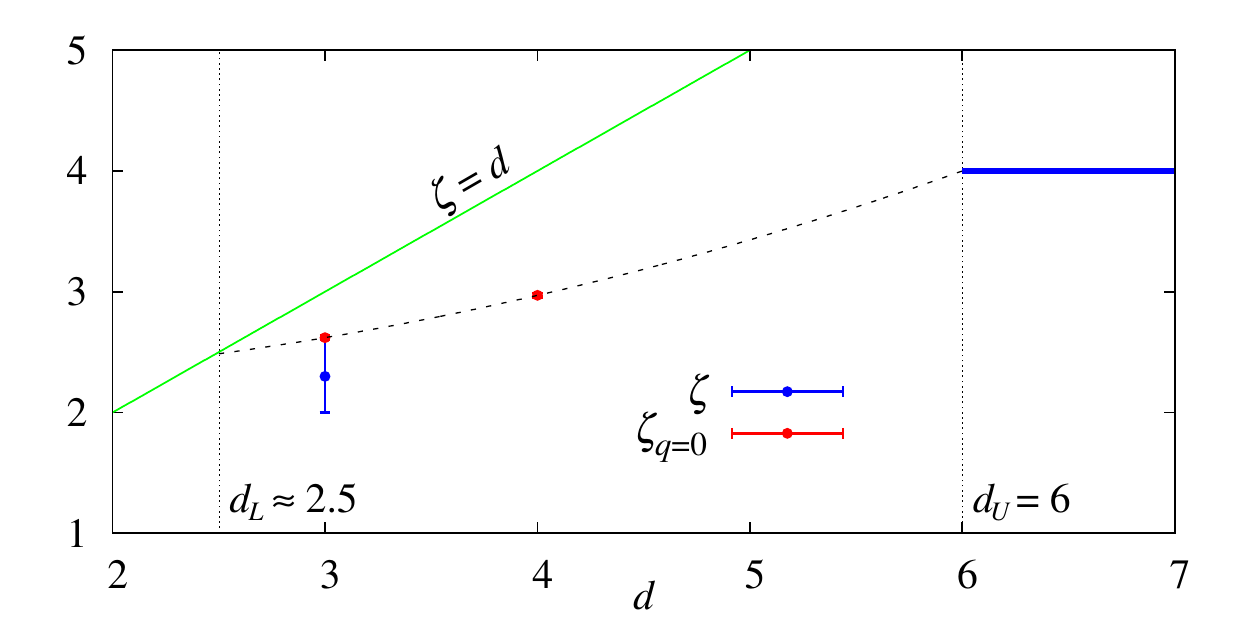}
\caption{The $\zeta$-exponent against dimensionality $d\equiv D$.\cite{billoire:17} In blue we have marked the known
  values (mean-field and $D=3$ computed in
  Ref.~[\refcite{billoire:17}]). In red we have marked the replicon
  exponent computed in numerical simulations at different
  dimensionalities (the Read conjecture is $\zeta=D-\theta(0)$).
  In green the droplet prediction for a non disperse
  metastate $D=\zeta$. Finally, we have drawn  vertical
  lines at the lower and upper critical dimensions ($D_l\simeq 2.5$ and $D_u=6$).\cite{billoire:17}
\label{fig:AW5}}
\end{figure} 

 Finally, we present in Fig.~\ref{fig:AW5} the dependence of the Read
 $\zeta$-exponent on the  space dimension, marking the mean-field value (the horizontal value for $D\ge D_U=6$), and the values
 obtained in numerical works.\cite{billoire:17}

 Therefore, the numerical implementation of the metastate approach
 unveil a structure of the low temperature phase with a disperse
 metastate (i.e. $D>\zeta$). Only chaotic pairs and the RSB theory
 satisfy the property to have a dispersed metastate and not the
 droplets.
 
\subsection{~Phase transition at $h\neq 0$} 
\label{sec:eqH}
\index{phase transition!$h\neq0$}

Since the work of de Almeida and Thouless\cite{dealmeida:78} forty
years ago, a large amount of work has been devoted to understand the
behavior of spin glasses in presence of a magnetic field (see Sec.~\ref{sec:infield}). In
particular during this long period of time a huge number of numerical
simulations have been
performed\cite{caracciolo:90,huse:91,caracciolo:91,ciria:93,parisi:98b,marinari:98e,marinari:98g,marinari:98h,houdayer:99,marinari:00d,houdayer:00,cruz:03,young:04,leuzzi:08,leuzzi:09,leuzzi:11,janus:12,larson:13,janus:14b,janus:14c},
though, a complete understanding of the model in a field is still
lacking. One of the reasons is that strong finite size corrections are
present in these systems which mask the infinite volume behavior.

\begin{figure}[ht]
\centering
\includegraphics[width=0.65\columnwidth,angle=270]{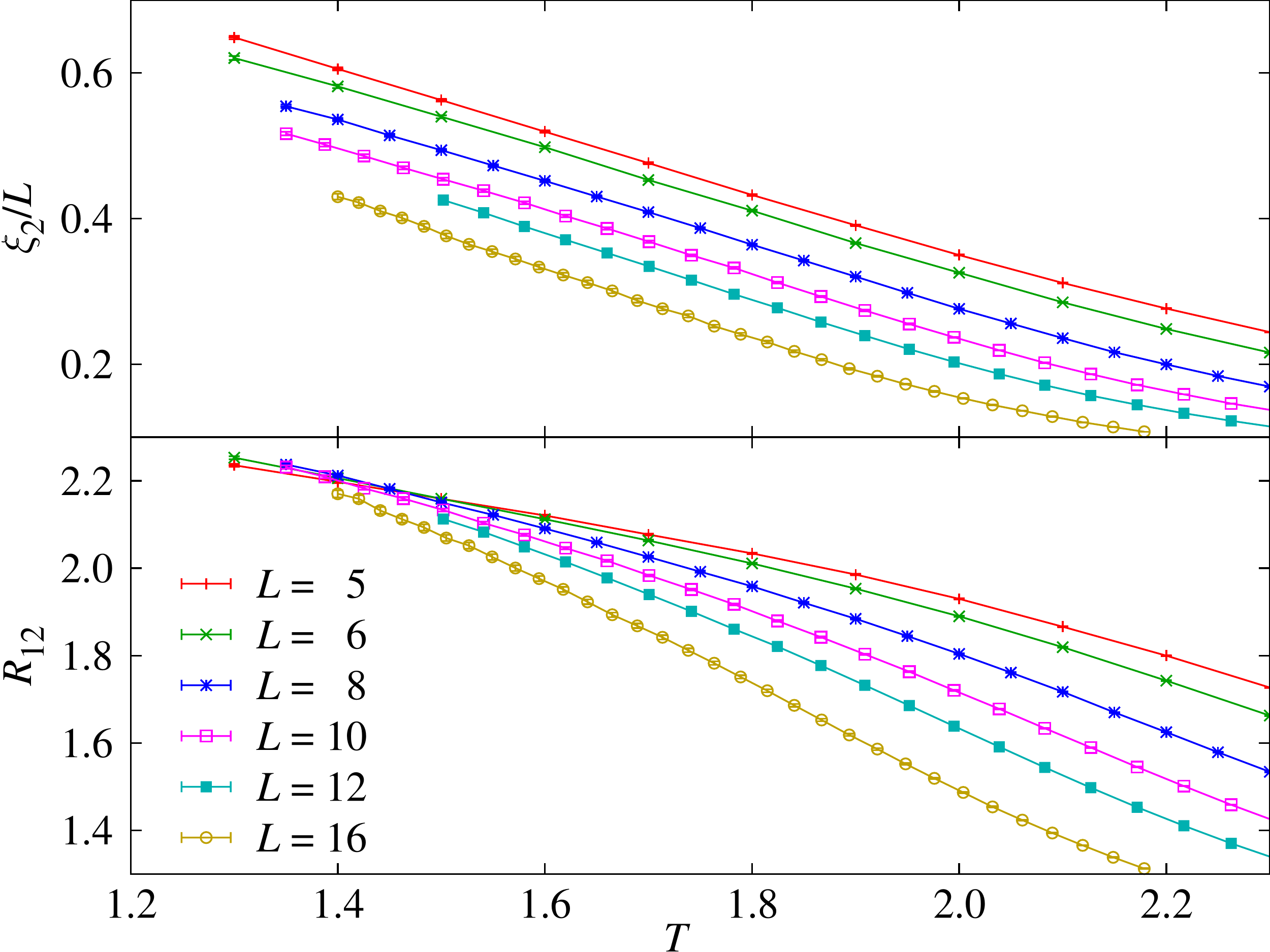}
\caption{ In the top panel we show $\xi_2/L$ for the simulated lattice sizes
  at $h=0.15$. Notice the lack of crossing points of this
  cumulant. The absence of these crossing points has been interpreted as
  a clear signature of a paramagnatic phase in the whole region of
  positive temperatures. However, as explained in the main text, this
  behavior can be associated with the presence of the zero mode in the
  definition of the correlation length.  In the bottom panel, we show
  the different curves for the $R_{12}$ cumulant, which avoids the
  zero mode in its definition. This observable presents crossing
  points for the different lattices, therefore, it is possible to
  characterize a second order phase transition, including the critical
  temperature and exponents.\cite{janus:12}
\label{fig:xi-R12}}
\end{figure}

\begin{figure}[ht]
\centering
\includegraphics[angle=0, width=0.9\columnwidth]{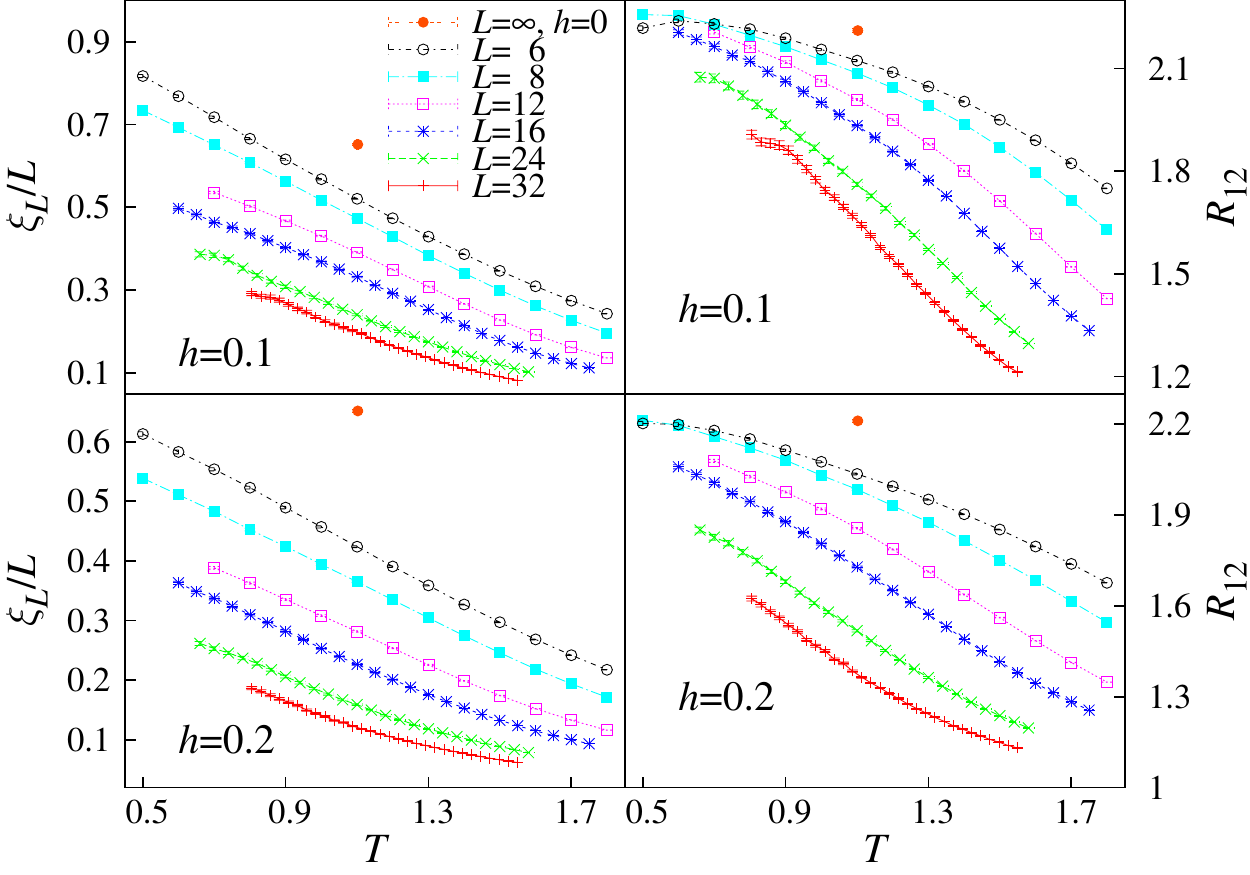}
\caption{We plot in the left panels $\xi/L$ (and $R_{12}$ in the right
  panels) as a function of the temperature for $h=0.1$ and $h=0.2$. As
  happens for $h=0.15$, see Fig.~\ref{fig:xi-R12}, the $R_{12}$-curves
  cross but those of $\xi/L$ ratio do not.\cite{janus:12}}
\label{fig:xi-R12H}
\end{figure}

As in the case of the $h=0$ numerical simulations, we have initially
based our analysis on the study of the behavior of the correlation
length in units of the lattice size $R_\xi$. To do that, we need to
compute a critical correlation function. In particular, in presence of
a magnetic field, one can extract the critical behavior from two
different correlation functions
\begin{eqnarray}
  \label{eq:corrH}
G_1(\boldsymbol r)&=& \frac{1}{L^4} \sum_{\boldsymbol{x}} \overline{\bigl(\langle s_{\boldsymbol{x}} s_{\boldsymbol{x}+\boldsymbol r}\rangle -\langle s_{\boldsymbol{x}}\rangle\langle s_{\boldsymbol{x}+\boldsymbol r}\rangle\bigr)^2}\,,\\
G_2(\boldsymbol r)&=& \frac{1}{L^4} \sum_{\boldsymbol{x}} \overline{\bigl(\langle s_{\boldsymbol{x}} s_{\boldsymbol{x}+\boldsymbol r}\rangle^2 -\langle s_{\boldsymbol{x}}\rangle^2\langle s_{\boldsymbol{x}+\boldsymbol r}\rangle^2\bigr)}\,.
\end{eqnarray}
Both correlation functions have the same critical behavior. The
associated correlation lengths are computed in the usual way by calling
Eq.~(\ref{eq:xi-second-moment}).

This correlation length in units of the lattice size has worked pretty
well in characterizing the phase transition at $h=0$ (see
Sec.~\ref{sec:exponentsh0}). However, in presence of a magnetic field,
it fails to identify a phase transition via the usual crossing point
of the different curves computed with different lattice sizes, see the
top panels of Fig.~\ref{fig:xi-R12} and the left panels of
Fig.~\ref{fig:xi-R12H}. This lack of crossing on the $\xi/L$-curves
has been interpreted in the past as a clear signal for a stable
paramagnetic phase for all positive temperatures (i.e. there is no
phase transition).

However, in some models based on random graphs, the phase transition
has not been found numerically even in models where the phase
transition has already been characterized
analytically.\cite{mezard:01,takahashi2010finite}

\begin{figure}[ht]
\centering
\includegraphics[angle=0,width=0.75\columnwidth]{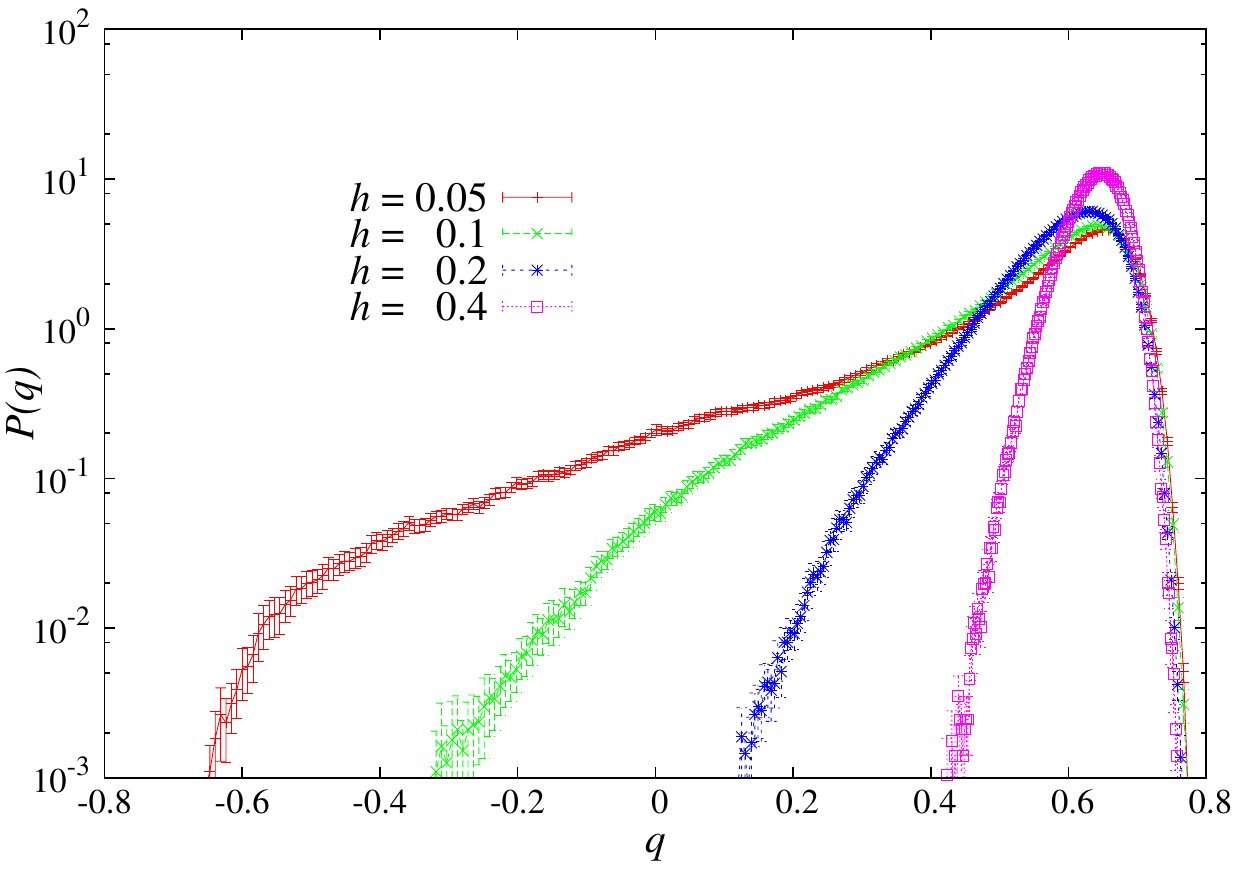}
\includegraphics[width=0.75\columnwidth]{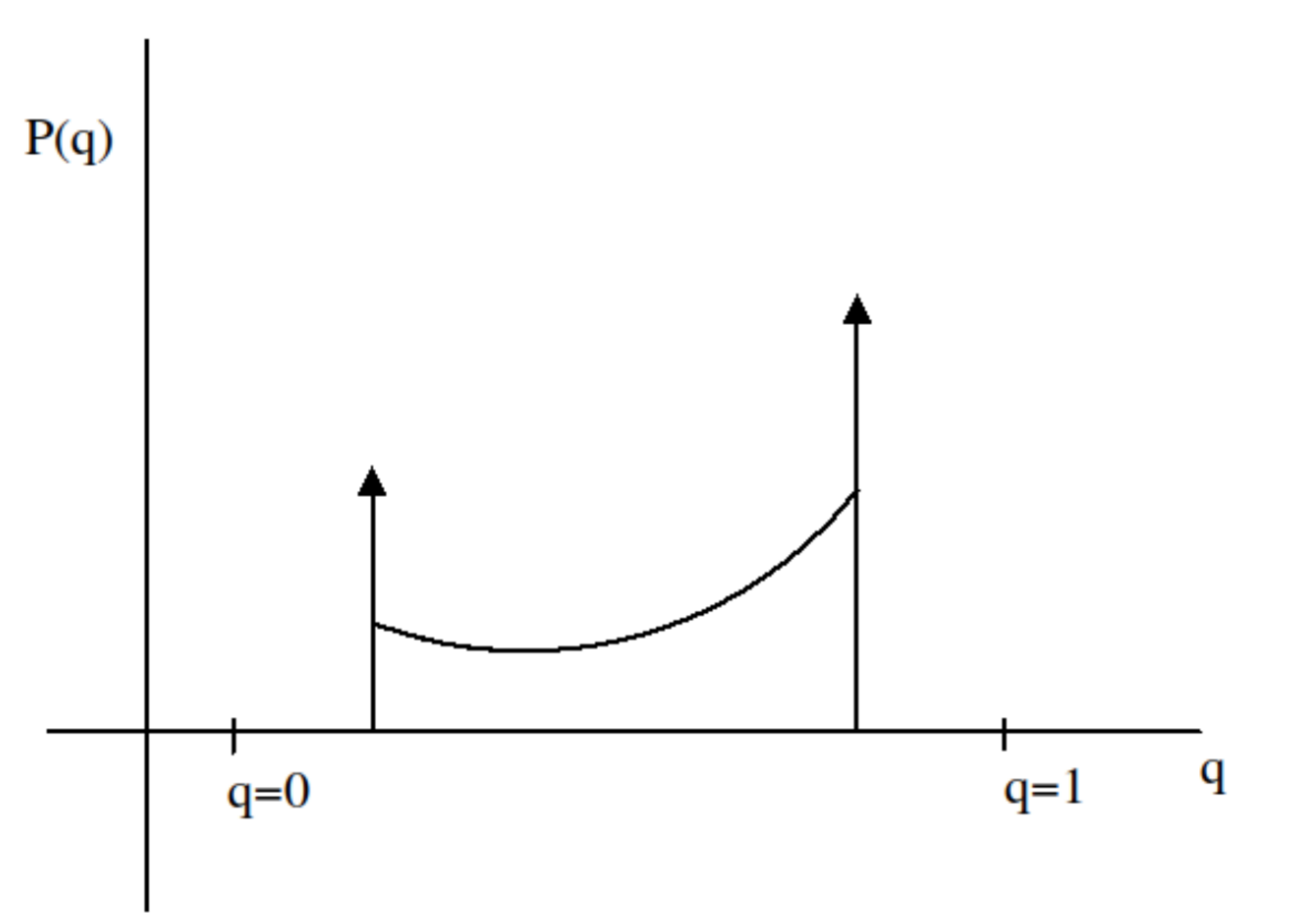}
\caption{$P(q)$ from numerical simulations (top) and RSB prediction (bottom) for the probability
  distribution of the overlap in the spin glass phase in presence
  of a magnetic field. The big arrows in the bottom panel represent
delta functions. The RSB $P(q)$ function is different from zero only for positive overlaps.\cite{janus:12}}
\label{fig:pqsimnum}
\end{figure}

In order to analyze the origin of these strong finite size effects,
which could spoil the crossing of the cumulants, we can compare the
mean-field probability density function of the overlap, $P(q)$, with
the one computed in a numerical simulation working on finite
systems. In Fig.~\ref{fig:pqsimnum}-top we show the numerical $P(q)$
for different values of a magnetic field, and in the bottom panel the
mean-field prediction. Notice that the support of the analytical
$P(q)$ is fully contained in the positive overlap axis and the same
happens for the $P(q)$ in the droplet theory.  Instead, the numerical
$P(q)$ still shows large tails in the negative overlap region. These
tails in the negative overlap region bias the correlation length,
mainly via the zero mode used in its definition. Furthermore, the spin
glass susceptibility, which is the Fourier transform of the
correlation function computed a zero momentum, strongly suffers from
the existence of these tails in $P(q)$. Another way to understand this
phenomena is, following
Refs.~[\refcite{janus:14c,parisi2012numerical}], realizing that the
final results are dominated by atypical measurements. Focusing on
typical measurements will improve the final description of these
systems.

\begin{table}[ht]
  \centering
    \tbl{Summary of the critical exponents and the critical temperatures for three magnetic fields.\cite{janus:12}}
{
 \begin{tabular}{c|c|c|c}
\hline \hline
 Parameters  & $h=0.3$ & $h=0.15$ & $h=0.075$\\
\hline
$T_\text{c}(h)$ & 0.906(40)[3] & 1.229(30)[2] &  1.50(7)  \\ 
$\nu$            & \multicolumn{2}{c}{1.46(7)[6]} & ---\\
$\eta$           & \multicolumn{2}{c}{$-0.30(4)[1]$} & ---\\
 \hline \hline
 \end{tabular}
 \label{tab:parametersH}
}
\end{table}

To avoid the strong effects induced by the zero mode we defined a new
cumulant using the two smallest (and non-zero) momenta. The $R_{12}$
has been used and defined in the section devoted to $h=0$ (see
Eq.~(\ref{eq:R12-def})), but for the  commodity of the reader we repeat here
its definition particularizing in four dimensions
\begin{equation}\label{eq:R12}
R_{12} = \frac{\widetilde G (\boldsymbol k_1)}{\widetilde G(\boldsymbol k_2)},
\end{equation}
where $\boldsymbol k_1=(2\pi/L,0,0,0)$ and $\boldsymbol k_2 = (2\pi/L,
2\pi/L, 0,0)$ (plus permutations) are the two non-zero smallest
momenta allowed by the periodic boundary conditions imposed at the
system.\cite{janus:12}

Fig.~\ref{fig:xi-R12}-bottom shows that the cumulant $R_{12}$
signals the phase transition. Its value at the critical point, as for
other cumulants, is universal. Using these $R_{12}$ crossing points it
has been possible to characterize the phase transition in four
dimension. Furthermore, it is possible to show that the critical
exponents are independent of the strength of the magnetic field
(taking into account corrections to scaling). In Table
\ref{tab:parametersH} we report the values computed for the critical
exponents and the critical temperatures. The analysis of the numerical data  was
  performed working at constant coupling, see Sec.~\ref{sec:quotient}
  for a description of this method.

We can analyze the scaling behavior of these critical temperature
 performing a
test of the Fisher-Sompolinsky relation:\cite{fisher:85}
\begin{equation}
  h^2(T_c)\simeq A
|T_c(h)-T_c(0)|^{\beta^{(0)}+\gamma^{(0)}}\;,
\end{equation}
where $T_c(h)$ is the critical temperature in a field, and the symbols
with a zero as superscript refer to the critical properties of the
model in zero magnetic field: critical temperature and exponents.

\begin{figure}[ht]
\centering
\includegraphics[width=0.9\columnwidth]{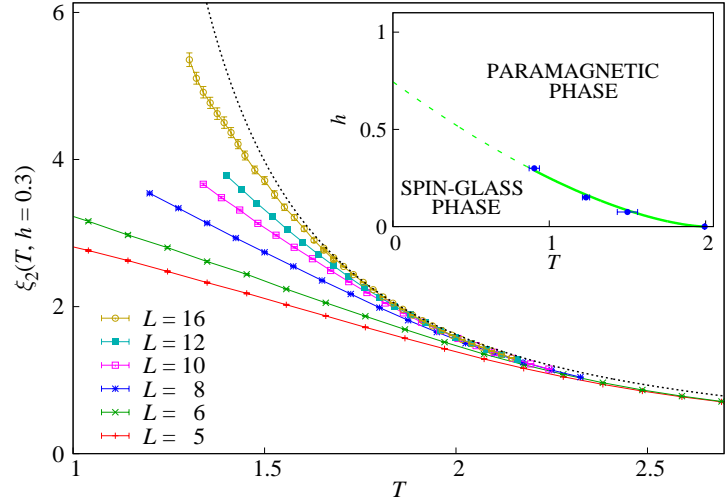}
\caption{In the inset we plot the Fisher-Sompolinsky relation for a
  four-dimensional spin glass with binary couplings: $h^2(T_c)\simeq A
  |T_c(h)-T_c(0)|^{\beta^{(0)}+\gamma^{(0)}}$. Notice that there is
  only one free parameter in the fit. In the main plot, we present
  behavior of the correlation length for different lattice sizes
  with the infinite volume extrapolation at $h=0.3$ (continuous dotted line).\cite{janus:12}}
\label{fig:FS}
\end{figure}

In the inset of Fig.~\ref{fig:FS} we report this analysis finding a
very good agreement between the numerical results (for the computed
critical temperatures) and the previous relation. In addition, the
fit has only one free parameter.

However, using this methodology, no traces of a phase transition in
the thee dimensional model in a field has been found. The 
simplest explanation is that the lower critical dimension of the model
in a magnetic field ($3<D_l<4$) is different of that in $h=0$
($D_l=2.5$), as happens in the random field Ising model case (the lower
critical dimension in $h=0$ is one and in presence of a random
magnetic field is two).

\section{Conclusions}

We have presented an overview of different analytical approaches to
finite dimensional Ising spin glasses. The analytical approach based
on field theory is still incomplete both for $h=0$ and $h\neq 0$. In
presence of a magnetic field, recent analytical work points out to a
complicated structure of the critical behavior below the upper
critical dimension.

Moreover,  we have presented numerical evidence for existence of the
spin glass phase in absence of a magnetic field. The
properties of this low temperature phase fit very well in the
framework of the RSB theory. Nextly, we have studied in
detailed the replicon propagator and we have shown how a disperse
metastate has been found. The situation in presence of a
magnetic field is not clear. In particular, there is no
evidence of a phase transition in three dimensions, though in four
dimensions it has been found.

\section*{Acknowledgments}

This work was partially supported by Ministerio de Econom\'{\i}a y
Competitividad (Spain) through Grant No.\ FIS2016-76359-P, by Junta de
Extremadura (Spain) through Grants No.\ GRU18079 and IB16013
(partially funded by FEDER).

I have enjoyed interesting and fruitful discussions on spin glasses
and numerical simulations with M. Baity-Jesi, A. Billoire, A. Cruz,
L.A. Fernandez, A. Gordillo-Guerrero, D. I\~niguez, R. Kenna,
A. Lasanta, L. Leuzzi, A. Maiorano, E. Marinari, V. Martin-Mayor,
J. Monforte, A. Mu\~noz-Sudupe, D. Navarro, G. Parisi,
S. Perez-Gaviro, F. Ricci-Tersenghi, B. Seoane, A. Tarancon,
R. Tripiccione and D. Yllanes.

I warmly thank Yu. Holovatch and E. Ruiz Espejo for a careful reading
of the manuscript.

Finally, I would like to thank M. Dudka and Yu. Holovatch for inviting
me to present these results in the 2019 Lviv Ising Lectures.

\begin{appendix}
\section{~Characterization of a phase transition: Quotient and fixed coupling methods}
\label{sec:quotient}

Let us consider a  quantity $O(\beta,L)$ which scales in
the thermodynamic limit as $\xi^{x_O/\nu}$. We can study the behavior
of this observable by computing it at $L$ and $2L$, ${\mathcal
  Q}_O=O_{2L}/O_L$, at the crossing point $\beta_\text{cross}(L, 2L)$
of $R_\xi$ or $U_4$. In the case of a dimensionless observable the
exponent $x_O=0$. We will denote as $g$ all the dimensionless
quantities.\cite{amit:05}

Hence, one gets
\begin{equation}\label{eq:QO}
{\mathcal Q}_O^{\,\mathrm{cross}}=2^{x_O/\nu}+\mathcal{O}(L^{-\omega})\,,\
\end{equation}
or
\begin{equation}\label{eq:QOB}
g^{\,\mathrm{cross}}=g^* +\mathcal{O}(L^{-\omega})\,,\
\end{equation}
where $x_O/\nu$, $g^*$ and the correction-to-scaling exponent
$\omega$ are universal quantities.  Examples of dimensionless quantities are
$R_\xi$ and the cumulants (e.g. $U_4$ and $U_{22}$). We could also  
consider dimensionful observables as the the susceptibility ($x_\chi=
\nu(2-\eta)$) and the $\beta$-derivatives of
$R_\xi$ and $U_4$ ($x=1$ for both).

The behavior of the crossing points of the inverse
temperature ($\beta_\text{cross}(L,2L)$) are given by
\begin{equation}\label{eq:Tc}
\beta_\text{cross}(L,2L) = \beta_\text{c} + A_{\beta_\text{c},g} L^{-\omega -
  1/\nu}+\ldots,
\end{equation}
where in our case $g=R_\xi$ or $U_4$.

In order to study the leading correction-to-scaling exponent we can build the
quotient of a given dimensionless quantity $g$ 
 $$\mathcal{Q}_g=g_{2L}/g_L$$
at $\beta_{\text{cross}}(L,2L)$. This quotient
behaves as
\begin{equation}\label{eq:dimensionless-quotients}
\mathcal Q^{\text{cross}}_g(L) = 1 + A_g L^{-\omega} + B_g L^{-2\omega} + \ldots.
\end{equation} 

In the fixed coupling method the analysis is slightly different. For
instance we have a fixed value of a dimensionless observable $g=g_f$ near the universal one (for
example a given value of $R_\xi$) and we compute the value of $\beta(L)$ at which
\begin{equation}
g_f=g(\beta(g_f, L), L)\,.
\end{equation}
At this value of the inverse temperature we can study scaling of
the derivatives of different observables (e.g. susceptibility,
derivatives of $R_\xi$ and Binder cumulant, etc.) which allows to extract the critical exponents via
\begin{equation}
O(\beta(g_f,L),L)=A(g_f) L^{x_O/\nu}\left (1 + O\left(\frac{1}{L^\omega}\right)\right)\,.
\end{equation}

\section{~The Janus supercomputers}
\label{sec:janus}
Most of the numerical simulations presented in this chapter were
obtained using the Janus I and II supercomputers. These computers were
built to take advantage of the powerful integer arithmetic and a high
number of processor units of the FPGA (Fast Programmable
Devices). They were designed and used by a scientific collaboration
composed by researchers of two Italian Universities, Ferrara and Roma
I ``La Sapienza'' and three Spanish ones: Complutense de Madrid,
Zaragoza and Extremadura.

The physics obtained with the help of the  Janus
supercomputers\cite{janus:08b,janus:09b,janus:09c, janus:10,
  janus:11, janus:12, janus:13, janus:14b,
  janus:14c,janus:16,janus:17b,janus:18, janus:19} has covered a wide
variety of spin models, mainly Ising spin glass but also Potts glass
models. These discrete models allow the supercomputers to achieve
their maximum performance.

\begin{figure}[ht]
\centering
\includegraphics[width=0.9\columnwidth]{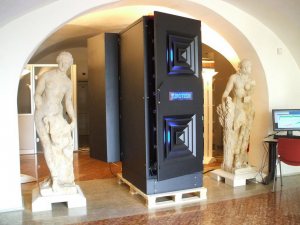}
\caption{External view of the Janus I supercomputer during its presentation in Italy.}
\label{fig:janus1_1}
\end{figure} 

The Janus I computer, see Fig.~\ref{fig:janus1_1}, entered in production mode in
2008.\cite{janus:06,janus:08,janus:09,janus:12b} Its most important
features are:
\begin{itemize}
\item It is composed by 16 boards of 16 FPGAs each (Virtex 4).
\item For Ising models, Janus I is equivalent to 10000 PC.
\item High degree of parallelization inside the boards.
\item Janus allows us to simulate in the 0.1 s time region. Usually, 
  experimental times range from 1 s to 3000 s and previous
    numerical simulations simulated the $10^{-5}$ s region (SSUE).
\end{itemize}

The next generation of Janus computer, the Janus II one, was built in
2015 and its principal characteristic are:\cite{janus:14}

\begin{itemize}
\item  Janus II is $5$ times most powerful than Janus.
\item It is still a dedicated computer optimized to simulate a wide
  variety of spin models.
\item It presents a more flexible topology.
\item It is formed by 16 boards of 16 FPGAs each (one IOP and PC integrated on each board)
  (Virtex 7).
\item Janus II allows  to simulate in the 1 second time region. 
\end{itemize}

\end{appendix}

\bibliographystyle{ws-rv-van}

\begin{thebibliography}{149}
\providecommand{\natexlab}[1]{#1}
\providecommand{\url}[1]{\texttt{#1}}
\expandafter\ifx\csname urlstyle\endcsname\relax
  \providecommand{\doi}[1]{doi: #1}\else
  \providecommand{\doi}{doi: \begingroup \urlstyle{rm}\Url}\fi

\bibitem{mydosh:93}
J.~A. Mydosh, \emph{Spin Glasses: an Experimental Introduction}. Taylor and
  Francis, London  (1993).

\bibitem{kotliar:83}
G.~Kotliar, P.~W. Anderson, and D.~L. Stein, One-dimensional spin-glass model
  with long-range random interactions, \emph{Phys. Rev. B}. {\bf 27}, \penalty0
  602  (1983).
\newblock \doi{10.1103/PhysRevB.27.602}.

\bibitem{leuzzi:99}
L.~Leuzzi, Critical behaviour and ultrametricity of {Ising} spin-glass with
  long-range interactions, \emph{Journal of Physics A: Mathematical and
  General}. {\bf 32}\penalty0 (8), \penalty0 1417--1426  (jan, 1999).
\newblock \doi{10.1088/0305-4470/32/8/010}.
\newblock URL \url{https://doi.org/10.1088%2F0305-4470%2F32%2F8%2F010}.

\bibitem{leuzzi:09}
L.~Leuzzi, G.~Parisi, F.~Ricci-Tersenghi, and J.~J. Ruiz-Lorenzo, {Ising}
  spin-glass transition in a magnetic field outside the limit of validity of
  mean-field theory, \emph{Phys. Rev. Lett.} {\bf 103}, \penalty0 267201
  (2009).
\newblock \doi{10.1103/PhysRevLett.103.267201}.

\bibitem{leuzzi:11}
L.~Leuzzi, G.~Parisi, F.~Ricci-Tersenghi, and J.~Ruiz-Lorenzo, Bond diluted
  {Levy} spin-glass model and a new finite-size scaling method to determine a
  phase transition, \emph{Philosophical Magazine}. {\bf 91}\penalty0 (13-15),
  \penalty0 1917--1925  (2011).
\newblock \doi{10.1080/14786435.2010.534741}.
\newblock URL \url{https://doi.org/10.1080/14786435.2010.534741}.

\bibitem{larson:13}
D.~Larson, H.~G. Katzgraber, M.~A. Moore, and A.~P. Young, Spin glasses in a
  field: Three and four dimensions as seen from one space dimension,
  \emph{Phys. Rev. B}. {\bf 87}, \penalty0 024414  (2013).
\newblock \doi{10.1103/PhysRevB.87.024414}.

\bibitem{leuzzi:15}
L.~Leuzzi, G.~Parisi, F.~Ricci-Tersenghi, and J.~J. Ruiz-Lorenzo, Infinite
  volume extrapolation in the one-dimensional bond diluted {Levy} spin-glass
  model near its lower critical dimension, \emph{Phys. Rev. B}. {\bf 91},
  \penalty0 064202  (Feb, 2015).
\newblock \doi{10.1103/PhysRevB.91.064202}.
\newblock URL \url{https://link.aps.org/doi/10.1103/PhysRevB.91.064202}.

\bibitem{dilucca:20}
M.~{Dilucca}, L.~{Leuzzi}, G.~{Parisi}, F.~{Ricci-Tersenghi}, and J.~J.
  {Ruiz-Lorenzo}, {Spin glasses in a field show a phase transition varying the
  distance among real replicas (and how to exploit it to find the critical line
  in a field)}, \emph{Entropy}. {\bf 22(2)}, \penalty0 250  (2020).

\bibitem{katzgraber:03}
H.~Katzgraber and A.~P. Young, {Monte Carlo} studies of the one-dimensional
  ising spin glass with power-law interactions, \emph{Phys. Rev. B}. {\bf 67},
  \penalty0 134410  (2003).
\newblock \doi{10.1103/PhysRevB.67.134410}.

\bibitem{katzgraber:05b}
H.~G. Katzgraber and A.~P. Young, Probing the {Almeida-Thouless} line away from
  the mean-field model, \emph{Phys. Rev. B}. {\bf 72}, \penalty0 184416  (Nov,
  2005).
\newblock \doi{10.1103/PhysRevB.72.184416}.
\newblock URL \url{https://link.aps.org/doi/10.1103/PhysRevB.72.184416}.

\bibitem{katzgraber:12}
H.~G. Katzgraber, T.~J\"org, F.~Krzk{a}kal{}a, and A.~K. Hartmann, Ultrametric
  probe of the spin-glass state in a field, \emph{Phys. Rev. B}. {\bf 86},
  \penalty0 184405  (Nov, 2012).
\newblock \doi{10.1103/PhysRevB.86.184405}.
\newblock URL \url{https://link.aps.org/doi/10.1103/PhysRevB.86.184405}.

\bibitem{katzgraber:09}
H.~G. Katzgraber, D.~Larson, and A.~P. Young, Study of the de
  {Almeida--Thouless} line using power-law diluted one-dimensional {Ising} spin
  glasses, \emph{Phys. Rev. Lett.} {\bf 102}, \penalty0 177205  (Apr, 2009).
\newblock \doi{10.1103/PhysRevLett.102.177205}.
\newblock URL \url{https://link.aps.org/doi/10.1103/PhysRevLett.102.177205}.

\bibitem{katzgraber:09b}
H.~G. Katzgraber and A.~K. Hartmann, Ultrametricity and clustering of states in
  spin glasses: A one-dimensional view, \emph{Phys. Rev. Lett.} {\bf 102},
  \penalty0 037207  (Jan, 2009).
\newblock \doi{10.1103/PhysRevLett.102.037207}.
\newblock URL \url{https://link.aps.org/doi/10.1103/PhysRevLett.102.037207}.

\bibitem{aspelmeier:16}
T.~Aspelmeier, H.~G. Katzgraber, D.~Larson, M.~A. Moore, M.~Wittmann, and
  J.~Yeo, Finite-size critical scaling in {Ising} spin glasses in the
  mean-field regime, \emph{Phys. Rev. E}. {\bf 93}, \penalty0 032123  (Mar,
  2016).
\newblock \doi{10.1103/PhysRevE.93.032123}.
\newblock URL \url{https://link.aps.org/doi/10.1103/PhysRevE.93.032123}.

\bibitem{binder:86}
K.~Binder and A.~P. Young, Spin glasses: Experimental facts, theoretical
  concepts, and open questions, \emph{Rev. Mod. Phys.} {\bf 58}, \penalty0
  801--976  (Oct, 1986).
\newblock \doi{10.1103/RevModPhys.58.801}.
\newblock URL \url{http://link.aps.org/doi/10.1103/RevModPhys.58.801}.

\bibitem{young:98}
A.~P. Young, \emph{Spin Glasses and Random Fields}. World Scientific, Singapore
   (1998).
\newblock \doi{10.1142/3517}.

\bibitem{dedominicis:06}
C.~de~Dominicis and I.~Giardina, \emph{{Random {F}ields and {S}pin {G}lasses}:
  a field theory approach}. Cambridge University Press, Cambridge, England
  (2006).

\bibitem{fischer:93}
K.~H. Fischer and J.~A. Hertz, \emph{Spin Glasses}. Cambridge University Press
  (1993).

\bibitem{ruderman:54}
M.~Ruderman and C.~Kittel, Indirect exchange coupling of nuclear magnetic
  moments by conduction electrons, \emph{Phys. Rev.} {\bf 96}, \penalty0 99
  (1954).

\bibitem{kasuya:56}
T.~Kasuya, A theory of metallic ferro- and antiferromagnetism on zener's model,
  \emph{Prog. Theor. Phys.} {\bf 16}, \penalty0 45  (1956).

\bibitem{yosida:57}
K.~Yosida, Magnetic properties of {Cu-Mn} alloys, \emph{Phys. Rev.} {\bf 106},
  \penalty0 893--898  (Jun, 1957).
\newblock \doi{10.1103/PhysRev.106.893}.
\newblock URL \url{http://link.aps.org/doi/10.1103/PhysRev.106.893}.

\bibitem{guchhait:14}
S.~Guchhait and R.~Orbach, Direct dynamical evidence for the spin glass lower
  critical dimension $2<d<3$, \emph{Phys. Rev. Lett.} {\bf 112}, \penalty0
  126401  (Mar, 2014).
\newblock \doi{10.1103/PhysRevLett.112.126401}.
\newblock URL \url{http://link.aps.org/doi/10.1103/PhysRevLett.112.126401}.

\bibitem{guchhait:15a}
S.~Guchhait, G.~G. Kenning, R.~L. Orbach, and G.~F. Rodriguez, Spin glass
  dynamics at the mesoscale, \emph{Phys. Rev. B}. {\bf 91}, \penalty0 014434
  (Jan, 2015).
\newblock \doi{10.1103/PhysRevB.91.014434}.
\newblock URL \url{http://link.aps.org/doi/10.1103/PhysRevB.91.014434}.

\bibitem{guchhait:15b}
S.~Guchhait and R.~L. Orbach, Temperature chaos in a {Ge:Mn} thin-film spin
  glass, \emph{Phys. Rev. B}. {\bf 92}, \penalty0 214418  (Dec, 2015).
\newblock \doi{10.1103/PhysRevB.92.214418}.
\newblock URL \url{http://link.aps.org/doi/10.1103/PhysRevB.92.214418}.

\bibitem{guchhait:17}
S.~Guchhait and R.~L. Orbach, Magnetic field dependence of spin glass free
  energy barriers, \emph{Phys. Rev. Lett.} {\bf 118}, \penalty0 157203  (Apr,
  2017).
\newblock \doi{10.1103/PhysRevLett.118.157203}.
\newblock URL \url{https://link.aps.org/doi/10.1103/PhysRevLett.118.157203}.

\bibitem{joh:99}
Y.~G. Joh, R.~Orbach, G.~G. Wood, J.~Hammann, and E.~Vincent, Extraction of the
  spin glass correlation length, \emph{Phys. Rev. Lett.} {\bf 82}, \penalty0
  438--441  (Jan, 1999).
\newblock \doi{10.1103/PhysRevLett.82.438}.
\newblock URL \url{http://link.aps.org/doi/10.1103/PhysRevLett.82.438}.

\bibitem{zhai:17}
Q.~Zhai, D.~C. Harrison, D.~Tennant, E.~D. Dalhberg, G.~G. Kenning, and R.~L.
  Orbach, Glassy dynamics in {CuMn} thin-film multilayers, \emph{Phys. Rev. B}.
  {\bf 95}, \penalty0 054304  (Feb, 2017).
\newblock \doi{10.1103/PhysRevB.95.054304}.
\newblock URL \url{https://link.aps.org/doi/10.1103/PhysRevB.95.054304}.

\bibitem{janus:17b}
M.~Baity-Jesi, E.~Calore, A.~Cruz, L.~A. Fernandez, J.~M. Gil-Narvion,
  A.~Gordillo-Guerrero, D.~I\~niguez, A.~Maiorano, E.~Marinari,
  V.~Martin-Mayor, J.~Monforte-Garcia, A.~Mu\~noz Sudupe, D.~Navarro,
  G.~Parisi, S.~Perez-Gaviro, F.~Ricci-Tersenghi, J.~J. Ruiz-Lorenzo, S.~F.
  Schifano, B.~Seoane, A.~Tarancon, R.~Tripiccione, and D.~Yllanes, Matching
  microscopic and macroscopic responses in glasses, \emph{Phys. Rev. Lett.}
  {\bf 118}, \penalty0 157202  (Apr, 2017).
\newblock \doi{10.1103/PhysRevLett.118.157202}.
\newblock URL \url{https://link.aps.org/doi/10.1103/PhysRevLett.118.157202}.

\bibitem{janus:18}
M.~Baity-Jesi, E.~Calore, A.~Cruz, L.~A. Fernandez, J.~M. Gil-Narvion,
  A.~Gordillo-Guerrero, D.~I\~niguez, A.~Maiorano, E.~Marinari,
  V.~Martin-Mayor, J.~Moreno-Gordo, A.~Mu\~noz Sudupe, D.~Navarro, G.~Parisi,
  S.~Perez-Gaviro, F.~Ricci-Tersenghi, J.~J. Ruiz-Lorenzo, S.~F. Schifano,
  B.~Seoane, A.~Tarancon, R.~Tripiccione, and D.~Yllanes, Aging rate of spin
  glasses from simulations matches experiments, \emph{Phys. Rev. Lett.} {\bf
  120}, \penalty0 267203  (Jun, 2018).
\newblock \doi{10.1103/PhysRevLett.120.267203}.
\newblock URL \url{https://link.aps.org/doi/10.1103/PhysRevLett.120.267203}.

\bibitem{fernandez:19b}
L.~A. Fernandez, E.~Marinari, V.~Martin-Mayor, I.~Paga, and J.~J. Ruiz-Lorenzo,
  Dimensional crossover in the aging dynamics of spin glasses in a film
  geometry, \emph{Phys. Rev. B}. {\bf 100}, \penalty0 184412  (Nov, 2019).
\newblock \doi{10.1103/PhysRevB.100.184412}.
\newblock URL \url{https://link.aps.org/doi/10.1103/PhysRevB.100.184412}.

\bibitem{edwards:75}
S.~F. Edwards and P.~W. Anderson, Theory of spin glasses, \emph{Journal of
  Physics F: Metal Physics}. {\bf 5}, \penalty0 965  (1975).
\newblock \doi{10.1088/0305-4608/5/5/017}.
\newblock URL \url{http://stacks.iop.org/0305-4608/5/i=5/a=017}.

\bibitem{mezard:87}
M.~M{\'e}zard, G.~Parisi, and M.~Virasoro, \emph{Spin-Glass Theory and Beyond}.
  World Scientific, Singapore  (1987).
\newblock \doi{10.1142/0271}.

\bibitem{kirkpatrick:78}
S.~Kirkpatrick and D.~Sherrington, Infinite-ranged models of spin-glasses,
  \emph{Phys. Rev. B}. {\bf 17}, \penalty0 4384--4403  (Jun, 1978).
\newblock \doi{10.1103/PhysRevB.17.4384}.
\newblock URL \url{http://link.aps.org/doi/10.1103/PhysRevB.17.4384}.

\bibitem{dotsenko:01}
V.~Dotsenko, \emph{Introduction to the Replica Theory of Disordered Statistical
  Systems}. Cambridge University Press, Cambridge, England  (2001).

\bibitem{cardy:96}
J.~Cardy, \emph{Scaling and Renormalization in Statistical Physics}. vol.~5,
  \emph{Lecture notes in physics}, Cambridge University Press, Cambridge
  (1996).
\newblock ISBN 0521499593.

\bibitem{amit:05}
D.~J. Amit and V.~Mart\'{i}n-Mayor, \emph{Field Theory, the Renormalization
  Group and Critical Phenomena}, third edn. World Scientific, Singapore
  (2005).
\newblock \doi{10.1142/9789812775313_bmatter}.
\newblock URL \url{http://www.worldscientific.com/worldscibooks/10.1142/5715}.

\bibitem{parisi:88}
G.~Parisi, \emph{Statistical Field Theory}. Addison-Wesley  (1988).

\bibitem{itzykson:89}
C.~Itzykson and J.~M. Drouffe, \emph{Statistical Field Theory}. Cambridge
  University Press  (1989).

\bibitem{parisi:79}
G.~Parisi, Infinite number of order parameters for spin-glasses, \emph{Phys.
  Rev. Lett.} {\bf 43}, \penalty0 1754--1756  (Dec, 1979).
\newblock \doi{10.1103/PhysRevLett.43.1754}.
\newblock URL \url{http://link.aps.org/doi/10.1103/PhysRevLett.43.1754}.

\bibitem{parisi:79b}
G.~Parisi, Toward a mean-field theory for spin glasses, \emph{Phys. Lett.} {\bf
  73A}, \penalty0 203  (1979).
\newblock \doi{10.1016/0375-9601(79)90708-4}.

\bibitem{parisi:80}
G.~Parisi, The order parameter for spin glasses: a function on the interval
  0-1, \emph{J. Phys. A: Math. Gen.} {\bf 13}, \penalty0 1101  (1980).
\newblock \doi{10.1088/0305-4470/13/3/042}.

\bibitem{parisi:80b}
G.~Parisi, A sequence of approximated solutions to the {S-K} model for spin
  glasses, \emph{J. Phys. A: Math. Gen.} {\bf 13}, \penalty0 L115--L121
  (1980).
\newblock ISSN 0305-4470.
\newblock \doi{10.1088/0305-4470/13/4/009}.

\bibitem{parisi:80c}
G.~Parisi, Magnetic properties of spin glasses in a new mean-field theory,
  \emph{Journal of Physics A: Mathematical and General}. {\bf 13}\penalty0 (5),
  \penalty0 1887  (1980).
\newblock \doi{10.1088/0305-4470/13/5/047}.
\newblock URL \url{http://stacks.iop.org/0305-4470/13/i=5/a=047}.

\bibitem{parisi:83}
G.~Parisi, Order parameter for spin glasses, \emph{Phys. Rev. Lett.} {\bf 50},
  \penalty0 1946  (1983).
\newblock \doi{10.1103/PhysRevLett.50.1946}.
\newblock URL
  \url{http://journals.aps.org/prl/abstract/10.1103/PhysRevLett.50.1946}.

\bibitem{monforte:13}
J.~Monforte.
\newblock \emph{Rugged Free-Energy Landscapes in Disordered Spin Systems}.
\newblock PhD thesis, Universidad de Zaragoza  (2013).
\newblock URL
  \url{http://zaguan.unizar.es/record/11743/files/TESIS-2013-077.pdf}.

\bibitem{mezard:84}
M.~M{\'e}zard, G.~Parisi, N.~Sourlas, G.~Toulouse, and M.~Virasoro, Nature of
  the spin-glass phase, \emph{Phys. Rev. Lett.} {\bf 52}, \penalty0 1156
  (1984).
\newblock \doi{10.1103/PhysRevLett.52.1156}.

\bibitem{mezard:84b}
M.~M{\'e}zard, G.~Parisi, N.~Sourlas, G.~Toulouse, and M.~Virasoro, Replica
  symmetry breaking and the nature of the spin glass phase, \emph{J. Phys.
  France}. {\bf 45}, \penalty0 843--854  (1984).
\newblock \doi{10.1051/jphys:01984004505084300}.

\bibitem{mezard:85}
M.~M{\'e}zard and M.~Virasoro, On the microstructure of ultrametricity,
  \emph{J. Physique}. {\bf 46}, \penalty0 1293--1307  (1985).
\newblock \doi{10.1051/jphys:019850046080129300}.

\bibitem{mezard:85b}
M.~M{\'e}zard, G.~Parisi, and M.~Virasoro, Random free energies in spin
  glasses, \emph{J. Physique Lett.} {\bf 46}, \penalty0 217--222  (1985).
\newblock \doi{10.1051/jphyslet:01985004606021700}.

\bibitem{mezard:86}
M.~M{\'e}zard, G.~Parisi, and M.~Virasoro, Sk model: The replica solution
  without replicas, \emph{Europhys. Lett.} {\bf 1}, \penalty0 77  (1986).
\newblock \doi{10.1209/0295-5075/1/2/006}.

\bibitem{marinari:00}
E.~Marinari, G.~Parisi, F.~Ricci-Tersenghi, J.~J. Ruiz-Lorenzo, and F.~Zuliani,
  Replica symmetry breaking in short-range spin glasses: Theoretical
  foundations and numerical evidences, \emph{J. Stat. Phys.} {\bf 98},
  \penalty0 973  (2000).
\newblock \doi{10.1023/A:1018607809852}.

\bibitem{guerra:03}
F.~Guerra, Broken replica symmetry bounds in the mean-field spin glass model,
  \emph{Comm. Math. Phys.} {\bf 233}, \penalty0 1--12  (2003).
\newblock \doi{10.1007/s00220-002-0773-5}.

\bibitem{talagrand:11}
M.~Talagrand, \emph{Mean Field Models for Spin Glasses}. Springer-Verlag,
  Berlin  (2011).

\bibitem{panchenko:13b}
D.~Panchenko, \emph{The Sherrington-Kirkpatrick Model}. Springer-Verlag, Berlin
   (2013).

\bibitem{rammal:86}
R.~Rammal, G.~Toulouse, and M.~A. Virasoro, Ultrametricity for physicists,
  \emph{Rev. Mod. Phys.} {\bf 58}, \penalty0 765--788  (Jul, 1986).
\newblock \doi{10.1103/RevModPhys.58.765}.
\newblock URL \url{http://link.aps.org/doi/10.1103/RevModPhys.58.765}.

\bibitem{dealmeida:78}
J.~R.~L. de~Almeida and D.~J. Thouless, Stability of the
  {S}herrington-{K}irkpatrick solution of a spin glass model, \emph{J. Phys. A:
  Math. Gen.} {\bf 11}, \penalty0 983  (1978).
\newblock \doi{10.1088/0305-4470/11/5/028}.
\newblock URL \url{http://stacks.iop.org/0305-4470/11/i=5/a=028}.

\bibitem{wilson:74}
K.~G. Wilson and J.~Kogut, The renormalization group and the
  $\epsilon$-expansion, \emph{Physics Reports}. {\bf 12}\penalty0 (2),
  \penalty0 75 -- 199  (1974).
\newblock ISSN 0370-1573.
\newblock \doi{10.1016/0370-1573(74)90023-4}.
\newblock URL
  \url{http://www.sciencedirect.com/science/article/pii/0370157374900234}.

\bibitem{wilson:75}
K.~G. Wilson, The renormalization group: Critical phenomena and the {Kondo}
  problem, \emph{Rev. Mod. Phys.} {\bf 47}, \penalty0 773--840  (Oct, 1975).
\newblock \doi{10.1103/RevModPhys.47.773}.
\newblock URL \url{http://link.aps.org/doi/10.1103/RevModPhys.47.773}.

\bibitem{bray:87}
A.~J. Bray and M.~A. Moore.
\newblock Scaling theory of the ordered phase of spin glasses.
\newblock In eds. J.~L. van Hemmen and I.~Morgenstern, \emph{Heidelberg
  Colloquium on Glassy Dynamics}, number 275 in Lecture Notes in Physics.
  Springer, Berlin  (1987).

\bibitem{bray:87b}
A.~J. Bray and M.~A. Moore, Chaotic nature of the spin-glass phase, \emph{Phys.
  Rev. Lett.} {\bf 58}, \penalty0 57--60  (Jan, 1987).
\newblock \doi{10.1103/PhysRevLett.58.57}.
\newblock URL \url{https://link.aps.org/doi/10.1103/PhysRevLett.58.57}.

\bibitem{fisher:85}
D.~S. Fisher and H.~Sompolinsky, Scaling in spin-glasses, \emph{Phys. Rev.
  Lett.} {\bf 54}, \penalty0 1063  (1985).
\newblock \doi{10.1103/PhysRevLett.54.1063}.

\bibitem{fisher:86}
D.~S. Fisher and D.~A. Huse, Ordered phase of short-range {Ising} spin-glasses,
  \emph{Phys. Rev. Lett.} {\bf 56}, \penalty0 1601  (Apr, 1986).
\newblock \doi{10.1103/PhysRevLett.56.1601}.
\newblock URL \url{http://link.aps.org/doi/10.1103/PhysRevLett.56.1601}.

\bibitem{fisher:88}
D.~S. Fisher and D.~A. Huse, Nonequilibrium dynamics of spin glasses,
  \emph{Phys. Rev. B}. {\bf 38}, \penalty0 373--385  (Jul, 1988).
\newblock \doi{10.1103/PhysRevB.38.373}.
\newblock URL \url{https://link.aps.org/doi/10.1103/PhysRevB.38.373}.

\bibitem{fisher:88b}
D.~S. Fisher and D.~A. Huse, Equilibrium behavior of the spin-glass ordered
  phase, \emph{Phys. Rev. B}. {\bf 38}, \penalty0 386  (1988).
\newblock \doi{10.1103/PhysRevB.38.386}.

\bibitem{harris:76}
A.~B. Harris, T.~C. Lubensky, and J.-H. Chen, Critical properties of
  spin-glasses, \emph{Phys. Rev. Lett.} {\bf 36}, \penalty0 415--418  (Feb,
  1976).
\newblock \doi{10.1103/PhysRevLett.36.415}.
\newblock URL \url{http://link.aps.org/doi/10.1103/PhysRevLett.36.415}.

\bibitem{alcantara:81}
O.~F. de~Alcantara~Bonfirm, J.~E. Kirkham, and A.~J. McKane, Critical exponents
  for the percolation problem and the yang-lee edge singularity, \emph{Journal
  of Physics A: Mathematical and General}. {\bf 14}\penalty0 (9), \penalty0
  2391--2413  (sep, 1981).
\newblock \doi{10.1088/0305-4470/14/9/034}.
\newblock URL \url{https://doi.org/10.1088%2F0305-4470%2F14%2F9%2F034}.

\bibitem{green:85}
J.~E. Green, $\epsilon$-expansion for the critical exponents of a vector spin
  glass, \emph{Journal of Physics A: Mathematical and General}. {\bf
  18}\penalty0 (1), \penalty0 L43--L47  (jan, 1985).
\newblock \doi{10.1088/0305-4470/18/1/008}.
\newblock URL \url{https://doi.org/10.1088%2F0305-4470%2F18%2F1%2F008}.

\bibitem{ruizlorenzo:98}
J.~J. Ruiz-Lorenzo, Logarithmic corrections for spin glasses, percolation and
  lee-yang singularities in six dimensions, \emph{J. Phys. A: Math. Gen.} {\bf
  31}, \penalty0 8773  (1998).

\bibitem{kenna:06}
R.~Kenna, D.~A. Johnston, and W.~Janke, Scaling relations for logarithmic
  corrections, \emph{Phys. Rev. Lett.} {\bf 96}, \penalty0 115701  (Mar, 2006).
\newblock \doi{10.1103/PhysRevLett.96.115701}.
\newblock URL \url{https://link.aps.org/doi/10.1103/PhysRevLett.96.115701}.

\bibitem{kenna:06b}
R.~Kenna, D.~A. Johnston, and W.~Janke, Self-consistent scaling theory for
  logarithmic-correction exponents, \emph{Phys. Rev. Lett.} {\bf 97}, \penalty0
  155702  (Oct, 2006).
\newblock \doi{10.1103/PhysRevLett.97.155702}.
\newblock URL \url{https://link.aps.org/doi/10.1103/PhysRevLett.97.155702}.

\bibitem{kenna:13}
R.~Kenna.
\newblock Universal scaling relations for logarithmic-correction exponents.
\newblock In ed. Yu.~Holovatch, \emph{\emph{Order, Disorder and Criticality},
  vol.~3, pp. 1-46}. World Scientific, Singapore  (2013).

\bibitem{ruizlorenzo:17}
J.~J. Ruiz-Lorenzo, Revisiting (logarithmic) scaling relations using
  renormalization group, \emph{Condens. Matter Phys.} {\bf 20}, \penalty0 13601
   (2017).

\bibitem{franz:92}
S.~Franz, G.~Parisi, and M.~A. Virasoro, The replica method on and off
  equilibrium, \emph{J. Phys. I (France)}. {\bf 2}, \penalty0 1869  (1992).
\newblock \doi{10.1051/jp1:1992115}.

\bibitem{boettcher:05}
S.~Boettcher, Stiffness of the {Edwards-Anderson} model in all dimensions,
  \emph{Phys. Rev. Lett.} {\bf 95}, \penalty0 197205  (Nov, 2005).
\newblock \doi{10.1103/PhysRevLett.95.197205}.
\newblock URL \url{http://link.aps.org/doi/10.1103/PhysRevLett.95.197205}.

\bibitem{maiorano:18}
A.~Maiorano and G.~Parisi, Support for the value 5/2 for the spin glass lower
  critical dimension at zero magnetic field, \emph{Proceedings of the National
  Academy of Sciences}. {\bf 115}\penalty0 (20), \penalty0 5129--5134  (2018).
\newblock ISSN 0027-8424.
\newblock \doi{10.1073/pnas.1720832115}.
\newblock URL \url{https://www.pnas.org/content/115/20/5129}.

\bibitem{dedominicis:93}
C.~de~Dominicis, I.~Kondor, and T.~Temesv{\'a}ri, Ising spin glass: recent
  progress in the field theory approach, \emph{Int. J. Mod. Phys. B}. {\bf 7},
  \penalty0 986  (1993).
\newblock \doi{10.1142/S0217979293002134}.

\bibitem{dedominicis:98}
C.~de~Dominicis, I.~Kondor, and T.~Temesv{\'a}ri.
\newblock {Beyond the {S}herrington-{K}irkpatrick model}.
\newblock In ed. A.~P. Young, \emph{{Spin {G}lasses and {R}andom {F}ields}}.
  World Scientific, Singapore  (1998).

\bibitem{iniguez:96}
D.~I{\~n}iguez, G.~Parisi, and J.~J. Ruiz-Lorenzo, Simulation of $3d$ {Ising}
  spin glass model using three replicas: study of binder cumulants, \emph{J.
  Phys. A: Math. and Gen.} {\bf 29}, \penalty0 4337  (1996).
\newblock \doi{10.1088/0305-4470/29/15/009}.

\bibitem{parisi:00}
G.~Parisi and F.~Ricci-Tersenghi, On the origin of ultrametricity, \emph{J.
  Phys. A: Math. Gen.} {\bf 33}, \penalty0 113  (2000).
\newblock \doi{10.1088/0305-4470/33/1/307}.

\bibitem{janus:09b}
F.~Belletti, A.~Cruz, L.~A. Fernandez, A.~Gordillo-Guerrero, M.~Guidetti,
  A.~Maiorano, F.~Mantovani, E.~Marinari, V.~Mart\'{i}n-Mayor, J.~Monforte,
  A.~Mu{\~n}oz~Sudupe, D.~Navarro, G.~Parisi, S.~Perez-Gaviro, J.~J.
  Ruiz-Lorenzo, S.~F. Schifano, D.~Sciretti, A.~Tarancon, R.~Tripiccione, and
  D.~Yllanes, An in-depth look at the microscopic dynamics of {I}sing spin
  glasses at fixed temperature, \emph{J. Stat. Phys.} {\bf 135}, \penalty0 1121
   (2009).
\newblock \doi{10.1007/s10955-009-9727-z}.

\bibitem{janus:10}
R.~Alvarez~Ba{\~n}os, A.~Cruz, L.~A. Fernandez, J.~M. Gil-Narvion,
  A.~Gordillo-Guerrero, M.~Guidetti, A.~Maiorano, F.~Mantovani, E.~Marinari,
  V.~Mart\'{i}n-Mayor, J.~Monforte-Garcia, A.~Mu{\~n}oz~Sudupe, D.~Navarro,
  G.~Parisi, S.~Perez-Gaviro, J.~J. Ruiz-Lorenzo, S.~F. Schifano, B.~Seoane,
  A.~Tarancon, R.~Tripiccione, and D.~Yllanes, Nature of the spin-glass phase
  at experimental length scales, \emph{J. Stat. Mech.} {\bf 2010}, \penalty0
  P06026  (2010).
\newblock \doi{10.1088/1742-5468/2010/06/P06026}.

\bibitem{guerra:97}
F.~Guerra, About the overlap distribution in mean-field spin glass models,
  \emph{Int. J. Mod. Phys. B}. {\bf 10}, \penalty0 1675  (1996).

\bibitem{ghirlanda:98}
S.~Ghirlanda and F.~Guerra, General properties of overlap probability
  distributions in disordered spin systems. towards {Parisi} ultrametricity,
  \emph{J. Phys. A: Math. Gen.} {\bf 31}, \penalty0 9149  (1998).
\newblock \doi{10.1088/0305-4470/31/46/006}.

\bibitem{contucci:07b}
P.~Contucci, C.~Giardin{\`a}, C.~Giberti, G.~Parisi, and C.~Vernia,
  Ultrametricity in the {E}dwards-{A}nderson model, \emph{Phys. Rev. Lett}.
  {\bf 99}, \penalty0 057206  (2007).
\newblock \doi{10.1103/PhysRevLett.99.057206}.

\bibitem{parisi:99b}
G.~Parisi, F.~Ricci-Tersenghi, and J.~J. Ruiz-Lorenzo, Generalized
  off-equilibrium fluctuation-dissipation relations in random {Ising} systems,
  \emph{Eur. Phys. J.} {\bf 11}, \penalty0 317--325  (1999).
\newblock \doi{10.1007/s100510050942}.

\bibitem{bray:11}
A.~J. Bray and M.~A. Moore, Disappearance of the de {A}lmeida-{T}houless line
  in six dimensions, \emph{Phys. Rev. B}. {\bf 83}, \penalty0 224408  (2011).
\newblock \doi{10.1103/PhysRevB.83.224408}.

\bibitem{bray:80}
A.~J. Bray and S.~A. Roberts, Renormalisation-group approach to the spin glass
  transition in finite magnetic fields, \emph{J. Phys. C: Solid St. Phys.} {\bf
  13}, \penalty0 5405  (1980).
\newblock \doi{10.1088/0022-3719/13/29/019}.

\bibitem{charbonneau:17}
P.~Charbonneau and S.~Yaida, Nontrivial critical fixed point for
  replica-symmetry-breaking transitions, \emph{Phys. Rev. Lett.} {\bf 118},
  \penalty0 215701  (May, 2017).
\newblock \doi{10.1103/PhysRevLett.118.215701}.
\newblock URL \url{https://link.aps.org/doi/10.1103/PhysRevLett.118.215701}.

\bibitem{delamotte:07}
B.~Delamotte.
\newblock Introduction to the non-perturbative renormalization group.
\newblock In ed. Yu.~Holovatch, \emph{\emph{Order, Disorder and Criticality},
  vol.~2, pp. 1--77}. World Scientific, Singapore  (2007).

\bibitem{dedominicis:02}
I.~R. Pimentel, T.~Temesv\'ari, and C.~De~Dominicis, Spin-glass transition in a
  magnetic field: A renormalization group study, \emph{Phys. Rev. B}. {\bf 65},
  \penalty0 224420  (Jun, 2002).
\newblock \doi{10.1103/PhysRevB.65.224420}.
\newblock URL \url{https://link.aps.org/doi/10.1103/PhysRevB.65.224420}.

\bibitem{temesvari:08}
T.~Temesv{\'a}ri, {Almeida-Thouless} transition below six dimensions,
  \emph{Phys. Rev. B}. {\bf 78}, \penalty0 220401  (2008).

\bibitem{parisi:12}
G.~Parisi and T.~Temesv\'ari, Replica symmetry breaking in and around six
  dimensions, \emph{Nucl. Phys. B}. {\bf 858}, \penalty0 293  (2012).

\bibitem{holler:19}
J.~H\"oller and N.~Read, {One-step replica-symmetry-breaking phase below the de
  {Almeida-Thouless} line in low-dimensional spin glasses}, \emph{arXiv
  e-prints}. art. arXiv:1909.03284  (2019).

\bibitem{ruelle:69}
D.~Ruelle, \emph{Statistical Mechanics}. Benjamin  (1969).

\bibitem{parisi:94}
G.~Parisi, \emph{Field Theory, Disorder and Simulations}. World Scientific
  (1994).

\bibitem{newman:92}
C.~M. Newman and D.~L. Stein, Multiple states and thermodynamic limits in
  short-ranged {Ising} spin-glass models, \emph{Phys. Rev. B}. {\bf 46},
  \penalty0 973--982  (Jul, 1992).
\newblock \doi{10.1103/PhysRevB.46.973}.

\bibitem{newman:96b}
C.~M. Newman and D.~L. Stein, Spatial inhomogeneity and thermodynamic chaos,
  \emph{Phys. Rev. Lett.} {\bf 76}, \penalty0 4821--4824  (Jun, 1996).
\newblock \doi{10.1103/PhysRevLett.76.4821}.

\bibitem{newman:98}
C.~M. Newman and D.~L. Stein, Simplicity of state and overlap structure in
  finite-volume realistic spin glasses, \emph{Phys. Rev. E}. {\bf 57},
  \penalty0 1356--1366  (Feb, 1998).
\newblock \doi{10.1103/PhysRevE.57.1356}.
\newblock URL \url{http://link.aps.org/doi/10.1103/PhysRevE.57.1356}.

\bibitem{billoire:17}
A.~Billoire, L.~A. Fernandez, A.~Maiorano, E.~Marinari, V.~Martin-Mayor,
  J.~Moreno-Gordo, G.~Parisi, F.~Ricci-Tersenghi, and J.~J. Ruiz-Lorenzo,
  Numerical construction of the {Aizenman-Wehr} metastate, \emph{Phys. Rev.
  Lett.} {\bf 119}, \penalty0 037203  (Jul, 2017).
\newblock \doi{10.1103/PhysRevLett.119.037203}.
\newblock URL \url{https://link.aps.org/doi/10.1103/PhysRevLett.119.037203}.

\bibitem{aizenman:90}
M.~Aizenman and J.~Wehr, Rounding effects of quenched randomness on first-order
  phase transitions, \emph{Communications in Mathematical Physics}. {\bf
  130}\penalty0 (3), \penalty0 489--528  (1990).
\newblock ISSN 0010-3616.
\newblock \doi{10.1007/BF02096933}.
\newblock URL \url{http://dx.doi.org/10.1007/BF02096933}.

\bibitem{read:14}
N.~Read, Short-range {Ising} spin glasses: The metastate interpretation of
  replica symmetry breaking, \emph{Phys. Rev. E}. {\bf 90}, \penalty0 032142
  (Sep, 2014).
\newblock \doi{10.1103/PhysRevE.90.032142}.
\newblock URL \url{https://link.aps.org/doi/10.1103/PhysRevE.90.032142}.

\bibitem{hukushima:96}
K.~Hukushima and K.~Nemoto, {E}xchange {M}onte {C}arlo method and application
  to spin glass simulations, \emph{J. Phys. Soc. Japan}. {\bf 65}, \penalty0
  1604  (1996).
\newblock \doi{10.1143/JPSJ.65.1604}.

\bibitem{marinari:98b}
E.~Marinari.
\newblock {O}ptimized {M}onte {C}arlo methods.
\newblock In eds. J.~Kerst\'esz and I.~Kondor, \emph{Advances in Computer
  Simulation}. Springer-Verlag  (1998).
\newblock \doi{10.1007/BFb0105459}.

\bibitem{marinari:98i}
E.~Marinari, G.~Parisi, and J.~J. Ruiz-Lorenzo.
\newblock Numerical simulations of spin glass systems.
\newblock In ed. A.~P. Young, \emph{Spin Glasses and Random Fields}. World
  Scientific, Singapore  (1998).

\bibitem{janke:13}
W.~Janke.
\newblock {M}onte {C}arlo simulations in statistical physics -- from basic
  principles to advanced applications.
\newblock In ed. Yu.~Holovatch, \emph{\emph{Order, Disorder and Criticality},
  vol.~3, pp. 93-166}. World Scientific, Singapore  (2013).

\bibitem{machta:10}
J.~Machta, Population annealing with weighted averages: A {Monte Carlo} method
  for rough free-energy landscapes, \emph{Phys. Rev. E}. {\bf 82}, \penalty0
  026704  (Aug, 2010).
\newblock \doi{10.1103/PhysRevE.82.026704}.
\newblock URL \url{https://link.aps.org/doi/10.1103/PhysRevE.82.026704}.

\bibitem{wang:15b}
W.~Wang, J.~Machta, and H.~G. Katzgraber, Comparing {Monte Carlo} methods for
  finding ground states of {Ising} spin glasses: Population annealing,
  simulated annealing, and parallel tempering, \emph{Phys. Rev. E}. {\bf 92},
  \penalty0 013303  (Jul, 2015).
\newblock \doi{10.1103/PhysRevE.92.013303}.
\newblock URL \url{https://link.aps.org/doi/10.1103/PhysRevE.92.013303}.

\bibitem{wang:15c}
W.~Wang, J.~Machta, and H.~G. Katzgraber, Population annealing: Theory and
  application in spin glasses, \emph{Phys. Rev. E}. {\bf 92}, \penalty0 063307
  (Dec, 2015).
\newblock \doi{10.1103/PhysRevE.92.063307}.
\newblock URL \url{https://link.aps.org/doi/10.1103/PhysRevE.92.063307}.

\bibitem{janus:13}
M.~Baity-Jesi, R.~A. Ba\~{n}os, A.~Cruz, L.~A. Fernandez, J.~M. Gil-Narvion,
  A.~Gordillo-Guerrero, D.~Iniguez, A.~Maiorano, F.~Mantovani, E.~Marinari,
  V.~Mart\'{i}n-Mayor, J.~Monforte-Garcia, A.~Mu{\~n}oz~Sudupe, D.~Navarro,
  G.~Parisi, S.~Perez-Gaviro, M.~Pivanti, F.~Ricci-Tersenghi, J.~J.
  Ruiz-Lorenzo, S.~F. Schifano, B.~Seoane, A.~Tarancon, R.~Tripiccione, and
  D.~Yllanes, Critical parameters of the three-dimensional {I}sing spin glass,
  \emph{Phys. Rev. B}. {\bf 88}, \penalty0 224416  ({2013}).
\newblock \doi{10.1103/PhysRevB.88.224416}.

\bibitem{cooper:82}
F.~Cooper, B.~Freedman, and D.~Preston, Solving {$\phi_{1,2}^4$} field theory
  with {M}onte {C}arlo, \emph{Nucl. Phys. B}. {\bf 210}, \penalty0 210  (1982).
\newblock \doi{10.1016/0550-3213(82)90240-1}.

\bibitem{palassini:99}
M.~Palassini and S.~Caracciolo, Universal finite-size scaling functions in the
  {3D} {I}sing spin glass, \emph{Phys. Rev. Lett.} {\bf 82}, \penalty0
  5128--5131  (1999).
\newblock \doi{10.1103/PhysRevLett.82.5128}.

\bibitem{jorg:08}
T.~J{\"o}rg and H.~G. Katzgraber, Evidence for universal scaling in the
  spin-glass phase, \emph{Phys. Rev. Lett.} {\bf 101}, \penalty0 197205
  (2008).
\newblock \doi{10.1103/PhysRevLett.101.197205}.

\bibitem{jorg:08b}
T.~J{\"o}rg, H.~G. Katzgraber, and F.~Krzakala, Behavior of {I}sing spin
  glasses in a magnetic field, \emph{Phys. Rev. Lett.} {\bf 100}, \penalty0
  197202  (2008).
\newblock \doi{10.1103/PhysRevLett.100.197202}.

\bibitem{jorg:08c}
T.~J{\"o}rg and H.~G. Katzgraber, Universality and universal finite-size
  scaling functions in four-dimensional {I}sing spin glasses, \emph{Phys. Rev.
  B}. {\bf 77}, \penalty0 214426  (2008).
\newblock \doi{10.1103/PhysRevB.77.214426}.

\bibitem{bernardi:95}
L.~Bernardi and I.~A. Campbell, Violation of universality for {Ising}
  spin-glass transitions, \emph{Phys. Rev. B}. {\bf 52}, \penalty0 12501--12504
   (Nov, 1995).
\newblock \doi{10.1103/PhysRevB.52.12501}.
\newblock URL \url{http://link.aps.org/doi/10.1103/PhysRevB.52.12501}.

\bibitem{marinari:96}
E.~Marinari, G.~Parisi, J.~Ruiz-Lorenzo, and F.~Ritort, Numerical evidence for
  spontaneously broken replica symmetry in 3d spin glasses, \emph{Phys. Rev.
  Lett.} {\bf 76}, \penalty0 843--846  (Jan, 1996).
\newblock \doi{10.1103/PhysRevLett.76.843}.
\newblock URL \url{http://link.aps.org/doi/10.1103/PhysRevLett.76.843}.

\bibitem{marinari:98d}
E.~Marinari, G.~Parisi, and J.~J. Ruiz-Lorenzo, On the phase structure of the
  $3d$ {Edwards-Anderson} spin glass, \emph{Phys. Rev. B}. {\bf 58}, \penalty0
  14852  (1998).
\newblock \doi{10.1103/PhysRevB.58.14852}.

\bibitem{janus:10b}
R.~Alvarez~Ba{\~n}os, A.~Cruz, L.~A. Fernandez, J.~M. Gil-Narvion,
  A.~Gordillo-Guerrero, M.~Guidetti, A.~Maiorano, F.~Mantovani, E.~Marinari,
  V.~Mart\'{i}n-Mayor, J.~Monforte-Garcia, A.~Mu{\~n}oz~Sudupe, D.~Navarro,
  G.~Parisi, S.~Perez-Gaviro, J.~J. Ruiz-Lorenzo, S.~F. Schifano, B.~Seoane,
  A.~Tarancon, R.~Tripiccione, and D.~Yllanes, Static versus dynamic
  heterogeneities in the {$D = 3$} {E}dwards-{A}nderson-{I}sing spin glass,
  \emph{Phys. Rev. Lett.} {\bf 105}, \penalty0 177202  (2010).
\newblock \doi{10.1103/PhysRevLett.105.177202}.

\bibitem{contucci:09}
P.~Contucci, C.~Giardin{\`a}, C.~Giberti, G.~Parisi, and C.~Vernia, Structure
  of correlations in three dimensional spin glasses, \emph{Phys. Rev. Lett}.
  {\bf 103}, \penalty0 017201  (2009).
\newblock \doi{10.1103/PhysRevLett.103.017201}.

\bibitem{caracciolo:90}
{S. Caracciolo}, {G. Parisi}, {S. Patarnello}, and {N. Sourlas}, Low
  temperature behaviour of 3-d spin glasses in a magnetic field, \emph{J. Phys.
  France}. {\bf 51}\penalty0 (17), \penalty0 1877--1895  (1990).
\newblock \doi{10.1051/jphys:0199000510170187700}.
\newblock URL \url{https://doi.org/10.1051/jphys:0199000510170187700}.

\bibitem{huse:91}
{D. A. Huse} and {D. S. Fisher}, On the behavior of {Ising} spin glasses in a
  uniform magnetic field, \emph{J. Phys. I France}. {\bf 1}\penalty0 (5),
  \penalty0 621--625  (1991).
\newblock \doi{10.1051/jp1:1991157}.
\newblock URL \url{https://doi.org/10.1051/jp1:1991157}.

\bibitem{caracciolo:91}
{S. Caracciolo}, {G. Parisi}, {S. Patarnello}, and {N. Sourlas}, On computer
  simulations for spin glasses to test mean-field predictions, \emph{J. Phys. I
  France}. {\bf 1}\penalty0 (5), \penalty0 627--628  (1991).
\newblock \doi{10.1051/jp1:1991158}.
\newblock URL \url{https://doi.org/10.1051/jp1:1991158}.

\bibitem{ciria:93}
{J.C. Ciria}, {G. Parisi}, {F. Ritort}, and {J.J. Ruiz-Lorenzo}, The de
  {Almeida-Thouless} line in the four dimensional {Ising} spin glass, \emph{J.
  Phys. I France}. {\bf 3}\penalty0 (11), \penalty0 2207--2227  (1993).
\newblock \doi{10.1051/jp1:1993241}.
\newblock URL \url{https://doi.org/10.1051/jp1:1993241}.

\bibitem{parisi:98b}
G.~Parisi, F.~Ricci-Tersenghi, and J.~J. Ruiz-Lorenzo, Dynamics of the
  four-dimensional spin glass in a magnetic field, \emph{Phys. Rev. B}. {\bf
  57}, \penalty0 13617  (1998).
\newblock \doi{10.1103/PhysRevB.57.13617}.

\bibitem{marinari:98e}
E.~Marinari, G.~Parisi, and F.~Zuliani, Four-dimensional spin glasses in a
  magnetic field have a mean-field-like phase, \emph{J. Phys. A: Math. and
  Gen.} {\bf 31}, \penalty0 1181  (1998).
\newblock \doi{doi:10.1088/0305-4470/31/4/008}.

\bibitem{marinari:98g}
E.~Marinari, C.~Naitza, and G.~Parisi, Critical behavior of the 4d spin glass
  in magnetic field, \emph{Journal of Physics A: Mathematical and General}.
  {\bf 31}, \penalty0 6355  (1998).
\newblock \doi{10.1088/0305-4470/31/30/005}.

\bibitem{marinari:98h}
E.~Marinari, C.~Naitza, F.~Zuliani, G.~Parisi, M.~Picco, and F.~Ritort, General
  method to determine replica symmetry breaking transitions, \emph{Phys. Rev.
  Lett.} {\bf 81}, \penalty0 1698--1701  (Aug, 1998).
\newblock \doi{10.1103/PhysRevLett.81.1698}.
\newblock URL \url{https://link.aps.org/doi/10.1103/PhysRevLett.81.1698}.

\bibitem{houdayer:99}
J.~Houdayer and O.~C. Martin, {Ising} spin glasses in a magnetic field,
  \emph{Phys. Rev. Lett.} {\bf 82}, \penalty0 4934--4937  (Jun, 1999).
\newblock \doi{10.1103/PhysRevLett.82.4934}.
\newblock URL \url{https://link.aps.org/doi/10.1103/PhysRevLett.82.4934}.

\bibitem{marinari:00d}
E.~Marinari, G.~Parisi, and F.~Zuliani, Comment on ``{Ising} spin glasses in a
  magnetic field'', \emph{Phys. Rev. Lett.} {\bf 84}, \penalty0 1056--1056
  (Jan, 2000).
\newblock \doi{10.1103/PhysRevLett.84.1056}.
\newblock URL \url{https://link.aps.org/doi/10.1103/PhysRevLett.84.1056}.

\bibitem{houdayer:00}
J.~Houdayer and O.~C. Martin, Houdayer and {Martin} reply:, \emph{Phys. Rev.
  Lett.} {\bf 84}, \penalty0 1057--1057  (Jan, 2000).
\newblock \doi{10.1103/PhysRevLett.84.1057}.
\newblock URL \url{https://link.aps.org/doi/10.1103/PhysRevLett.84.1057}.

\bibitem{cruz:03}
A.~Cruz, L.~A. Fern\'andez, S.~Jim\'enez, J.~J. Ruiz-Lorenzo, and
  A.~Taranc\'on, Off-equilibrium fluctuation-dissipation relations in the $3d$
  {Ising} spin glass in a magnetic field, \emph{Phys. Rev. B}. {\bf 67},
  \penalty0 214425  (Jun, 2003).
\newblock \doi{10.1103/PhysRevB.67.214425}.
\newblock URL \url{http://link.aps.org/doi/10.1103/PhysRevB.67.214425}.

\bibitem{young:04}
A.~P. Young and H.~G. Katzgraber, Absence of an {A}lmeida-{T}houless line in
  three-dimensional spin glasses, \emph{Phys. Rev. Lett.} {\bf 93}, \penalty0
  207203  (2004).
\newblock \doi{10.1103/PhysRevLett.93.207203}.

\bibitem{leuzzi:08}
L.~Leuzzi, G.~Parisi, F.~Ricci-Tersenghi, and J.~J. Ruiz-Lorenzo, Dilute
  one-dimensional spin glasses with power law decaying interactions,
  \emph{Phys. Rev. Lett.} {\bf 101}, \penalty0 107203  (Sep, 2008).
\newblock \doi{10.1103/PhysRevLett.101.107203}.
\newblock URL \url{http://link.aps.org/doi/10.1103/PhysRevLett.101.107203}.

\bibitem{janus:12}
R.~A. Ba\~{n}os, A.~Cruz, L.~A. Fernandez, J.~M. Gil-Narvion,
  A.~Gordillo-Guerrero, M.~Guidetti, D.~Iniguez, A.~Maiorano, E.~Marinari,
  V.~Mart\'{i}n-Mayor, J.~Monforte-Garcia, A.~Mu{\~n}oz~Sudupe, D.~Navarro,
  G.~Parisi, S.~Perez-Gaviro, J.~J. Ruiz-Lorenzo, S.~F. Schifano, B.~Seoane,
  A.~Tarancon, P.~Tellez, R.~Tripiccione, and D.~Yllanes, {Thermodynamic glass
  transition in a spin glass without time-reversal symmetry}, \emph{Proc. Natl.
  Acad. Sci. USA}. {\bf {109}}, \penalty0 6452  ({2012}).
\newblock \doi{10.1073/pnas.1203295109}.

\bibitem{janus:14b}
M.~Baity-Jesi, R.~A. Ba\~{n}os, A.~Cruz, L.~A. Fernandez, J.~M. Gil-Narvion,
  A.~Gordillo-Guerrero, D.~Iniguez, A.~Maiorano, M.~F., E.~Marinari,
  V.~Mart\'{i}n-Mayor, J.~Monforte-Garcia, A.~Mu{\~n}oz~Sudupe, D.~Navarro,
  G.~Parisi, S.~Perez-Gaviro, M.~Pivanti, F.~Ricci-Tersenghi, J.~J.
  Ruiz-Lorenzo, S.~F. Schifano, B.~Seoane, A.~Tarancon, R.~Tripiccione, and
  D.~Yllanes, Dynamical transition in the d=3 {Edwards-Anderson} spin glass in
  an external magnetic field, \emph{Phys. Rev. E}. {\bf 89}, \penalty0 032140
  (2014).
\newblock \doi{10.1103/PhysRevE.89.032140}.

\bibitem{janus:14c}
M.~Baity-Jesi, R.~A. Ba\~{n}os, A.~Cruz, L.~A. Fernandez, J.~M. Gil-Narvion,
  A.~Gordillo-Guerrero, D.~Iniguez, A.~Maiorano, M.~F., E.~Marinari,
  V.~Mart\'{i}n-Mayor, J.~Monforte-Garcia, A.~Mu{\~n}oz~Sudupe, D.~Navarro,
  G.~Parisi, S.~Perez-Gaviro, M.~Pivanti, F.~Ricci-Tersenghi, J.~J.
  Ruiz-Lorenzo, S.~F. Schifano, B.~Seoane, A.~Tarancon, R.~Tripiccione, and
  D.~Yllanes, The three dimensional {Ising} spin glass in an external magnetic
  field: the role of the silent majority, \emph{J. Stat. Mech.} {\bf 2014},
  \penalty0 P05014  (2014).
\newblock \doi{10.1088/1742-5468/2014/05/P05014}.

\bibitem{mezard:01}
M.~M{\'e}zard and G.~Parisi, The {Bethe} lattice spin glass revisited,
  \emph{Eur. Phys. J. B}. {\bf 20}, \penalty0 217  (2001).
\newblock \doi{10.1007/PL00011099}.

\bibitem{takahashi2010finite}
H.~Takahashi, F.~Ricci-Tersenghi, and Y.~Kabashima, Finite-size scaling of the
  de {Almeida--Thouless} instability in random sparse networks, \emph{Physical
  Review B}. {\bf 81}\penalty0 (17), \penalty0 174407  (2010).

\bibitem{parisi2012numerical}
G.~Parisi and F.~Ricci-Tersenghi, A numerical study of the overlap probability
  distribution and its sample-to-sample fluctuations in a mean-field model,
  \emph{Philosophical Magazine}. {\bf 92}\penalty0 (1-3), \penalty0 341--352
  (2012).

\bibitem{janus:08b}
F.~Belletti, M.~Cotallo, A.~Cruz, L.~A. Fernandez, A.~Gordillo-Guerrero,
  M.~Guidetti, A.~Maiorano, F.~Mantovani, E.~Marinari, V.~Mart\'{i}n-Mayor,
  A.~M. Sudupe, D.~Navarro, G.~Parisi, S.~Perez-Gaviro, J.~J. Ruiz-Lorenzo,
  S.~F. Schifano, D.~Sciretti, A.~Tarancon, R.~Tripiccione, J.~L. Velasco, and
  D.~Yllanes, Nonequilibrium spin-glass dynamics from picoseconds to one tenth
  of a second, \emph{Phys. Rev. Lett.} {\bf 101}, \penalty0 157201  (2008).
\newblock \doi{10.1103/PhysRevLett.101.157201}.

\bibitem{janus:09c}
A.~Cruz, L.~A. Fernandez, A.~Gordillo-Guerrero, M.~Guidetti, A.~Maiorano,
  F.~Mantovani, E.~Marinari, V.~Mart\'{i}n-Mayor, A.~Mu{\~n}oz~Sudupe,
  D.~Navarro, G.~Parisi, S.~Perez-Gaviro, J.~J. Ruiz-Lorenzo, S.~F. Schifano,
  D.~Sciretti, A.~Tarancon, R.~Tripiccione, D.~Yllanes, and A.~P. Young, The
  spin glass phase in the four-state, three-dimensional {P}otts model,
  \emph{Phys. Rev. B}. {\bf 79}, \penalty0 184408  (2009).
\newblock \doi{10.1103/PhysRevB.79.184408}.

\bibitem{janus:11}
R.~A. Ba\~nos, A.~Cruz, L.~A. Fernandez, J.~M. Gil-Narvion,
  A.~Gordillo-Guerrero, M.~Guidetti, D.~I\~niguez, A.~Maiorano, F.~Mantovani,
  E.~Marinari, V.~Mart\'{i}n-Mayor, J.~Monforte-Garcia, A.~Mu\~noz Sudupe,
  D.~Navarro, G.~Parisi, S.~Perez-Gaviro, F.~Ricci-Tersenghi, J.~J.
  Ruiz-Lorenzo, S.~F. Schifano, B.~Seoane, A.~Taranc\'on, R.~Tripiccione, and
  D.~Yllanes, Sample-to-sample fluctuations of the overlap distributions in the
  three-dimensional {Edwards-Anderson} spin glass, \emph{Phys. Rev. B}. {\bf
  84}, \penalty0 174209  (Nov, 2011).
\newblock \doi{10.1103/PhysRevB.84.174209}.
\newblock URL \url{http://link.aps.org/doi/10.1103/PhysRevB.84.174209}.

\bibitem{janus:16}
M.~Baity-Jesi, E.~Calore, A.~Cruz, L.~A. Fernandez, J.~M. Gil-Narvi\'on,
  A.~Gordillo-Guerrero, D.~Iñiguez, A.~Maiorano, E.~Marinari, V.~Martin-Mayor,
  J.~Monforte-Garcia, A.~Muñoz~Sudupe, D.~Navarro, G.~Parisi, S.~Perez-Gaviro,
  F.~Ricci-Tersenghi, J.~J. Ruiz-Lorenzo, S.~F. Schifano, B.~Seoane,
  A.~Taranc\'on, R.~Tripiccione, and D.~Yllanes, A statics-dynamics equivalence
  through the fluctuation–dissipation ratio provides a window into the
  spin-glass phase from nonequilibrium measurements, \emph{Proceedings of the
  National Academy of Sciences}. {\bf 114}\penalty0 (8), \penalty0 1838--1843
  (2017).
\newblock \doi{10.1073/pnas.1621242114}.
\newblock URL \url{http://www.pnas.org/content/114/8/1838.abstract}.

\bibitem{janus:19}
M.~Baity-Jesi, E.~Calore, A.~Cruz, L.~A. Fernandez, J.~M. Gil-Narvi{\'o}n,
  A.~Gordillo-Guerrero, D.~I{\~n}iguez, A.~Lasanta, A.~Maiorano, E.~Marinari,
  V.~Martin-Mayor, J.~Moreno-Gordo, A.~Mu{\~n}oz~Sudupe, D.~Navarro, G.~Parisi,
  S.~Perez-Gaviro, F.~Ricci-Tersenghi, J.~J. Ruiz-Lorenzo, S.~F. Schifano,
  B.~Seoane, A.~Taranc{\'o}n, R.~Tripiccione, and D.~Yllanes, The {Mpemba}
  effect in spin glasses is a persistent memory effect, \emph{Proceedings of
  the National Academy of Sciences}. {\bf 116}\penalty0 (31), \penalty0
  15350--15355  (2019).
\newblock ISSN 0027-8424.
\newblock \doi{10.1073/pnas.1819803116}.
\newblock URL \url{https://www.pnas.org/content/116/31/15350}.

\bibitem{janus:06}
F.~Belletti, F.~Mantovani, G.~Poli, S.~F. Schifano, R.~Tripiccione, I.~Campos,
  A.~Cruz, D.~Navarro, S.~Perez-Gaviro, D.~Sciretti, A.~Tarancon, J.~L.
  Velasco, P.~Tellez, L.~A. Fernandez, V.~Mart\'{i}n-Mayor,
  A.~Mu{\~n}oz~Sudupe, S.~Jimenez, A.~Maiorano, E.~Marinari, and J.~J.
  Ruiz-Lorenzo, Ianus: And adaptive fpga computer, \emph{Computing in Science
  and Engineering}. {\bf 8}, \penalty0 41  (2006).

\bibitem{janus:08}
F.~Belletti, M.~Cotallo, A.~Cruz, L.~A. Fernandez, A.~Gordillo, A.~Maiorano,
  F.~Mantovani, E.~Marinari, V.~Mart\'{i}n-Mayor, A.~Mu{\~n}oz~Sudupe,
  D.~Navarro, S.~Perez-Gaviro, J.~J. Ruiz-Lorenzo, S.~F. Schifano, D.~Sciretti,
  A.~Tarancon, R.~Tripiccione, and J.~L. Velasco, Simulating spin systems on
  {IANUS}, an {FPGA}-based computer, \emph{Comp. Phys. Comm.} {\bf 178},
  \penalty0 208--216  (2008).
\newblock \doi{10.1016/j.cpc.2007.09.006}.

\bibitem{janus:09}
F.~Belletti, M.~Guidetti, A.~Maiorano, F.~Mantovani, S.~F. Schifano,
  R.~Tripiccione, M.~Cotallo, S.~Perez-Gaviro, D.~Sciretti, J.~L. Velasco,
  A.~Cruz, D.~Navarro, A.~Tarancon, L.~A. Fernandez, V.~Mart\'{i}n-Mayor,
  A.~Mu{\~n}oz-Sudupe, D.~Yllanes, A.~Gordillo-Guerrero, J.~J. Ruiz-Lorenzo,
  E.~Marinari, G.~Parisi, M.~Rossi, and G.~Zanier, Janus: An {FPGA}-based
  system for high-performance scientific computing, \emph{Computing in Science
  and Engineering}. {\bf 11}, \penalty0 48  (2009).
\newblock \doi{10.1109/MCSE.2009.11}.

\bibitem{janus:12b}
M.~Baity-Jesi, R.~A. Ba\~{n}os, A.~Cruz, L.~A. Fernandez, J.~M. Gil-Narvion,
  A.~Gordillo-Guerrero, M.~Guidetti, D.~Iniguez, A.~Maiorano, F.~Mantovani,
  E.~Marinari, V.~Mart\'{i}n-Mayor, J.~Monforte-Garcia, A.~Munoz~Sudupe,
  D.~Navarro, G.~Parisi, M.~Pivanti, S.~Perez-Gaviro, F.~Ricci-Tersenghi, J.~J.
  Ruiz-Lorenzo, S.~F. Schifano, B.~Seoane, A.~Tarancon, P.~Tellez,
  R.~Tripiccione, and D.~Yllanes, {Reconfigurable computing for Monte Carlo
  simulations: Results and prospects of the Janus project}, \emph{Eur. Phys. J.
  Special Topics}. {\bf {210}}, \penalty0 {33}  ({AUG}, {2012}).
\newblock \doi{10.1140/epjst/e2012-01636-9}.

\bibitem{janus:14}
M.~Baity-Jesi, R.~A. Ba\~{n}os, A.~Cruz, L.~A. Fernandez, J.~M. Gil-Narvion,
  A.~Gordillo-Guerrero, D.~Iniguez, A.~Maiorano, F.~Mantovani, E.~Marinari,
  V.~Mart\'{i}n-Mayor, J.~Monforte-Garcia, A.~Mu{\~n}oz~Sudupe, D.~Navarro,
  G.~Parisi, S.~Perez-Gaviro, M.~Pivanti, F.~Ricci-Tersenghi, J.~J.
  Ruiz-Lorenzo, S.~F. Schifano, B.~Seoane, A.~Tarancon, R.~Tripiccione, and
  D.~Yllanes, Janus {II}: a new generation application-driven computer for
  spin-system simulations, \emph{Comp. Phys. Comm}. {\bf 185}, \penalty0
  550--559  (2014).
\newblock \doi{10.1016/j.cpc.2013.10.019}.

\end{thebibliography}

\printindex

\end{document}